\documentclass[a4paper,11pt,xcolor=dvipsnames]{article}

\pdfoutput=1

\usepackage{jheppub}
\usepackage{amsmath}
\usepackage{amssymb}
\usepackage{amsthm}
\usepackage{mathrsfs}
\usepackage{xspace}
\usepackage{times}
\usepackage{etoolbox}
\usepackage{braket}
\usepackage{multirow}
\usepackage{slashed}
\usepackage{bm}
\usepackage{mathtools}
\usepackage{xcolor}

\renewcommand{\vec}[1]{\mathbf{#1}}
\newcommand{\C}{\textit{C}\xspace}
\newcommand{\CP}{\textit{CP}\xspace}
\renewcommand{\Re}{\mbox{Re}}
\renewcommand{\Im}{\mbox{Im}}
\newcommand{\Tr}{{\rm Tr}}
\newcommand{\sign}{{\rm sign}}
\newcommand{\adj}{{\rm adj}}
\newcommand{\D}{{\mathrm{d}}}
 
\makeatletter
\patchcmd{\maketitle}{\@fpheader}{}{}{}
\makeatother
  
\preprint{MPP-2015-235; TUM-HEP-1018-15}

\title{Lepton asymmetry from mixing and oscillations}

\author{A. Kartavtsev$^{a}$,}
\emailAdd{alexander.kartavtsev@mpp.mpg.de}

\author{P. Millington$^{b,c}$  and}
\emailAdd{p.millington@nottingham.ac.uk}

\author{H. Vogel$^{a}$}
\emailAdd{hvogel@mpp.mpg.de}

\affiliation[a]{Max-Planck-Institut f\"ur Physik,\\ F\"ohringer Ring 6, 80805 M\"unchen, Germany}
\affiliation[b]{Physik Department T70, Technische Universit\"at M\"unchen,\\ James-Franck-Stra\ss e, 85748 Garching, Germany}
\affiliation[c]{School of Physics and Astronomy, University of Nottingham,\\ Nottingham NG7 2RD, United Kingdom}

\arxivnumber{1601.03086}

\abstract{We show how the two physically-distinct sources of \CP-asymmetry relevant to scenarios of leptogenesis: (i) resonant 
mixing and (ii) oscillations between different flavours can be unambiguously identified 
within the Kadanoff-Baym formalism. These contributions are isolated by analyzing the 
spectral structure of the non-equ\-i\-librium propagators without relying on the definition 
of particle number densities. The mixing source is associated with the usual mass shells, 
whereas the oscillation source is identified with a third intermediate shell. In addition, we 
identify terms lying on the oscillation shell that can be interpreted as the \emph{destructive} 
interference between mixing and oscillation. We confirm that identical shell structure is 
obtained in both the Heisenberg- and interaction-picture realizations of the Kadanoff-Baym 
formalism. In so doing, we illustrate the self-consistency and complementarity of these 
two approaches. The interaction-picture approach in particular has the advantage that 
it may be used to analyze all forms of mass spectra from quasi-degenerate through to 
hierarchical.}

\begin{document}

\maketitle
\flushbottom
\newpage

\section{\label{sec:intro}Introduction}

In the presence of \CP violation, particle mixing and oscillations can provide two physically-distinct sources of \CP-asymmetry. In the quark sector, mixing arises due to the misalignment of the weak and Yukawa eigenbases, which gives rise to the CKM matrix of the Standard Model, whose complex entries provide the \CP violation observed in the $K$, $B$ and $B_s$ systems~\cite{Agashe:2014kda}. Oscillations, on the other hand, arise due to the formation of coherences in populations of particles with the same quantum numbers. These coherences are of particular interest in medium, leading for instance to oscillations via regeneration, as occurs for the $K_0$--$\bar{K}_0$ system in the presence of nuclear matter~\cite{Pais:1955sm}. A similar distinction between mixing and oscillations can be identified in the cascade decays of heavy particles~\cite{Boyanovsky:2006yg,Boyanovsky:2014uya,Boyanovsky:2014lqa,Boyanovsky:2014una}. In extensions of the Standard Model, the physical relevance of these two sources of \CP-
asymmetry has also been identified in the context of leptogenesis (see e.g.~refs.~\cite{Garbrecht:2011aw,Garbrecht:2014aga,Dev:2014laa,Dev:2015wpa,Dev:2014wsa}), where, in certain scenarios, it is acknowledged that both effects must be accounted for in order to obtain quantitatively accurate predictions of the baryon asymmetry of the universe.

Leptogenesis~\cite{Fukugita:1986hr} (for an overview, see e.g.~ref.~\cite{Blanchet:2012bk}) provides a potential explanation for the observed baryon asymmetry of the universe. It relies upon the existence of heavy right-handed Majorana neutrinos, whose out-of-equilibrium decays in the early universe are able to produce a net lepton number. This lepton asymmetry is subsequently converted to the observed baryon asymmetry through the sphaleron processes of the standard electroweak theory~\cite{Kuzmin:1985mm}. Whereas it is widely accepted that the source of \CP-asymmetry provided by the mixing of different heavy-neutrino flavours is important for all mass spectra, the relative importance 
of the source provided by coherent oscillations between populations of heavy-neutrino flavours
is still under debate. Even so, one would anticipate that oscillations are most relevant  when the mass spectrum of the heavy neutrinos is quasi-degenerate, where it has long been recognized that flavour effects play a significant role both from the heavy-neutrino~\cite{Pilaftsis:1997jf,Pilaftsis:1998pd,Ellis:2002eh,Endoh:2003mz,Pilaftsis:2004xx,Pilaftsis:2005rv,Asaka:2005pn,Vives:2005ra,Deppisch:2010fr,Blanchet:2011xq} and charged-lepton sectors~\cite{Barbieri:1999ma,Abada:2006fw,Nardi:2006fx,Abada:2006ea,Blanchet:2006be,Blanchet:2006ch,Pascoli:2006ci,Branco:2006hz,DeSimone:2006nrs}. Thus, one would expect flavour oscillations to provide a significant source of \CP-asymmetry in scenarios of resonant leptogenesis~\cite{Pilaftsis:1997dr,Pilaftsis:1997jf,Pilaftsis:2003gt}. In such models,
heavy-neutrino self-energy effects dominate~\cite{Flanz:1994yx,Covi:1996fm,Covi:1996wh,Buchmuller:1997yu} and provide a resonant enhancement of the leptonic \CP-asymmetry, when the mass difference of at least two of the heavy neutrinos is comparable to their decay widths~\cite{Pilaftsis:1997dr,Pilaftsis:1997jf}.  In this context, it has recently been observed that the mixing and oscillation sources of lepton asymmetry can be of equal magnitude and the same sign in the strong-washout regime~\cite{Dev:2014laa,Dev:2015wpa,Dev:2014wsa} (for a summary, see ref.~\cite{Dev:2015dka}). This leads to a factor of two enhancement in the final lepton asymmetry, when both sources, rather than only one, are included, thereby expanding the viable parameter space for successful leptogenesis. However, it remains an open question as to what extent these two distinct phenomena and the interplay between them are captured by competing approaches.

In order to determine the asymmetry generated in scenarios of leptogenesis, it is necessary to solve systems of transport equations, akin to the classical Boltzmann equations (see e.g.~ref.~\cite{Kolb:1979qa,Luty:1992un}), that describe the time evolution of particle number densities. The impact of flavour oscillations can be accounted for through the quantum improvement of the classical Boltzmann equations  by promoting the number densities of individual flavours to a \emph{matrix of densities}~\cite{Sigl:1992fn}, thereby allowing for flavour coherences. This approach yields the so-called \emph{density matrix formalism}, which has been applied extensively to scenarios of leptogenesis~\cite{Akhmedov:1998qx,Asaka:2005pn,Abada:2006ea,Shaposhnikov:2008pf,Gagnon:2010kt,Asaka:2011wq,Blanchet:2011xq,Canetti:2012kh,Shuve:2014zua,Dev:2014laa, Dev:2015wpa}. On the other hand, a semi-classical treatment of mixing is possible through the inclusion of effective Yukawa couplings~\cite{Pilaftsis:1997jf,
Pilaftsis:2003gt}, which can account for the $\varepsilon$- and $\varepsilon'$-type \CP violation, arising respectively from self-energy and vertex effects. Recently, there has been much progress in the literature~\cite{Buchmuller:2000nd,Prokopec:2003pj,Prokopec:2004ic,DeSimone:2007rw,DeSimone:2007pa,
Cirigliano:2007hb,Anisimov:2008dz,Garny:2009rv,Garny:2009qn,Cirigliano:2009yt,Anisimov:2010aq,Garny:2010nj,Beneke:2010wd,Beneke:2010dz,Garbrecht:2010sz,Anisimov:2010dk,Garbrecht:2011aw,Garny:2011hg,Drewes:2012ma,Garbrecht:2012pq,Frossard:2012pc,Drewes:2013gca,Garbrecht:2013iga,Hohenegger:2014cpa,Iso:2013lba,Iso:2014afa,Hohenegger:2014cpa,Garbrecht:2014aga,Dev:2014wsa} aiming to go beyond these semi-classical treatments and obtain `first-principles' field-theoretic analogues of the Boltzmann equation. Often, these quantum transport equations are derived by means of the Kadanoff-Baym (KB) formalism~\cite{Baym:1961zz,KB:1962} (for reviews, see~refs.~\cite{Blaizot:2001nr,Berges:2004yj}), itself embedded in the Schwinger-Keldysh~\cite{Schwinger:1960qe,Keldysh:1964ud} closed-time path formalism (see also refs.~\cite{Jordan:1986ug,Calzetta:1986cq,Calzetta:1986ey}) of non-equilibrium thermal field theory. These approaches have the advantage that all quantum-mechanical effects are in principle accounted for 
consistently 
and systematically. However, an outstanding difficulty of such approaches is in the approximations needed to make the solution tractable and to extract physically-meaningful observables. As a result, it is often not straightforward to compare directly the results of different analyses or to ascertain to what extent relevant physical effects are accounted for.

In this article, we illustrate how the mixing and oscillation sources of lepton asymmetry can be
identified unambiguously in the Kadanoff-Baym formalism by means of the spectral structure of the resummed heavy-neutrino propagators and independently of the definition of particle number densities. In the context of a toy two-flavour model, we will show that this spectral structure contains \emph{three} distinct shells: two of these shells correspond to mixing and can be associated 
with the quasi-particle mass shells, whereas the third, which can be identified with 
oscillations, lies at an intermediate energy. In addition, we identify terms lying on the 
oscillation shell that can be interpreted as
the interference between oscillation and mixing. 
In so doing, we provide a further illustration of the interplay of these two effects 
in the generation of lepton asymmetry. Most significantly, we find that this interference is 
destructive. With respect to the ``benchmark'' of the Boltzmann approximation (effective Yukawa couplings but flavour-diagonal number densities), this destructive interference can be viewed as a suppression of the oscillation source. Conversely, with respect to the ``benchmark'' of the density matrix approximation (tree-level Yukawa couplings but flavour-off-diagonal number densities), this destructive interference can be viewed as a suppression of the mixing source. This observation may in part
account for apparent discrepancies between competing approaches and is anticipated to be 
of relevance to scenarios of resonant leptogenesis. Nevertheless, in spite of this destructive interference and in the weak-washout regime, we find that the oscillation and mixing sources 
can be of equal magnitude and contribute additively to the final asymmetry  in agreement with the conclusions of refs.~\cite{Dev:2014laa, Dev:2015wpa, Dev:2014wsa}.

Aside from illustrating the interplay of these sources of \CP-asymmetry, we perform the 
calculations in two very different approaches, namely the Heisenberg- and interaction-picture
realizations of non-equilibrium quantum field theory. In contrast to earlier approaches, the interaction-picture description introduced in ref.~\cite{Millington:2012pf} (see ref.~\cite{Millington:2013isa} for a summary) enables one to proceed in a perturbative loop-wise fashion without encountering so-called pinch singularities~\cite{Weldon:1991ek,Altherr:1994fx,Altherr:1994jc,Bedaque:1994di,Dadic:1998yd,Greiner:1998ri,Garbrecht:2011xw} or secular terms~\cite{Berges:2004yj} thought previously to spoil such approaches to non-equilibrium field theory. Quite remarkably, we find exact 
agreement between these two formulations, illustrating the self-consistency and complementarity of these two approaches. Working in the interaction picture has the particular advantage that all forms of mass spectra can be analyzed using a single method, from quasi-degenerate through to 
hierarchical.

In order to reduce the technical complications to a minimum and yet to include 
all qualitatively important effects  for the generation of the asymmetry, we 
consider a simple toy model studied previously in refs.~\cite{Garny:2009rv,Garny:2009qn,
Garny:2010nj,Garny:2010nz,Hohenegger:2013zia,Hohenegger:2014cpa}. The model 
contains one complex ($b$) and two real scalar fields ($\psi_i$):
\begin{align}
   \label{Lagrangian}
   {\cal L} \ =\ \frac12\partial^\mu\psi_{i}\,\partial_\mu\psi_{i}
             - \frac12\psi_{i} M^2_{ij} \psi_{j} 
             + \partial^\mu \bar{b}\,\partial_\mu b
             -m^2\, \bar{b} b 
             - \frac{\lambda}{2!2!}(\bar b b)^2
             - \frac{h_{i}}{2!}\psi_{i} bb 
             - \frac{h^*_{i}}{2!}\psi_{i}\bar{b}\bar{b}\,,
\end{align}
where $\bar{b}$ denotes the Hermitian conjugate of $b$.
Here and in the following, we assume summation over repeated 
indices, unless otherwise specified.
Despite its simplicity, the model incorporates all features relevant 
for leptogenesis. The real scalar fields $\psi_i$ imitate the (two lightest) 
heavy right-handed neutrinos, whereas the complex scalar field $b$ models 
the leptons. The $U(1)$ symmetry, which we use to define ``lepton'' number, 
is explicitly broken by the presence of the last two terms, just as the 
$B-L$ symmetry is explicitly broken by Majorana mass terms in phenomenological 
models. Thus, the first Sakharov condition~\cite{Sakharov:1967dj} is fulfilled. The couplings $h_i$ 
model the complex Yukawa couplings of the right-handed neutrinos to the charged-leptons 
and the Higgs doublet. By rephasing the complex scalar field, at least one of the 
couplings $h_i$ can be made real. If ${\rm arg}(h_1)\neq {\rm arg}(h_2)$,
the other one remains complex and there is \C-violation, as is required 
by the second Sakharov condition.  

The remainder of this article is organized as follows. Using the Heisenberg-picture 
realization  of the Kadanoff-Baym formalism, we 
confirm in section~\ref{sec:heisenberg} that the mixing and oscillation between different flavours indeed provide 
two distinct sources of lepton asymmetry. In section~\ref{sec:interaction}, 
we repeat the analysis in the interaction picture, finding identical results. Subsequently, in section~\ref{sec:densitymatrix}, we make comparison with the density matrix approach and, in section~\ref{sec:vacuumcp}, describe the inclusion of mixing effects via effective Yukawa couplings. Finally,  in section~\ref{sec:pheno}, we 
discuss the phenomenological implications of these two sources of lepton asymmetry, as well as their interference,
and present numerical results. Our conclusions are presented in section~\ref{sec:conc}. 
In appendix~\ref{sec:noneq}, we provide a brief outline of the details of the 
Kadanoff-Baym formalism pertinent to the analysis of this article. In addition, 
we summarize our definitions and notational conventions, making comparison with those that 
appear in the literature. In appendix~\ref{sec:syms}, we discuss the transformation 
properties of the model under generalized discrete symmetries and emphasize the need to specify  \C-symmetric initial conditions in the weak-washout  regime. 
A derivation of the rate equations in an expanding universe, relevant to the study of leptogenesis in the strong-washout regime, is presented in 
appendix~\ref{sec:rate}.

\section{\label{sec:heisenberg}Shell structure for two-particle mixing in the Heisenberg picture}

In this section, we show that the \emph{mixing}  and  
\emph{oscillation between different flavours} provide two distinct sources 
of lepton 
asymmetry, in agreement with  arguments presented in refs.~\cite{Dev:2014laa,Dev:2015wpa,Dev:2014wsa}.
Whereas the standard mixing contributions~\cite{Pilaftsis:1997dr,Pilaftsis:2003gt} are 
associated with the mass shells $\omega_i$ ($i=1,2$) of the corresponding quasi-particles, the 
oscillation contribution is associated with an intermediate shell at $\bar{\omega}=(\omega_1+\omega_2)/2$, 
which we will refer to as the \emph{oscillation shell} in the remainder of this paper. In order to identify this structure, we first analyze the
generation of the lepton asymmetry using the
Kadanoff-Baym equations in the Heisenberg picture, as were previously applied to the toy model from eq.~\eqref{Lagrangian} in ref.~\cite{Hohenegger:2014cpa}.

\paragraph{Asymmetry in the absence of washout.}
Following refs.~\cite{Garny:2011hg, Hohenegger:2014cpa}, we assume that the complex scalar 
field forms a thermal bath with temperature $T$ and neglect the back-reaction. The system begins its 
evolution at $t=-\,\infty$ in an equilibrium state. At $t=0$, the real scalars are 
brought out of equilibrium by an external source, thereby fulfilling the third Sakharov condition. This 
leads to the production of an asymmetry between the number densities of $b$ and $\bar b$. As time goes by, 
this asymmetry is erased by washout processes. Finally, at $t=\infty$, the system again reaches thermal 
equilibrium.

The expression for the produced asymmetry can be derived by considering the volume integral of the conserved 
Noether current:
\begin{align}
\label{Current}
J_\mu(x)\ &=\ \langle\, j_\mu(x) \rangle  
\ =\ i\lim_{y\rightarrow x}\bigl[\partial_{\mu_x} S_<(x,y)-\partial_{\mu_y} S_>(x,y)\bigr]\,,
\end{align}
where $S_>(x,y)\equiv \braket{b(x)\bar{b}(y)}$ and $S_<(x,y)\equiv \braket{\bar{b}(y)b(x)}$ are the Wightman propagators of the complex scalar field. Using the Kadanoff-Baym equations for $S_\gtrless$, which are similar to those that we will discuss below for the real scalar fields [see eq.~\eqref{KBeqs_real}], and taking the equal-time limit $x_0=y_0=t$, we obtain the kinetic equation for the produced asymmetry, which takes 
into account quantum corrections to both the source and washout terms. Details of the derivation, together with 
the discussion of the approximations used, can be found in ref.~\cite{Hohenegger:2014cpa}.

Washout processes are physically very important and must be taken into account in a 
phenomenological analysis. On the other hand, in the analysis limited to the source term 
alone, one can neglect them, as was previously done in refs.~\cite{Garny:2011hg,Hohenegger:2014cpa}. 
In this approximation, the produced asymmetry is given by  
\begin{align}
\label{SourceTerm}
\eta(t) \ \equiv\ \int\D^3\mathbf{x}\;\braket{j_0(t,\mathbf{x})}\ &=\  -\,2\,\Im H_{12}\int\limits^{t}_{-\infty}  \!\D x^0 \int\limits^{t}_{-\infty}  \!\D y^0
\int_{\mathbf{q}} \nonumber\\
&\times\ i\Bigl[G^{12}_<(x^0,y^0,\vec{q})\,\Pi_>(y^0,x^0,\vec{q})\:-\:
G^{12}_>(x^0,y^0,\vec{q})\,\Pi_<(y^0,x^0,\vec{q})\Bigr]\,, 
\end{align} 
where  $H_{ij}\equiv h_i h^*_j$, the functions $G^{12}_{\gtrless}$ are 
components of the Wightman propagators of the real scalar fields $\psi_i$, 
and $\Pi_\gtrless$ are the self-energies with the couplings ($h_i$) ``amputated.'' We use the shorthand notation $\int_{\mathbf{q}}\equiv\int\!\frac{\mathrm{d}^3\mathbf{q}}{(2\pi)^3}$. A comprehensive summary 
of the various propagators and self-energies, their definitions and useful identities, 
as well as the differing nomenclature used throughout the literature is provided in 
appendix~\ref{sec:noneq}. The expression for the asymmetry in eq.~\eqref{SourceTerm} is entirely equivalent to the one obtained 
via the definition of particle number densities used in the interaction-picture approach to the 
Kadanoff-Baym formalism, developed in ref.~\cite{Millington:2012pf}.  In 
section~\ref{sec:pheno}, we present numerical solutions of eq.~\eqref{SourceTerm} for 
\C-symmetric initial conditions.

\paragraph{Solution of the Kadanoff-Baym equations.} The Wightman
propagators in eq.~\eqref{SourceTerm} are solutions to the Kadanoff-Baym equations 
for the mixing fields $\psi_i$. In the absence of external sources, these 
transport equations read \cite{Garny:2009qn} 
\begin{align}
\label{KBeqs_real}
      [\square_x\delta_{ik}+M_{ik}^2]G_{\gtrless}^{kj}(x,y) &\ =\
      \int\limits^{y^0}_{-\infty} \!\D^4z\;
      \,\Pi_{\gtrless}^{ik}(x,z)\,G_{\rho}^{kj}(z,y)
      - \int\limits^{x^0}_{-\infty} \!\D^4z\;
      \,\Pi_{\rho}^{ik}(x,z)\,G_{\gtrless}^{kj}(z,y)\,,
\end{align}
where $M_{ij}$ are the mass parameters of the renormalized Lagrangian, $G^{ij}_\rho$ is the 
spectral function, and  $\Pi^{ij}_{\gtrless}$ and $\Pi^{ij}_\rho$ are the Wightman and spectral 
self-energies, respectively. Using the definitions of the retarded and advanced propagators,
and the self-energies in appendix~\ref{sec:noneq}, we can rewrite eq.~\eqref{KBeqs_real} in 
a form more convenient for the analysis that follows:
\begin{align}
\label{GFrhoeq}
 [\square_x\delta_{ik}+M_{ik}^2]G_{\gtrless}^{kj}(x,y) \ = \  &  -\int_z\;
\Bigl[\Pi_{R}^{ik}(x,z)\,G_{\gtrless}^{kj}(z,y)\:
     +\: \Pi_{\gtrless}^{ik}(x,z)\,G_{A}^{kj}(z,y)\Bigr]\,.
\end{align}
We have also made use of the fact that
\begin{equation}
\int_{-\infty}^{x_0}\mathrm{d}z_0\int\mathrm{d}^3\mathbf{z}\ =\ \int_{-\infty}^{+\infty}\mathrm{d}z_0\int\mathrm{d}^3\mathbf{z}\;\theta(x_0-z_0)\ \equiv\ \int_z \theta(x_0-z_0)\;.
\end{equation}
The Kadanoff-Baym equations for the retarded and advanced propagators can be derived 
from eq.~\eqref{KBeqs_real}:
\begin{align}
   \label{GRAQK}
   [\square_x\delta_{ik}+M_{ik}^2] & G_{R(A)}^{kj}(x,y)\ =\ 
   -\int_z\; \Pi_{R(A)}^{ik}(x,z)\,G_{R(A)}^{kj}(z,y)\:+\:\delta^{4}(x-y)\delta^{ij}\,.
\end{align}

For our purposes, 
it is sufficient to know that, at the one-loop level to 
which we limit ourselves here, the self-energies of the real scalar fields are translationally 
invariant in the thermal bath. This implies, in particular, that eq.~\eqref{GRAQK} admits a 
translationally-invariant solution. Using eq.~\eqref{GRAQK}, one can readily check that 
\begin{align}
\label{GFrhoEqSol}
G_{\gtrless}^{ij}(x,y)\ =\ -\int_{u,v}\; G_R^{im}(x,u)\,\Pi_{\gtrless}^{mn}(u,v)\,G_A^{nj}(v,y)
\end{align}
is a solution to eq.~\eqref{GFrhoeq}. Since, as is discussed above, the self-energies, as well as
the retarded and advanced propagators on the rhs of eq.~\eqref{GFrhoEqSol} are 
translationally invariant, the lhs of eq.~\eqref{GFrhoEqSol} is also translationally 
invariant. In other words, eq.~\eqref{GFrhoEqSol} is an equilibrium solution for the Wightman 
propagators. 

In the setup considered here, the system is assumed to be brought out of equilibrium instantaneously by an external 
source at $t=0$. While it is hard to imagine a physically-motivated scenario that would 
generate such an initial condition, this assumption will allow us to solve the equations 
analytically and access qualitative features of the solution important also for phenomenologically-viable initial conditions.  The source can be considered as a bi-local contribution to the self-energy. 
Following refs.~\cite{Garny:2011hg,Hohenegger:2014cpa}, we consider an external source that leaves 
the spectral function unperturbed. Thus, both of the Wightman self-energies are ``perturbed'' in the same 
way, $\Pi^{mn}_{\gtrless}(u,v) \rightarrow \Pi^{mn}_{\gtrless}(u,v)-K^{mn}(u,v)$, with  
$K^{mn}(u,v)=\delta(u^0)\delta(v^0)\mathcal{K}^{mn}(\vec{u}-\vec{v})$. The translational invariance 
of the one-loop self-energies in the thermal bath renders the Kadanoff-Baym equations linear, i.e.~a
sum of two solutions is also a solution. Using this linearity, we obtain the following equation 
for the non-equilibrium part  $G_{\!\delta\gtrless}^{kj}\subset G^{kj}_{\gtrless}$ of the Wightman propagators induced by the external source:
\begin{align}
\label{GFrhoeqSource}
 [\square_x\delta_{ik}+M_{ik}^2]G_{\!\delta\gtrless}^{kj}(x,y)\  =\ &  -\int_z\;
\Bigl[\Pi_{R}^{ik}(x,z)\,G_{\!\delta\gtrless}^{kj}(z,y)\:-\:K^{ik}(x,z)\,G_{A}^{kj}(z,y)\Bigr]\,.
\end{align}
Using eq.~\eqref{GRAQK}, one can readily check that \cite{Garny:2011hg,Hohenegger:2014cpa} 
\begin{align}
\label{DevFromEq}
G^{ij}_{\!\delta\gtrless}(x,y)\ =\ \int_{u,v}\; G_R^{im}(x,u)\,K^{mn}(u,v)\,G_A^{nj}(v,y)
\end{align}
is a solution to eq.~\eqref{GFrhoeqSource}. In the absence of spacetime expansion, we are only interested in 
the non-equilibrium part of the resummed Wightman propagators, which are common to both the 
positive- and negative-frequency components, $G^{ij}_{\!\delta>}=G^{ij}_{\!\delta<}\equiv 
G^{ij}_{\!\delta}$. The sum of the equilibrium and non-equilibrium parts [eqs.~\eqref{GFrhoEqSol} and 
\eqref{DevFromEq}] gives the full solution of the Kadanoff-Baym equations in the thermal bath, 
as was studied in detail in ref.~\cite{Hohenegger:2014cpa}.
 
\paragraph{Shell structure of the non-equilibrium solution.}
The equilibrium solution in eq.~\eqref{GFrhoEqSol}  does not contribute to the asymmetry 
in agreement with the third Sakharov condition  (see ref.~\cite{Hohenegger:2014cpa}
for more details) and will not be considered further.
In order to unravel the shell structure of the non-equili\-brium solution in eq. \eqref{DevFromEq},
we perform a Wigner trans\-form (see appendix~\ref{sec:noneq}). Using the relation between 
the double-momentum and Wigner representations [see eq.~\eqref{WignerFromDouble}] and neglecting 
sub-leading off-shell contributions (see e.g.~section 6 of ref.~\cite{Hohenegger:2014cpa}), we obtain
\begin{align}
\label{GKWignerTrafo}
G^{ij}_{\!\delta}(t,q_0>0)\ &\approx\ \int\limits_0^\infty\frac{\D p_0}{2\pi}\int\limits_0^\infty\frac{\D p'_0}{2\pi}
\,2\pi\,\delta\!\left(q_0-\textstyle{\frac12}(p_0+p'_0)\right)\,e^{-i(p_0-p'_0)t}\, G^{im}_R(p_0)\,\mathcal{K}^{mn}\,G^{nj}_A(p'_0)\;.
\end{align}
We will later analytically continue the real variable $q_0$ to the complex plane in order to apply Cauchy's theorem. The notation $q_0>0$ is therefore understood throughout this article to mean $\mathrm{Re}\,q_0>0$.
In addition, we omit all dependence on the common three-momentum 
$\vec{q}$ when no ambiguity results. The explicit forms 
of the Wigner transforms of  
the retarded and advanced propagators can be inferred from eq.~\eqref{GRAQK} using 
translational-invariance of the self-energies:
\begin{align}
\label{GRWT}
G^{ij}_{R(A)}(q_0)\ =\ -\,\frac{\adj D_{R(A)}^{ij}(q_0)}{\det \bm{D}_{R(A)}(q_0)} \,,
\end{align}
where 
\begin{align}
\label{DRdef}
D^{ij}_{R(A)}(q_0)\ \equiv\ q^2\delta^{ij}-M^2_{ij}-\Pi^{ij}_{R(A)}(q_0)\,,
\end{align}
and we use boldface for matrices in flavour space. Having not needed to employ the gradient expansion (cf.~refs.~\cite{Fidler:2011yq,Garbrecht:2011xw}),
the leading self-energy corrections to the spectral structure of the non-equilibrium part of the propagator, 
specifically the shifts of the poles in the real and imaginary directions, have been taken into account.

The imaginary parts of the retarded (advanced) self-energies are odd under $q_0\to -\,q_0$, i.e.~$\mathrm{Im}\,\Pi_{R(A)}^{ij}(q_0)=-\,\mathrm{Im}\,\Pi^{ij}_{R(A)}(-\,q_0)$, such that all four poles of $G^{ij}_{R(A)}(q_0)$ lie in the lower-half complex plane. These four poles correspond to the zeros of $\mathrm{det}\,\bm{D}_R(q_0)$ and lie at $q_0=\Omega_i$ and $q_0=-\,\Omega_i^*$, where
\begin{align}
\label{eq:omegaidef}
\Omega_{i}\ =\ \omega_i\:-\:\frac{i}{2}\,\Gamma_i\;.
\end{align}
The real and imaginary parts of $\Omega_i$ correspond to the in-medium frequency $\omega_i$ and width $\Gamma_i$, respectively. In the neighbourhood of the poles with $q_0>0$, we can approximate~\cite{Hohenegger:2014cpa}
\begin{align}
\label{DRapprox}
\mathrm{det}\,\bm{D}_R(q_0>0)\ \approx\ (q_0^2-\Omega^2_1)(q_0^2-\Omega^2_2)\;,
\end{align}
where it is assumed that the self-energies are slowly varying functions of $q_0$ for $q_0\sim \Omega_{i}$.
We would like to emphasize that eq.~\eqref{DRapprox} is not only applicable, but actually 
becomes exact in the degenerate limit. The implications of this pole approximation for the effective regulator of the lepton asymmetry
will be discussed later in the context of degeneracy symmetry limits (see e.g. ref.~\cite{Dev:2014laa}).  Instead, if we were interested in the poles with $q_0<0$, we could approximate
\begin{align}
\label{DRapprox2}
\mathrm{det}\,\bm{D}_R(q_0<0)\ \approx\ (q_0^2-\Omega^{*2}_1)(q_0^2-\Omega^{*2}_2)\;.
\end{align}

Using Cauchy's theorem to evaluate the integral in eq.~\eqref{GKWignerTrafo} approximately,  we arrive at the advertised
\emph{three-shell} structure
\begin{align}
\label{Gdeltaexpansion}
G^{ij}_{\!\delta}(t,q_0>0)\ &\approx\  
\frac{1}{|\Delta\Omega^2|^2}\,\Biggl[\;
\sum_{k\,=\,1}^{2} 2\pi\,\delta(q_0-\omega_{k}) e^{-\Gamma_{k} t}
\,\frac{\adj D_R^{im}(\omega_{k})}{2\omega_{k}}\,\mathcal{K}^{mn}\,
\frac{\adj D_A^{nj}(\omega_{k})}{2\omega_{k}}\nonumber\\
&-\ 2\pi\,\delta(q_0-\bar\omega)e^{-i(\omega_1-\omega_2)t} e^{-\bar\Gamma t}
\,\frac{\adj D_R^{im}(\omega_1)}{2\omega_1}\,\mathcal{K}^{mn}\,
\frac{\adj D_A^{nj}(\omega_2)}{2\omega_2}\nonumber\\
&-\ 2\pi\,\delta(q_0-\bar\omega)e^{-i(\omega_2-\omega_1) t} e^{-\bar\Gamma t}
\,\frac{\adj D_R^{im}(\omega_2)}{2\omega_2}\,\mathcal{K}^{mn}\,
\frac{\adj D_A^{nj}(\omega_1)}{2\omega_1}\Biggr]\,,
\end{align}
where we have defined the average in-medium decay width 
\begin{align}
\bar\Gamma\ =\ \frac12(\Gamma_1+\Gamma_2)\,,
\end{align}
and introduced
\begin{align}
\label{DOmega}
\Delta \Omega^2\ \equiv\ \Omega^2_2\:-\:\Omega^2_1\,.
\end{align}
In eq.~\eqref{Gdeltaexpansion}, the three distinct shells are identified by the frequencies 
$q_0=\omega_i$ ($i=1,2$) and $q_0=\bar\omega\equiv \frac12(\omega_1+\omega_2)$. The shells with frequencies
$q_0=\omega_i$ lie at the two poles of the retarded propagator, which can be associated with 
quasi-particle degrees of freedom. As such, these terms correspond to the contribution from 
\emph{mixing}. On the other hand, the additional intermediate shell with frequency 
$q_0=\bar \omega$ corresponds to the contribution from \emph{oscillations} 
and, as we will see, the \emph{interference} between mixing and oscillations. 
This three-shell structure matches that obtained in~ref.~\cite{Fidler:2011yq}, which makes use 
of a gradient expansion of the KB equations. Therein, the authors also find an additional fourth 
shell with frequency $q_0=\omega_{1}-\omega_{2}$  corresponding to particle-anti-particle 
coherences, which are not considered in the present analysis.

In order to gain a better understanding of the shell structure and to make comparisons 
with the existing literature, we will now consider the non-equilibrium part of the propagator [eq.~\eqref{Gdeltaexpansion}] to leading order 
in powers of the self-energies. Specifically, we will neglect terms higher than first-order in the self-energies in the products of adjugate matrices in eq.~\eqref{Gdeltaexpansion}. With regards to the lepton asymmetry, we are only interested in the 
off-diagonal components of this non-equilibrium part of the propagator. In the mass eigenbasis, 
in which the remainder of this article is understood, these components read
\begin{align}
\label{GF12Expansion}
G^{i\slashed{i}}_{\!\delta}(t,q_0 >0) \ &\approx\
2\pi\,\delta(q_0-\omega_i)\,\frac{1}{2\omega_i}\,e^{-\Gamma_it}\,\delta n^{ii}(0)\,
\Pi_A^{i\slashed{i}}(\omega_i)R_{i\slashed{i}}\nonumber\\
&-\ 2\pi\,\delta(q_0-\omega_\slashed{i})\,\frac{1}{2\omega_\slashed{i}}
\,e^{-\Gamma_\slashed{i}t}\,\delta n^{\slashed{i}\slashed{i}}(0)
\,\Pi_R^{i\slashed{i}}(\omega_\slashed{i})R_{i\slashed{i}}\nonumber\\
&+\ 2\pi\,\delta(q_0-\bar\omega)\,\frac{1}{(2\omega_i)^\frac12(2\omega_\slashed{i})^\frac12}\, 
e^{-i(\omega_i-\omega_\slashed{i})t}e^{-\bar\Gamma t}
\Bigl[\delta n^{i\slashed{i}}(0)\,\Delta M_{i\slashed{i}}^2\nonumber\\
&-\ \delta n^{ii}(0)\Pi_A^{i\slashed{i}}(\omega_\slashed{i})+
\delta n^{\slashed{i}\slashed{i}}(0)\Pi_R^{i\slashed{i}}(\omega_i)\Bigr]R_{i\slashed{i}}\,.
\end{align}
In eq.~\eqref{GF12Expansion},
$\Delta M^2_{i\slashed{i}}\ \equiv\ M^2_i\:-\:M_\slashed{i}^2$ is  the mass splitting, and 
we have employed the notation used in ref.~\cite{Dev:2014wsa}:
\begin{equation}
\slashed{i}\ \equiv\ \begin{cases}2\;,\qquad &i\ = \ 1\\ 1\;,\qquad &i\ = \ 2\;.\\ \end{cases}
\end{equation}
In addition, for later convenience, we have introduced the following notation for the initial 
deviation of ``particle number densities'' from equilibrium:
\begin{align}
\label{nijdef}
\delta n^{ij}(0)\ & \equiv\ \frac{\mathcal{K}^{ij}}{(2\omega_i)^\frac12(2\omega_j)^\frac12}\,.
\end{align}
Note, however, that the identification of the mixing and oscillation shells in eq.~\eqref{Gdeltaexpansion} is 
independent of this definition. Finally,
\begin{equation}
\label{RegulatorDef}
R_{i\slashed{i}}\ \equiv\ \frac{\Delta M^2_{i\slashed{i}}}{(\Delta M^2_{i\slashed{i}})^2
+(\omega_i\Gamma_i-\omega_\slashed{i}\Gamma_\slashed{i})^2}\; 
\end{equation}
is the  effective regulator. 

The first and second lines of eq.~\eqref{GF12Expansion} live on the mass shells 
and  describe the standard \emph{mixing} contributions to the 
asymmetry. On the other hand, the third and fourth lines describe the contribution 
from oscillations and the interference between mixing and oscillations. 
In section~\ref{sec:interaction}, we will make use of the interaction-picture approach in order to isolate
the interference terms from the pure oscillation terms.
We note that the regulator in eq.~\eqref{RegulatorDef} cannot be applied naively in the 
doubly-degenerate limit $M_2\rightarrow M_1$ and $\Gamma_2\rightarrow \Gamma_1$ (for a comparison 
of various regulators in degeneracy symmetry limits, see e.g.~refs.~\cite{Dev:2014laa,Garny:2011hg}). 
Nevertheless, the last two terms of eq.~\eqref{GF12Expansion} have structure similar to 
those of the $i$-th and  $\slashed{i}$-th  mass shell terms, respectively, but with opposite signs. Therefore, 
there is a partial cancellation of these contributions, an effect that becomes important in the 
maximally-resonant regime, where the interference between mixing and oscillations is 
anticipated to be  of most relevance. This cancellation has been analyzed in detail in 
refs.~\cite{Garny:2011hg,Hohenegger:2014cpa}, where it was demonstrated 
that, in the degenerate limit $\omega_2\rightarrow \omega_1$ and $\Gamma_2\rightarrow \Gamma_1$, 
back-reaction of mixing on the oscillation ensures exact cancellation in agreement with 
the physical expectations.

\paragraph{Mixing and oscillation sources of CP asymmetry.}
It remains for us to study how each term in the non-equilibrium propagator [eq.~\eqref{GF12Expansion}] contributes to the asymmetry 
by substituting it into the source term [eq.~\eqref{SourceTerm}].  As identified earlier, in the absence of cosmological expansion, we 
are only interested in  the non-equilibrium part of the resummed Wightman propagators for which
$G^{ij}_{\!\delta>}=G^{ij}_{\!\delta<}=G^{ij}_{\delta}$. In this case, the expression for the produced asymmetry [eq.~\eqref{SourceTerm}]
simplifies to 
\begin{align}
\label{SourceTerm1}
\eta(t)  \ &=\ -\,2\,\Im H_{12}\int\limits^{t}_{0}  \D x^0\! \int\limits^{t}_{0}  \D y^0\!
\int_{\vec q}\, G^{12}_\delta(x^0,y^0,\vec{q})\, \Pi_\rho(y^0,x^0,\vec{q})\,, 
\end{align} 
where we have taken into account that the system is brought out of equilibrium at $t=0$ in the lower limits of the time integration.
Next, we trade $x^0$ and $y^0$ for the central and relative coordinates $t\equiv 
\frac12(x^0+y^0)$ and $R^0\equiv x^0-y^0$. In addition, we use the Markovian approximation
\begin{align}
\label{eq:Markovian}
\int\limits^{2t}_{-2t} \D R^0\;\sin(R^0 q_0)\,\cos(R^0 p_0)\ =\ 0\,,\quad 
\int\limits^{2t}_{-2t} \D R^0\;\sin(R^0 q_0)\,\sin(R^0 p_0) \ \approx\ \pi\,\delta(q_0-p_0)\,.
\end{align}
In this way, we may rewrite eq.~\eqref{SourceTerm1} in the differential form 
\begin{align}
\label{qt}
\frac{\D \eta}{\D t}\ =\ 4\,\Im\, H_{12}\,
\int_{q_0,\,\vec{q}}\theta(q_0)\, \Im\, G^{12}_\delta(t,q_0,\vec{q}) \, \widetilde{\Pi}_\rho(q_0,\vec{q})\,,
\end{align}
where we have restored the common momentum $\vec q$,
\begin{align}
\label{PirhoDef}
\widetilde{\Pi}_\rho(q_0,\vec{q})\ \equiv\ -\,i\Pi_{\rho}(q_0,\mathbf{q}) \ =\ \frac{1}{8\pi}\, L_\rho(q_0,\vec{q})\;,
\end{align}
(see eq.~\ref{eq:pitilde} and appendix~\ref{sec:noneq}) and, in the $\overline{\mathrm{MS}}$ scheme,
\begin{align}
L_\rho(q_0,\vec{q})\ =\ 1\:+\:\frac{2T}{|\vec{q}|}\ln\biggl(\frac{1-e^{-(q_0+|\vec{q}|)/2T}}{1-e^{-(q_0-|\vec{q}|)/2T}}\biggr)
\end{align}
(see ref.~\cite{Hohenegger:2014cpa} for more details). Substituting the expression for $G^{12}_{\delta}$
from  eq.~\eqref{GF12Expansion} into eq.~\eqref{qt}, we obtain the following expression
for the time-derivative of the asymmetry:
\begin{align}
\label{dqdtapprox}
\frac{\D \eta}{\D t} \ & \approx\  2 \sum_i \int_{\vec q}\;\frac{M_i}{\omega_i} \, e^{-\Gamma_it}\, \delta n^{ii}(0,\vec{q}) 
\, \Gamma^{\rm med}_i(\omega_i,\vec{q})\, \epsilon_i^{\rm med}(\omega_i,\vec{q})\nonumber\\
&+\ 2\,\Im\,H_{12}\, \Im\,\int_{\vec q}\frac{1}{(\omega_1\omega_2)^\frac12}\, e^{-i(\omega_1-\omega_2)t}e^{-\bar\Gamma t}\,
\widetilde{\Pi}_\rho(\bar\omega,\mathbf{q})
\Bigl[\delta n^{12}(0,\vec{q})\,\Delta M^2_{12}\nonumber\\
&-\delta n^{11}(0,\vec{q})\Pi_A^{12}(\omega_2,\vec{q})+\delta n^{22}(0,\vec{q})\Pi_R^{12}(\omega_1,\vec{q})\Bigr] R_{12}\,.
\end{align}
The first line of eq.~\eqref{dqdtapprox} originates from the mass shell terms of 
eq.~\eqref{GF12Expansion} and describes the \emph{mixing source} of lepton asymmetry.  
The second and third lines stem from the oscillation-shell terms of 
eq.~\eqref{GF12Expansion} and contain the \emph{oscillation source} and the \emph{interference} between mixing and oscillations,
which will be isolated in section~\ref{sec:interaction}.
 
Before concluding this section, we comment in more detail on the physical content of 
eq.~\eqref{dqdtapprox}.  Firstly, the overall factor of two arises because, in the toy 
model [eq.~\eqref{Lagrangian}], each decay of the heavy real scalar violates 
``lepton'' number by two units. Secondly, we note that the mixing contribution has the 
standard structure \cite{Pilaftsis:1997dr,Pilaftsis:2003gt}.  
The asymmetry produced per unit time and unit volume is proportional to the departure from 
equilibrium $\delta n^{ii}$, the in-medium decay probability $\Gamma^{\rm med}_i(\omega_i,\vec{q})= 
\Gamma_i\,L_\rho(\omega_i,\vec{q})$ and the in-medium asymmetry produced in each decay
\begin{align}
\label{EpsilonMed}
\epsilon_i^{\rm med}\ =\ \Im\biggl(\frac{H_{i\slashed{i}}}{H^*_{i\slashed{i}}}\biggr)
\frac{(M_i^2-M_\slashed{i}^2)M_\slashed{i}\Gamma_\slashed{i}}{(M_i^2-M_\slashed{i}^2)^2
+(\omega_i\Gamma_i-\omega_\slashed{i}\Gamma_\slashed{i})^2}\,L_\rho(\omega_i,\vec{q})\,,
\end{align}
where the function $L_\rho(\omega_i,\vec{q})$ takes into account quantum-statistical 
corrections to the decay width and \C-violating parameter, respectively (see eq.~\eqref{PirhoDef} and refs.~\cite{Garny:2009rv,
Garny:2009qn}). Thirdly, the leading contribution to the oscillation term is proportional 
to the off-diagonal element of the matrix of densities $\delta n^{12}$, as one might expect,  sourcing asymmetry only in the presence of flavour coherences.

\section{\label{sec:interaction}Shell structure for two-particle mixing in the interaction picture}

In this section, we show how the shell structure identified above in the Heisenberg picture
is reproduced in the interaction picture.

\paragraph{Tree-level Wightman propagator.} In the interaction-picture approach, the tree-level Wightman propagator can 
be obtained straightforwardly by evaluating the ensemble expectation value (EEV) of field operators 
directly (see appendix~\ref{sec:noneq} and ref.~\cite{Millington:2012pf}). In the double-momentum representation and assuming spatial homogeneity, it takes the form
\begin{align}
\label{eq:freeprop}
G_<^{0,\,ij}(p,p',\tilde{t})\ &=\ 2\pi\big(2\,\mathrm{sign}(p_0)p_0\big)^{1/2}\delta(p^2-M^2_i)\,e^{i(p_0-p_0')\tilde{t}}\nonumber\\& \times\:\big[\theta(p_0)\theta(p_0')\,n^{ij}(t,\mathbf{p})\:+\:\theta(-p_0)\theta(-p_0')\big(\delta^{ij}+n^{ij*}(t,-\,\mathbf{p})\big)\big](2\pi)^3\delta^{3}(\mathbf{p}-\mathbf{p}'\big)\nonumber\\&  \times\:2\pi\big(2\,\mathrm{sign}(p_0')p_0'\big)^{1/2}\delta(p'{}^2-M^2_j)\;.
\end{align}
Since the system of interest is spatially isotropic, the number densities $n^{ij}(t,\mathbf{p})$ are functions only of $|\mathbf{p}|$, such that $n^{ij*}(t,-\,\mathbf{p})\ =\ n^{ij*}(t,\mathbf{p})$. Note that, in eq.~\eqref{eq:freeprop}, we have distinguished 
between a macroscopic time $t$ and a microscopic time $\tilde t$, as is necessary in the interaction picture (see appendix~\ref{sec:noneq}). In the end, the physical limit 
will be obtained at equal times $X^0=\tilde t$ (for more details, see ref.~\cite{Millington:2012pf}).

\paragraph{Dressed Wightman propagator.} In order to find an explicit form for the dressed Wightman propagator, we restrict ourselves to the inclusion of one-loop self-energies and make use of the Markovian approximation. The latter has the effect of restoring exact energy-momentum conservation [see~eq.~\eqref{eq:Markovian}]. Using in addition that the self-energies, and retarded and advanced propagators are translationally invariant (see section~\ref{sec:heisenberg}), the Schwinger-Dyson equation of the Wightman propagator reduces to
\begin{align}
G_<^{ij}(p,p',\tilde{t})\ &=\ G_<^{0,\,ij}(p,p',\tilde{t})\:-\:G^{0,\,ik}_{R}(p)\,\Pi_{<}^{kl}(p)(2\pi)^4\delta^{4}(p-p')\,G_{A}^{lj}(p')\nonumber\\& -\:G^{0,\,ik}_{R}(p)\,\Pi_{R}^{kl}(p)\,G_{<}^{lj}(p,p',\tilde{t})\:-\:G^{0,\,ik}_{<}(p,p',\tilde{t})\,\Pi_{A}^{kl}(p')\,G_{A}^{lj}(p')\;,
\end{align}
and that of the retarded (advanced) propagator to
\begin{align}
G_{R(A)}^{ij}(p)\ &=\ G_{R(A)}^{0,\,ij}(p)\:-\:G^{0,\,ik}_{R(A)}(p)\,\Pi_{R(A)}^{kl}(p)\,G^{lj}_{R(A)}(p)\;.
\end{align}
The result of the approximations described above is the elimination of convolution integrals over intermediate momenta in the Schwinger-Dyson equations. As was shown in ref.~\cite{Dev:2014wsa}, this system may then be solved analytically for the resummed Wightman propagator, and we find
\begin{align}
\label{eq:dressed}
& G_<^{ij}(p,p',\tilde{t})\ =\ F_R^{ik}(p)\,G_<^{0,\,kl}(p,p',\tilde{t})\,F^{lj}_A(p')\:-\:G_{R}^{ik}(p)\,\Pi_<^{kl}(p)(2\pi)^4\delta^{4}(p-p')\,G_{A}^{lj}(p')\;,
\end{align}
where we have defined
\begin{subequations}
\label{eq:FRA}
\begin{align}
F_R^{ij}\ &\equiv \ \sum_{n\,=\,0}^{\infty}\Big[\big(-G^{0}_{R}\cdot \Pi_{R}\big)^n\Big]^{ij}\ =\ -\,G_{R}^{ik}\,D_{R}^{0,\,kj}\ =\ \delta^{ij}\:-\:G^{ik}_{R}\,\Pi^{kj}_{R}\;,\\
\:F_A^{ij}\ &\equiv \ \sum_{n\,=\,0}^{\infty}\Big[\big(-\Pi_{A}\cdot G^{0}_{A}\big)^n\Big]^{ij}\ =\ -\,D_{A}^{0,\,ik}\,G_{A}^{kj}\ =\ \delta^{ij}\:-\:\Pi^{ik}_{A}\,G^{kj}_{A}\;.
\end{align}
\end{subequations}

The second term on the rhs of eq.~\eqref{eq:dressed} describes equilibrium $\Delta L =0$ and 
$\Delta L = 2$ scatterings. Instead, the part of interest to us is contained within the first 
term on the rhs of eq.~\eqref{eq:dressed}. In particular, we wish to study the part proportional 
to the deviation from equilibrium $\delta n^{ij}(t,\mathbf{p})$. Inserting the tree-level 
Wightman propagator from eq.~\eqref{eq:freeprop} into eq.~\eqref{eq:dressed}, this part is 
given by
\begin{align}
\label{eq:Gfull}
G_{\!\delta}^{ij}(p,p',\tilde{t})\big|_{p_0,p_0'\,>\,0}\ &=\ F^{ik}_R(p)\,2\pi(2p_0)^{1/2}\delta_+(p^2-M^2_k)e^{i(p_0-p_0')\tilde{t}}\delta n^{kl}(t,\mathbf{p})(2\pi)^3\delta^{3}(\mathbf{p}-\mathbf{p}')\nonumber\\&  \times \ 2\pi(2p_0')^{1/2}\delta_+(p'{}^2-M^2_l)F^{lj}_A(p')\;,
\end{align}
where
\begin{equation}
\label{eq:deltaplus}
2\pi\,\delta_+(p^2-M^2_i)\ \equiv\ 2\pi\,\theta(p_0)\delta(p^2-M_i^2)=\frac1{2E_i}\left[
\frac{i}{p_0-E_i+i\epsilon}-\frac{i}{p_0-E_i-i\epsilon}
\right]\;,
\end{equation}
and $E_i=(\vec{p}^2+M^2_i)^\frac12$.

\paragraph{On-shell approximation.} We will first illustrate that there is no contribution to the 
resummed non-equilibrium propagator in eq.~\eqref{eq:Gfull} from the tree-level on-shell modes 
$p^2 = M_i^2$. In so doing, we will also illustrate explicitly that eq.~\eqref{eq:Gfull} is 
free of pinch singularities, which would potentially arise from ill-defined products of Dirac delta functions with identical arguments.

For this purpose, it is convenient to work with the Wigner transform (see appendix~\ref{sec:noneq}) 
of the non-equilibrium part of the dressed Wightman propagator:
\begin{align}
\label{eq:GWigner}
G_{\!\delta}^{ij}(q_0>0,X,\tilde{t})\ &=\ \int_{Q_0}e^{-iQ_0(X^0-\tilde{t})}F_R^{ik}(q_0+Q_0/2)2\pi(2E_k)^{1/2}\delta_+\bigl((q_0+Q_0/2)^2-E^2_k\bigr)\nonumber\\&\qquad \times \: \delta n^{kl}(t,\mathbf{q})\,2\pi(2E_l)^{1/2}\delta_+\bigl((q_0-Q_0/2)^2-E^2_l\bigr)F_A^{lj}(q_0-Q_0/2)\;,
\end{align}
where the trivial $\mathbf{Q}$ integral has been performed. Hereafter, we omit three-momentum 
arguments for notational brevity. In order to perform the $Q_0$ integral, we will now  assume \emph{erroneously} that 
the only poles are those provided by the Dirac delta functions appearing explicitly in 
eq.~\eqref{eq:GWigner}. We emphasize that we should not anticipate that we will obtain the correct result, 
since $G^0_{R(A)}$ also contains poles.

By virtue of the properties of the Dirac delta function, we may show that
\begin{equation}
(2E_i)^{1/2}\delta\bigl((q_0\pm Q_0/2)^2-E_i^2\bigr)\ =\ 2(2E_i)^{-1/2}\sum_{s\,=\,\pm 1}\delta\bigl(Q_0\pm 2q_0\mp 2 s E_i\bigr)\;.
\end{equation}
Performing the integral over $Q_0$, we then find
\begin{align}
\label{eq:Gosfinal}
G_{\!\delta}^{ij}(q_0>0,X,\tilde{t})\ =\ e^{-i\Delta E_{kl}(X_0-\tilde{t})}\,2\pi\,\delta(q_0-E_{kl})\,F^{ik}_R(E_k)\, \frac{\delta n^{kl}(t,\mathbf{q})}{(2E_k)^{1/2}(2E_l)^{1/2}}\,F_A^{lj}(E_l)\;,
\end{align}
where $E_{kl}\equiv (E_k+E_l)/2$. Equation~\eqref{eq:Gosfinal} is well defined in a distributional sense and, in evaluating the tree-level poles, we have not 
encountered any singular behaviour, illustrating explicitly that the expression for the 
resummed propagator in eq.~\eqref{eq:Gfull} is free of pinch singularities. 

One might be tempted to consider the terms in eq.~\eqref{eq:FRA} proportional to 
$G^{im}_{\rm R}\, \Pi^{mk}_{\rm R}$ and $\Pi_{\rm A}^{ln}\,G^{nj}_{\rm A}$ as subleading. 
Were we to drop these contributions, we would find the following for the off-diagonal elements 
of eq.~\eqref{eq:Gosfinal}:
\begin{equation}
\label{eq:KBansatz}
G_{\!\delta}^{i\slashed{i}}(q_0>0,X,\tilde{t})\ =\ e^{-i\Delta E_{i\slashed{i}}(X_0-\tilde{t})}\,2\pi\,\delta(q_0-\bar{E})\,\frac{\delta n^{i\slashed{i}}(t,\mathbf{q})}{(2E_i)^{1/2}(2E_{\slashed{i}})^{1/2}}\;.
\end{equation}
Such an approximation for the resummed would-be heavy-neutrino propagator, when used in the equation for the asymmetry, discards the phenomenon of mixing, accounting only for 
oscillations between the two flavours, as identified in ref.~\cite{Dev:2014wsa}. In fact, as we will now show, 
the terms omitted in eq.~\eqref{eq:KBansatz} are of order unity, and $G_\delta^{ij}(q,X,\tilde{t})$ is identically zero due to the erroneous treatment of the pole structure in this 
on-shell approximation.

Considering the explicit form of the dressed retarded propagator in eq.~\eqref{eq:GWigner}, we may show that the factor
\begin{equation}
\label{eq:FRterms}
F_R^{ik}(E_k)\ =\ \delta^{ik}\:-\:G_R^{im}(E_k)\,\Pi_R^{mk}(E_k)\ =\ \delta^{ik}\:-\:\frac{\Delta M^2_{\slashed{i}k}\,\delta^{im}+[\mathrm{adj}\,\bm{\Pi}_R(E_k)]^{im}}{\mathrm{det}\,\bm{D}_R(E_k)}\,\Pi_R^{mk}(E_k)\;.
\end{equation}
The determinant in the denominator of eq.~\eqref{eq:FRterms} can be written as
\begin{equation}
\mathrm{det}\,\bm{D}_{R}(E_k)\ = \ \Delta M^2_{\slashed{k}k}\,\Pi_R^{kk}(E_k)\:+\:\mathrm{det}\,\bm{\Pi}_R(E_k)\;.
\end{equation}
Thus, we have
\begin{equation}
G_R^{im}(E_k)\,\Pi_R^{mk}(E_k)\ =\ \frac{\Delta M^2_{\slashed{i}k}\,\Pi^{ik}_R(E_k)+\delta^{ik}\,\mathrm{det}\,\bm{\Pi}_R(E_k)}{\Delta M^2_{\slashed{k}k}\,\Pi_R^{kk}(E_k)\:+\:\mathrm{det}\,\bm{\Pi}_R(E_k)}\;,
\end{equation}
where we have also used the fact that
\begin{equation}
[\mathrm{adj}\,\bm{\Pi}_R(E_k)]^{im}\,\Pi_R^{mk}(E_k)\ =\ \delta^{ik}\mathrm{det}\,\bm{\Pi}_R(E_k)\;,
\end{equation}
by definition of the adjugate matrix. We may then show that
\begin{equation}
\label{eq:onshell}
G_R^{im}(E_k)\,\Pi_R^{mk}(E_k)\ =\ \delta^{ik}\;,\qquad \Pi_{\rm A}^{ln}(E_l)\,G^{nj}_{\rm A}(E_l)\ =\ \delta^{lj}\;.
\end{equation}
Substituting eq.~\eqref{eq:onshell} into the expression for the resummed propagator in 
eq.~\eqref{eq:Gosfinal}, it immediately follows that it is identically zero. Clearly, 
this result is incorrect. As we will now show, this is a consequence of having neglected the 
poles in $G_{R(A)}^0$, whose contributions are in fact pivotal in determining the correct form of the resummed propagator.

\paragraph{Pole structure.} The tree-level retarded (advanced) propagator has the form
\begin{equation}
\label{eq:GRexp}
G^{0,\,ij}_{R(A)}(p)\ =\ -\:\frac{\delta^{ij}}{p^2-M_i^2\pm i\epsilon\,\mathrm{sign}(p_0)}\ =\ -\:\mathcal{P}\,\frac{\delta^{ij}}{p^2-M_i^2}\:\pm\:i\pi\delta^{ij}\mathrm{sign}(p_0)\delta(p^2-M_i^2)\;,
\end{equation}
where we have used the identity
\begin{equation}
\label{Pdelta}
\frac{1}{x\pm i\epsilon}\ =\ \mathcal{P}\,\frac{1}{x}\:\mp\:i\pi\delta(x)\,,
\end{equation}
in which $\mathcal{P}$ denotes the Cauchy principal value. Equation~\eqref{Pdelta} 
may readily be confirmed by using the limit representations
\begin{equation}
\label{eq:deltalim}
\delta(x)\ =\ \lim_{\epsilon\,\to\,0^+}\,\frac{1}{\pi}\,\frac{\epsilon}{x^2+\epsilon^2}\;,\qquad
\mathcal{P}\,\frac{1}{x}\ =\ \lim_{\epsilon\,\to\,0^+}\,\frac{x}{x^2+\epsilon^2}\;.
\end{equation}
Substituting eq.~\eqref{eq:GRexp} into the non-equilibrium part of the resummed propagator [eq.~\eqref{eq:Gfull}], it would appear that we have 
products of Dirac delta functions of identical arguments. However, we have seen already 
that eq.~\eqref{eq:Gfull} is free of pinch singularities. The reason for this is that 
these pinch singularities are resummed, and it is by performing this resummation that we 
will obtain the correct form for the Wigner representation of the resummed Wightman propagator.
In particular, both the equilibrium and non-equilibrium parts of the propagator acquire 
finite widths (cf. ref.~\cite{Garbrecht:2011xw}).

In order to understand the structure of this resummation, it is helpful to begin with the 
single-flavour case. Therein, we wish to evaluate the following structure:
\begin{align}
\label{eq:I1}
I_R\ &\equiv\ \sum_{n\,=\,0}^{\infty}\big(-G^{0}_{\rm R}\cdot \Pi_{\rm R}\big)^n2\pi\,\mathrm{sign}(p_0)\,\delta(p^2-M^2)\nonumber\\& =\ \sum_{n\,=\,0}^{\infty}\bigg(\frac{\Pi_R}{p^2-M^2+i\epsilon\,\sign(p_0)}\bigg)^{\!n}2\pi\,\mathrm{sign}(p_0)\,\delta(p^2-M^2)\;.
\end{align}
Note that we are free to insert the product $\mathrm{sign}(p_0)\mathrm{sign}(p_0')$ into the tree-level propagator eq.~\eqref{eq:freeprop}, since the signs of $p_0$ and $p_0'$ are equal in the absence of particle-antiparticle correlations. We proceed by performing the following partial-fractioning, using the limit representation of the Dirac delta function in eq.~\eqref{eq:deltalim}:
\begin{equation}
2\pi\,\mathrm{sign}(p_0)\,\delta(p^2-M^2)\ =\ \frac{i}{p^2-M^2+i\epsilon\,\mathrm{sign}(p_0)}\:-\:\frac{i}{p^2-M^2-i\epsilon\,\mathrm{sign}(p_0)}\;.
\end{equation}
We then decompose
\begin{equation}
I_R\ \equiv\ I^+_R\:-\:I^-_R\;,
\end{equation}
where
\begin{equation}
I^{\pm}_R\ =\ i\sum_{n\,=\,0}^{\infty}\bigg(\frac{\Pi_R}{p^2-M^2+i\epsilon\,\sign(p_0)}\bigg)^{\!n}\,\frac{1}{p^2-M^2\pm i\epsilon\,\mathrm{sign}(p_0)}\;.
\end{equation}
By employing the distributional identity (see e.g.~ref.~\cite{HoskinsPinto})
\begin{equation}
\label{eq:npowers}
\bigg(\frac{1}{x\pm i\epsilon}\bigg)^{n}\ = \ \mathcal{P}\,\frac{1}{x^n}\:\mp\:\frac{(-1)^{n-1}}{(n-1)!}\,i\pi\delta^{(n-1)}(x)\;,
\end{equation}
where $\delta^{(n)}(x)$ is the $n$-th derivative of the Dirac delta function, we find
\begin{equation}
I^+_R\ =\ i\sum_{n\,=\,0}^{\infty}\mathcal{P}\,\bigg(\frac{1}{p^2-M^2}\bigg)^{\!n+1}\,(\Pi_R)^n\:+\:\sum_{n\,=\,0}^{\infty}\frac{(-\Pi_R)^n}{n!}\,\pi\mathrm{sign}(p_0)\delta^{(n)}(p^2-M^2)\;.
\end{equation}
Transposing the identity in eq.~\eqref{eq:npowers}, we may also show that
\begin{equation}
\mathcal{P}\,\frac{1}{x^n}\ =\ \frac{(-1)^{n-1}}{(n-1)!}\,\mathcal{P}^{(n-1)}\,\frac{1}{x}\;,
\end{equation}
where $\mathcal{P}^{(n)}(x)$ is the $n$-th derivative of the Cauchy principal value. Hence, we obtain
\begin{equation}
\label{eq:Iplusfin}
I^+_R\ =\ i\sum_{n\,=\,0}^{\infty}\frac{(-\Pi_R)^n}{n!}\,\mathcal{P}^{(n)}\,\frac{1}{p^2-M^2}\:+\:\sum_{n\,=\,0}^{\infty}\frac{(-\Pi_R)^n}{n!}\,\pi\mathrm{sign}(p_0)\delta^{(n)}(p^2-M^2)\;.
\end{equation}
As we might expect from comparing $I_R^+$ to the usual Feynman-Dyson series, this result is proportional to the resummed retarded propagator:
\begin{equation}
I_R^+\ =\ i\sum_{n\,=\,0}^{\infty}\frac{(-\Pi_R)^n}{n!}\,\frac{\partial^n}{\partial (p^2)^n}\,\big(-G_R^0(p)\big)\ =\ -\,i\,G_R(p)\;.
\end{equation}

In the case of $I_R^-$, we instead have
\begin{equation}
\label{eq:IRminus}
I^-_R\ =\ i\sum_{n\,=\,0}^{\infty}\bigg(\frac{\Pi_R}{p^2-M^2+i\epsilon\,\mathrm{sign}(p_0)}\bigg)^{\!n}\,\frac{1}{p^2-M^2- i\epsilon\,\mathrm{sign}(p_0)}\;.
\end{equation}
This term would appear to suffer from pinch singularities, arising from the product of poles at $p^2=M^2+i\epsilon\,\mathrm{sign}(p_0)$ and $p^2=M^2-i\epsilon\,\mathrm{sign}(p_0)$. However, such pinch singularities arise here only at a finite order in perturbation theory. This is a consequence of having artificially restored energy-momentum conservation through the Markovian approximation. It can be shown~\cite{Millington:2012pf} that the perturbation series is in fact well defined so long as one takes into account finite-time effects and the microscopic violation of energy-momentum conservation. Performing the summation over $n$ in eq.~\eqref{eq:IRminus} first, we can make use of the fact that
\begin{equation}
\label{eq:standardFD}
\sum_{n\,=\,0}^{\infty}\bigg(\frac{\Pi_R}{p^2-M^2+i\epsilon\,\mathrm{sign}(p_0)}\bigg)^{\!n}\ \equiv\ \sum_{n\,=\,0}^{\infty}\bigg(\frac{\Pi_R}{p^2-M^2-i\epsilon\,\mathrm{sign}(p_0)}\bigg)^{\!n}\ =\ \frac{p^2-M^2}{p^2-M^2-\Pi_R}
\end{equation}
does not depend on the pole prescription when $|\mathrm{Im}\,\Pi_R|>\epsilon$. We may therefore write
\begin{align}
I^-_R\ &\equiv\ i\sum_{n\,=\,0}^{\infty}\bigg(\frac{\Pi_R}{p^2-M^2-i\epsilon\,\mathrm{sign}(p_0)}\bigg)^{\!n}\,\frac{1}{p^2-M^2- i\epsilon\,\mathrm{sign}(p_0)}\nonumber\\
& = \ i\sum_{n\,=\,0}^{\infty}\frac{(-\Pi_R)^n}{n!}\,\mathcal{P}^{(n)}\,\frac{1}{p^2-M^2}\:-\:\sum_{n\,=\,0}^{\infty}\frac{(-\Pi_R)^n}{n!}\,\pi\mathrm{sign}(p_0)\delta^{(n)}(p^2-M^2)\;.
\end{align}
We see that $I_R^-$ differs from $I_R^+$ in eq.~\eqref{eq:Iplusfin} by the sign of the second term. Hence, we arrive at the result
\begin{equation}
\label{eq:IRfin}
I_R\ =\ \sum_{n\,=\,0}^{\infty}\frac{(-\Pi_R)^n}{n!}\,2\pi\,\mathrm{sign}(p_0)\,\delta^{(n)}(p^2-M^2)\;.
\end{equation}

In order to understand the meaning of eq.~\eqref{eq:IRfin} and why this contribution has not cancelled between $I_R^+$ and $I_R^-$, we first write
\begin{equation}
2\pi\,\mathrm{sign}(p_0)\delta(p^2-M^2)\ \equiv\ \frac{i}{(p^2-M^2)^+}\:-\:\frac{i}{(p^2-M^2)^-}\;,
\end{equation}
replacing the explicit $\epsilon$ by the following equivalent (but more general) prescription for deforming the contour in the complex plane. The $+$ and $-$ indicate that we are to deform the contour of integration in $p_0$ away from the real axis such that we pass always above ($+$) or below ($-$) the poles at $p^2-M^2=0$. Returning to the generalized Taylor series expansion in eq.~\eqref{eq:IRfin}, we see that it effects a shift $p^2-M^2\to p^2-M^2-\Pi_R$. Since this includes a shift of the poles in the imaginary direction, we must simultaneously deform the contour of integration such that no poles cross the contour during this shift. In this way, we have
\begin{equation}
\label{eq:complexdelta2}
I_{R}\ =\ \frac{i}{(p^2-M^2-\Pi_R)^{+}}\:-\:\frac{i}{(p^2-M^2-\Pi_R)^-}\;.
\end{equation}
The rhs of eq.~\eqref{eq:complexdelta2} is the complex delta function (see e.g.~refs.~\cite{Nakanishi01061958,PETROSKY1991175}):
\begin{equation}
I_{R}\ =\ 2\pi\delta(p^2-M^2-\Pi_R)\;,
\end{equation}
which corresponds to the contribution from the poles of $i/(p^2-M^2-\Pi_R)$. Since the imaginary part of the retarded self-energy is odd under $p_0\to -\,p_0$, all of these poles lie in the lower-half complex plane. We can therefore write $I_R$ as
\begin{equation}
\label{eq:complexdelta3}
I_R\ =\ \frac{if(p_0)}{p^2-M^2-\Pi_R}\;,
\end{equation}
where $f(p_0)$ is a single-valued and analytic function chosen such that we must close the contour of integration in the lower-half complex plane. Specifically, we require~\cite{Nakanishi01061958}: (i) $f(p_0)\approx 1$ in the vicinity of the poles, (ii) $f(p_0)\approx 0$, effectively, on the real axis far away from the poles, (iii) $f(p_0)$ regular near the real axis, and (iv) $f(p_0)$ vanishing far away in the lower-half complex plane.

We may proceed analogously in the case of two flavours:
\begin{equation}
I^i_R\ =\ I^{ii}_R\:+\:I^{i\slashed{i}}_R\ \equiv\ \sum_{n\,=\,0}^{\infty}\Big[\big(-G_R^0\cdot\Pi_R\big)^n\Big]^{ij}2\pi\,\mathrm{sign}(p_0)\,\delta(p^2-M_j^2)\;.
\end{equation}
We can make sense of the resummation in $I_R^{ii}$ by considering the series for the resummed propagator (no summation over $i$ implied):
\begin{equation}
\label{eq:resummedii}
-\,G_R^{ii}\ =\ \sum_{n\,=\,0}^{\infty}\Big[\big(-G_R^0\cdot\Pi_R\big)^n\Big]^{ii}\big(-\,G_R^{0,\,ii}\big)\ =\ \Bigg[p^2-M_i^2-\Pi_R^{ii}-\frac{\Pi_R^{i\slashed{i}}\Pi_R^{\slashed{i}i}}{p^2-M_{\slashed{i}}^2-\Pi_R^{\slashed{i}\slashed{i}}}\Bigg]^{-1}\;,
\end{equation}
which differs from $I_R^{ii}$ by the replacement
\begin{equation}
\label{eq:deltarepl}
2\pi\,\mathrm{sign}(p_0)\,\delta(p^2-M_i^2)\ \longrightarrow\ -\,G_R^{0,\,ii}(p)\;.
\end{equation}
In the case of $I_R^{ii}$, we can resum all but insertions of $1/(p^2-M_i^2+i\epsilon\,\mathrm{sign}(p_0))$ straightforwardly and obtain
\begin{align}
\label{eq:IRii}
I_{\mathrm{R}}^{ii}\ &=\ \sum_{n\,=\,0}^{\infty}\bigg[\frac{1}{p^2-M_i^2+i\epsilon\,\mathrm{sign}(p_0)}\bigg(\Pi_R^{ii}\:+\:\frac{\Pi^{i\slashed{i}}_R\Pi^{\slashed{i}i}_R}{p^2-M_{\slashed{i}}^2-\Pi^{\slashed{i}\slashed{i}}_R}\bigg)\bigg]^{\! n}2\pi\,\mathrm{sign}(p_0)\,\delta(p^2-M_i^2)\,,
\end{align}
which gives eq.~\eqref{eq:resummedii} on making the replacement in eq.~\eqref{eq:deltarepl}.
The potential pinch singularities in eq.~\eqref{eq:IRii} are resummed in the same way as for the single-flavour case, yielding
\begin{align}
I_{\mathrm{R}}^{ii}\ &=\ \sum_{n\,=\,0}^{\infty}\frac{1}{n!}\bigg(-\Pi_R^{ii}\:-\:\frac{\Pi^{i\slashed{i}}_R\Pi^{\slashed{i}i}_R}{p^2-M_{\slashed{i}}^2-\Pi^{\slashed{i}\slashed{i}}_R}\bigg)^{\! n}\,2\pi\,\mathrm{sign}(p_0)\,\delta^{(n)}(p^2-M_i^2)\nonumber\\ &=\ 2\pi\,\delta(-\,[G_R^{-1}]^{ii})\;,
\end{align}
where $\delta$ is understood to be the complex delta function, giving the contribution from the poles of
\begin{equation}
i\Bigg[p^2-M_i^2-\Pi_R^{ii}-\frac{\Pi_R^{i\slashed{i}}\Pi_R^{\slashed{i}i}}{p^2-M_{\slashed{i}}^2-\Pi_R^{\slashed{i}\slashed{i}}}\Bigg]^{-1}\;.
\end{equation}
This is equal to the contribution from the poles of
\begin{equation}
i[\mathrm{adj}\,\bm{D}_R]^{ii}/\mathrm{det}\,\bm{D}_R\;,
\end{equation}
which occur at $\mathrm{det}\,\bm{D}_R\ =\ 0$. Hence, we have
\begin{equation}
I_{\mathrm{R}}^{ii}\ =\ 2\pi\,[\mathrm{adj}\,\bm{D}_R]^{ii}\delta(\mathrm{det}\,\bm{D}_R)\;.
\end{equation}

For the series $I_R^{i\slashed{i}}$, we are able to resum all but insertions of $i/(p^2-M_{\slashed{i}}^2+i\epsilon\,\mathrm{sign}(p_0))$ straightforwardly and obtain
\begin{align}
\label{eq:IRislashislash}
I_R^{i\slashed{i}}\ &=\ \frac{\Pi_R^{i\slashed{i}}}{p^2-M_{i}^2-\Pi_R^{ii}}\nonumber\\&\qquad \times\:\sum_{n\,=\,0}^{\infty}\bigg[\frac{1}{p^2-M_{\slashed{i}}^2+i\epsilon\,\mathrm{sign}(p_0)}\bigg(\Pi_R^{\slashed{i}\slashed{i}}\:+\:\frac{\Pi^{\slashed{i}i}_R\Pi^{i\slashed{i}}_R}{p^2-M_i^2-\Pi^{ii}_R}\bigg)\bigg]^{\! n}2\pi\,\mathrm{sign}(p_0)\,\delta(p^2-M_{\slashed{i}}^2)\nonumber\\ &=\ \frac{\Pi_R^{i\slashed{i}}}{p^2-M_{i}^2-\Pi_R^{ii}}\sum_{n\,=\,0}^{\infty}\frac{1}{n!}\bigg(-\Pi_R^{\slashed{i}\slashed{i}}\:-\:\frac{\Pi^{\slashed{i}i}_R\Pi^{i\slashed{i}}_R}{p^2-M_{i}^2-\Pi^{ii}_R}\bigg)^{\! n}\,2\pi\,\mathrm{sign}(p_0)\,\delta^{(n)}(p^2-M_{\slashed{i}}^2)\nonumber\\ &=\ \frac{\Pi_R^{i\slashed{i}}}{p^2-M_{i}^2-\Pi_R^{ii}}\,2\pi\,\delta(-\,[G_R^{-1}]^{\slashed{i}\slashed{i}})\;.
\end{align}
We may readily verify that this gives the resummed propagator $G_R^{i\slashed{i}}(p)$ on making the replacement
\begin{equation}
2\pi\,\mathrm{sign}(p_0)\,\delta(p^2-M_{\slashed{i}}^2)\ \longrightarrow\ -\,G_R^{0,\,\slashed{i}\slashed{i}}(p)\;.
\end{equation}
Equation \eqref{eq:IRislashislash} corresponds to the contribution from the poles of
\begin{equation}
-i\,G_R^{i\slashed{i}}\ =\ \frac{i\Pi_R^{i\slashed{i}}}{(p^2-M_i^2-\Pi_R^{ii})(p^2-M_{\slashed{i}}^2-\Pi_R^{\slashed{i}\slashed{i}})-\Pi_R^{i\slashed{i}}\Pi_R^{\slashed{i}i}}\ =\ i[\mathrm{adj}\,\bm{D}_R]^{i\slashed{i}}/\mathrm{det}\,\bm{D}_R\;,
\end{equation}
which again occur at $\mathrm{det}\,\bm{D}_R\ =\ 0$. Hence, we have
\begin{equation}
I_{\mathrm{R}}^{i\slashed{i}}\ =\ 2\pi\,[\mathrm{adj}\,\bm{D}_R]^{i\slashed{i}}\delta(\mathrm{det}\,\bm{D}_R)\;.
\end{equation}

Continuing similarly for the remaining components and the corresponding advanced series ($I_A\equiv I_R^*$), we obtain the complete expression for the non-equilibrium part of the resummed propagator:
\begin{align}
\label{GdeltaDoubleMom}
G_{\!\delta}^{ij}(p,p',\tilde{t})\ &= \ 2\pi \big(2\mathrm{sign}(p_0)p_0\big)^{1/2}[\mathrm{adj}\,\bm{D}_R(p)]^{ik}\delta(\mathrm{det}\,\bm{D}_R(p))\,e^{i(p_0-p_0')\tilde{t}}\nonumber\\& \times\:\big[\theta(p_0)\theta(p_0')\,\delta n^{kl}(t,\mathbf{p})\:+\:\theta(-p_0)\theta(-p_0')\,\delta n^{kl*}(t,-\,\mathbf{p})\big](2\pi)^3\delta^{3}(\mathbf{p}-\mathbf{p}'\big)\nonumber\\&\times\:2\pi \big(2\mathrm{sign}(p_0')p_0'\big)^{1/2}\delta(\mathrm{det}\,\bm{D}_A(p'))[\mathrm{adj}\,\bm{D}_A(p')]^{lj}\;.
\end{align}
In order to compare this result directly with the Heisenberg picture, we make use of the pole approximation in eqs.~\eqref{DRapprox} and \eqref{DRapprox2}. For $p_0>0$, the complex delta function $2\pi\delta(\mathrm{det}\,\bm{D}_R)$ corresponds to the contribution from the poles at $p_0=\Omega_i$. Instead, for $p_0<0$, the complex delta function $2\pi\delta(\mathrm{det}\,\bm{D}_R)$ corresponds to the contribution from the poles at $p_0=-\,\Omega_i^*$. Hence, we can write
\begin{align}
\label{eq:complexdeltaexp}
2\pi\delta\big(\mathrm{det}\,\bm{D}_R(p)\big)\ &\approx \ \frac{i}{\Delta \Omega^2}\bigg[\frac{1}{2\Omega_1}\,\frac{f_1(p_0)}{p_0-\Omega_1}-\:\frac{1}{2\Omega_2}\,\frac{f_2(p_0)}{p_0-\Omega_2}\bigg]\nonumber\\&\qquad -\:\frac{i}{\Delta \Omega^{*2}}\bigg[\frac{1}{2\Omega_1^*}\,\frac{f_1^*(-p_0)}{p_0+\Omega_1^*}\:-\:\frac{1}{2\Omega_2^*}\,\frac{f_2^*(-p_0)}{p_0+\Omega_2^*}\bigg]\;,
\end{align}
where the $f_i(p_0)$ satisfy the properties highlighted above [see eq.~\eqref{eq:complexdelta3}]. An appropriate choice for these functions (see~ref.~\cite{Nakanishi01061958}) is $f_i(p_0)=[\lambda_i^2/(p_0^2+\lambda_i^2)]\,e^{-i(p_0-\mathrm{Re}\,\Omega_i)/\kappa_i}$, where $\kappa_i\gg \Gamma_i/2$ and $\lambda_i\gg\mathrm{Re}\,\Omega_i>\Gamma_i$. The relative sign between the poles at $\Omega_1$ and $\Omega_2$ arises from the partial-fractioning of $1/[(q_0^2-\Omega_1^2)(q_0^2-\Omega_2^2)]$, and the relative sign between the positive- and negative-frequency poles results from the partial-fractioning of $1/(q_0^2-\Omega_i^2)$. In order to ensure that this partial-fractioning is consistent with the analytic structure of the retarded propagator, we first let $\Omega_{i}\equiv \Omega_i(q_0)=-\,\Omega_i^*(-q_0)$, before approximating $\Omega_i$ in the vicinity of the poles by eq.~\eqref{eq:omegaidef}. In the limit $\Pi_R\to 0$, $\kappa_i\to 0$ and $\lambda_i\to\infty$, we recover the standard 
Dirac delta function:
\begin{subequations}
\begin{align}
\frac{i f_i(p_0)}{p_0-\Omega_i}\ \longrightarrow\ 2\pi\delta(p_0-E_i)\;,\\
\frac{i f^*_i(-\,p_0)}{p_0+\Omega_i^*}\ \longrightarrow\ 2\pi\delta(p_0+E_i)\;.
\end{align}
\end{subequations}
In the same limit, we therefore find
\begin{align}
2\pi\delta(\mathrm{det}\,\bm{D}_R(p))\ &\longrightarrow \  \frac{2\pi}{|\Delta M^2|}\bigg[\frac{\delta(p_0-E_1)}{2E_1}\:-\:\frac{\delta(p_0-E_2)}{2E_2}\:-\:\frac{\delta(p_0+E_1)}{2E_1}\:+\:\frac{\delta(p_0+E_2)}{2E_2}\bigg]\nonumber\\ &=\ \frac{2\pi}{\Delta M^2}\,\mathrm{sign}(p_0)\,\big[\delta(p^2-M_1^2)-\delta(p^2-M_2^2)\big]\;,
\end{align}
and
\begin{align}
G_{\!\delta}^{ij}(p,p',\tilde{t})\ &\longrightarrow \ 2\pi \big(2\mathrm{sign}(p_0)p_0\big)^{1/2}\delta(p^2-M_i^2)e^{i(p_0-p_0')\tilde{t}}\nonumber\\& \times\:\big[\theta(p_0)\theta(p_0')\,\delta n^{ij}(t,\mathbf{p})\:+\:\theta(-p_0)\theta(-p_0')\delta n^{ij*}(t,-\,\mathbf{p})\big](2\pi)^3\delta^{(3)}(\mathbf{p}-\mathbf{p}'\big)\nonumber\\&\times\:2\pi \big(2\mathrm{sign}(p_0')p_0'\big)^{1/2}\delta(p'^2-M_j^2)\nonumber\\&=\ G_{\!\delta}^{0,\,ij}(p,p',\tilde{t})\;,
\end{align}
recovering the non-equilibrium part of the tree-level propagator [cf.~eq.~\eqref{eq:freeprop}], as we would expect.

By extracting the positive-frequency part of eq.~\eqref{eq:complexdeltaexp}, we can define a generalization of eq.~\eqref{eq:deltaplus}:
\begin{align}
\label{eq:deltapluscomp}
2\pi\delta_+(\mathrm{det}\,\bm{D}_R(p))\ &\equiv\ \frac{i}{\Delta \Omega^2}\bigg[\frac{1}{2\Omega_1}\,\frac{f_1(p_0)}{p_0-\Omega_1}-\:\frac{1}{2\Omega_2}\,\frac{f_2(p_0)}{p_0-\Omega_2}\bigg]\nonumber\\
&\equiv\ \frac{2\pi}{\Delta \Omega^2}\bigg[\frac{1}{2\Omega_1}\,\delta(p_0-\Omega_1)\:-\:\frac{1}{2\Omega_2}\,\delta(p_0-\Omega_2)\bigg]\;.
\end{align}
We may then write the positive-frequency, non-equilibrium part of the full propagator as
\begin{align}
\label{GdeltaDoubleMomplus}
&G_{\!\delta}^{ij}(p,p',\tilde{t})\big|_{p_0,p_0'\,>\,0} \ = \ 2\pi (2p_0)^{1/2}[\mathrm{adj}\,\bm{D}_R(p)]^{ik}\delta_+(\mathrm{det}\,\bm{D}_R(p))\,e^{i(p_0-p_0')\tilde{t}}\nonumber\\&\qquad\qquad\times\:\delta n^{kl}(t,\mathbf{p})(2\pi)^3\delta^{3}(\mathbf{p}-\mathbf{p}'\big)2\pi (2p_0')^{1/2}\delta_+(\mathrm{det}\,\bm{D}_A(p'))[\mathrm{adj}\,\bm{D}_A(p')]^{lj}\;,
\end{align}
where
\begin{equation}
\delta_+(\mathrm{det}\,\bm{D}_A(p))\ =\ \big[\delta_+(\mathrm{det}\,\bm{D}_R(p))\big]^*\;.
\end{equation}

We note that eqs.~\eqref{eq:complexdeltaexp} and \eqref{eq:deltapluscomp} cannot be applied naively in the doubly-degenerate
case, where the limit $\Omega_2\rightarrow \Omega_1$ must be taken before the integral over $p_0$. It is in the use of the pole approximation in eq.~\eqref{DRapprox} that the present analysis differs from that of ref.~\cite{Dev:2014wsa}, where instead an alternative procedure was employed based upon the resummation techniques developed in ref.~\cite{Pilaftsis:2003gt}. We note however that the approximation used there (see appendix A.1 of ref.~\cite{Dev:2014wsa}) cannot be used in the weak-washout regime, where one cannot guarantee that the off-diagonal number densities are of $\mathcal{O}(h^2)$, as is the case for the strong-washout regime considered in ref.~\cite{Dev:2014wsa}.

In the equal-time limit $X^0=\tilde t$, and using eq.~\eqref{eq:deltapluscomp}, the Wigner transform of eq.~\eqref{GdeltaDoubleMomplus} is
\begin{align}
\label{GdeltaInit}
G_{\!\delta}^{ij}(t,q_0>0)\ &=\ 2\pi\delta(q_0-\Omega_{ab})\,[\mathrm{adj}\,\bm{D}_R(\Omega_a)]^{ik}\,\frac{\epsilon_{ab}\,\delta n^{kl}(t,\mathbf{q})}{(2\Omega_a)^{1/2}(2\Omega_b^*)^{1/2}|\Delta \Omega^2|^2}\,[\mathrm{adj}\,\bm{D}_A(\Omega_b^*)]^{lj}\;.
\end{align}
Here, the sum over $a,b=1,2$ has been left implicit,
$\Omega_{ab}\ \equiv\ (\Omega_a\:+\:\Omega_b^*)/2$, and $\epsilon_{ab}=1$ if $a=b$ and $\epsilon_{ab}=-1$ if $a\neq b$. Finally, performing the 
summations over $a$ and $b$, we find
\begin{align}
\label{eq:Gpoles}
G_{\!\delta}^{ij}(t,q_0>0)\ &= \ 2\pi\delta(q_0-\omega_1)\,[\mathrm{adj}\,\bm{D}_R(\Omega_1)]^{ik}\,\frac{\delta n^{kl}(t,\mathbf{q})}{|2\Omega_1||\Delta \Omega^2|^2}\,[\mathrm{adj}\,\bm{D}_A(\Omega_1^*)]^{lj}\nonumber\\& +\ 2\pi\delta(q_0-\omega_2)\,[\mathrm{adj}\,\bm{D}_R(\Omega_2)]^{ik}\,\frac{\delta n^{kl}(t,\mathbf{q})}{|2\Omega_2||\Delta \Omega^2|^2}\,[\mathrm{adj}\,\bm{D}_A(\Omega_2^*)]^{lj}\nonumber\\& -\ 2\pi\delta(q_0-\bar{\Omega})\,[\mathrm{adj}\,\bm{D}_R(\Omega_1)]^{ik}\,\frac{\delta n^{kl}(t,\mathbf{q})}{(2\Omega_1)^{1/2}(2\Omega_2^*)^{1/2}|\Delta \Omega^2|^2}\,[\mathrm{adj}\,\bm{D}_A(\Omega_2^*)]^{lj}\nonumber\\& -\ 2\pi\delta(q_0-\bar{\Omega}^*)\,[\mathrm{adj}\,\bm{D}_R(\Omega_2)]^{ik}\,\frac{\delta n^{kl}(t,\mathbf{q})}{(2\Omega_1^*)^{1/2}(2\Omega_2)^{1/2}|\Delta \Omega^2|^2}\,[\mathrm{adj}\,\bm{D}_A(\Omega_1^*)]^{lj}\;,
\end{align}
where $\bar{\Omega}\ =\ \big(\Omega_1\:+\:\Omega_2^*\big)/2$. It is essential to emphasize
that the deviations from equilibrium  $\delta n^{ij}$  are the non-equilibrium parts of the 
physical \emph{dynamical} number densities, which appear as unknowns in the interaction-picture 
propagators. Moreover, these are the \emph{spectrally-free} number densities, which count 
excitations with energy $E_i$.

In order to compare with the Heisenberg-picture result in eq.~\eqref{GF12Expansion}, we now expand the interaction-picture result in eq.~\eqref{eq:Gpoles} above to 
first order in $\Pi_{R(A)}$. This gives the following result for the off-diagonal components:
\begin{align}
\label{eq:DeltaGint}
G_{\!\delta}^{i\slashed{i}}(t,q_0>0)\ &\approx\ 2\pi\,\delta(q_0-\omega_i)\,\frac{1}{2\omega_i}\,\delta n^{ii}(t)\,\Pi_A^{i\slashed{i}}(\omega_i)\,R_{i\slashed{i}}\nonumber\\
&-\: 2\pi\,\delta(q_0-\omega_\slashed{i})\,\frac{1}{2\omega_\slashed{i}}\,\delta n^{\slashed{i}\slashed{i}}(t)\,\Pi_R^{i\slashed{i}}(\omega_\slashed{i})\,R_{i\slashed{i}}\nonumber\\&+\ 2\pi\,\delta(q_0-\bar{\omega})\,\frac{1}{(2\omega_i)^{1/2}(2\omega_\slashed{i})^{1/2}}\,\Bigl[\delta n^{i\slashed{i}}(t)\,\Delta M^2_{i\slashed{i}}\nonumber\\
&\hspace{1cm}-\:\delta n^{ii}(t)\,\Pi_{A}^{i\slashed{i}}(\omega_\slashed{i})\:+\:\delta n^{\slashed{i}\slashed{i}}(t)\,\Pi^{i\slashed{i}}_{R}(\omega_i)\Bigr]R_{i\slashed{i}}\;.
\end{align}
Thus, we find for the time-derivative of the asymmetry
\begin{align}
\label{dqdttime}
\frac{\D\eta}{\D t} \ & \approx\  2 \sum_i \int_{\vec q}\frac{M_i}{\omega_i} \, \delta n^{ii}(t,\vec{q}) 
\, \Gamma^{\rm med}_i(\omega_i,\vec{q})\, \epsilon_i^{\rm med}(\omega_i,\vec{q})\nonumber\\
&+\ 2\,\Im\,H_{12}\, \Im\,\int_{\vec q}\frac{\widetilde{\Pi}_{\rho}(\bar{\omega},\mathbf{q})}{(\omega_1\omega_2)^\frac12}\,
\Bigl[\delta n^{12}(t,\vec{q})\,\Delta M^2_{12}\nonumber\\
&-\ \delta n^{11}(t,\vec{q})\Pi_A^{12}(\omega_2,\vec{q})+\delta n^{22}(t,\vec{q})\Pi_R^{12}(\omega_1,\vec{q})\Bigr] R_{12}\,.
\end{align}
This closely resembles the Heisenberg-picture result in eq.~\eqref{dqdtapprox} with the exception of the time-dependent phases; the Heisenberg-picture result is written in terms of the \emph{initial} conditions for the non-equilibrium parts of the number densites. Hence,
in order to show that the expressions for $G_{\delta}^{ij}$ in eqs.~\eqref{GF12Expansion} and \eqref{eq:DeltaGint} are in fact identical to 
first order in the self-energies, and by extension the expressions for the asymmetry in eqs.~\eqref{dqdttime} and \eqref{dqdtapprox},  we must now
find the functional form of $\delta n^{ij}(t)$ by solving the transport equations directly. This will allow us to write the interaction-picture result directly in terms of the initial conditions.

Before proceeding to do this, however, it is important to remark upon the role played by the interference 
terms. These interference terms may now be distinguished from the pure oscillation contribution. The former appear in the final line of eq.~\eqref{dqdttime}
and originate in the final line of eq.~\eqref{eq:DeltaGint}, lying on the oscillation shell but being proportional to the \emph{diagonal} components of the \emph{time-dependent} number densities. On the other hand, the oscillation contribution appears in the second line of eq.~\eqref{dqdttime} and originates in third line of eq.~\eqref{eq:DeltaGint}, and is proportional to the \emph{off-diagonal} components of the \emph{time-dependent} number densities. This identification of the pure oscillation contribution is in accord with the conventions of refs.~\cite{Dev:2014laa,Dev:2015wpa, Dev:2014wsa} and, as we will see below, it is subtly different to identifying the pure oscillation and interference contributions in terms of the components of the initial deviations from equilibrium, as appear in eq.~\eqref{GF12Expansion}, which makes sense only in the weak-washout regime. Note that, although it is possible to identify the mixing and oscillation shells unambiguously by means of the spectral structure of the 
resummed propagators, it is possible to identify the pure oscillation and interference terms only through a physically-meaningful definition of the particle number densities.

We proceed by expanding all but the regulator structure in 
eq.~\eqref{eq:DeltaGint} around $\Delta \omega_{i\slashed{i}} = \omega_i-\omega_{\slashed{i}}= 0$. 
At zeroth order, we find
\begin{equation}
G_{\delta}^{i\slashed{i}}(t,q_0>0)\ \supset\ 2\pi\delta(q_0-\bar{\omega})\frac{1}{2\bar{\omega}}\delta n^{i\slashed{i}}(t)\Delta M^2_{i\slashed{i}}R_{i\slashed{i}}\;,
\end{equation}
in which the mixing contributions have canceled and from which we see that the interference between 
mixing and oscillations is \emph{destructive}. At this point, one might be tempted to conclude that using the on-shell approximation for the heavy-neutrino propagator, cf.~eq.~\eqref{eq:KBansatz}, in the equation for the asymmetry is valid, and therefore that the approach of refs.~\cite{Dev:2014laa,Dev:2015wpa,Dev:2014wsa}, by including contributions from \emph{both} mixing and oscillation, double-counts the final asymmetry. However, this is not the case.
Continuing to the next order in the expansion, we find
\begin{align}
\label{Gexpanded}
G_{\delta}^{i\slashed{i}}(t,q)\ &\approx\ 2\pi\delta(q_0-\bar{\omega})\frac{1}{2\bar{\omega}}\delta n^{i\slashed{i}}(t)\Delta M^2_{i\slashed{i}}R_{i\slashed{i}}\nonumber\\&\qquad-2\pi\delta(q_0-\bar{\omega})\frac{1}{2\bar{\omega}}\left[\delta n^{ii}(t)\frac{\Pi_A^{i\slashed{i}}(\bar{\omega})}{4\bar{\omega}^2}\:+\:\delta n^{\slashed{i}\slashed{i}}(t)\frac{\Pi_R^{i\slashed{i}}(\bar{\omega})}{4\bar{\omega}^2}\right]\Delta M_{i\slashed{i}}^2 R_{i\slashed{i}}\;,
\end{align}
where the mixing terms are present but are now suppressed by an additional factor of $\Delta M_{12}^2$. Here, we have 
neglected terms proportional to the derivative of the delta function $\delta'(q_0-\bar{\omega})$ and the derivative of the self-energy $\Pi^{i\slashed{i}}{}'(\bar{\omega})$, which 
contribute sub-dominantly to the asymmetry under the assumption that the self-energies are slowly varying functions of $q_0$ for $q_0\sim\bar{\omega}$. This same assumption underlies the pole approximations in eqs.~\eqref{DRapprox} and \eqref{DRapprox2}, which can be verified numerically (see ref.~\cite{Hohenegger:2014cpa}). The asymmetry now takes the form
\begin{align}
\label{afterinterference}
\frac{\D\eta}{\D t}\ & \approx\  2 \sum_i \int_{\vec q}\frac{M_i}{\bar{\omega}} \, \delta n^{ii}(t,\vec{q}) 
\, \Gamma^{\rm med}_i(\bar{\omega},\vec{q})\, \tilde{\epsilon}_i^{\rm med}(\bar{\omega},\vec{q})\nonumber\\
&+\ 2\,\Im\,H_{12}\,\int_{\vec q}\frac{\widetilde{\Pi}_{\rho}(\bar{\omega},\mathbf{q})}{\bar{\omega}}\,
 \Im\,\delta n^{12}(t,\vec{q})\,\Delta M^2_{12}R_{12}\,,
\end{align}
where the usual \CP-violating parameter has been modified:
\begin{equation}
\label{epsilontilde}
\tilde{\epsilon}_i^{\,\rm med}(\bar{\omega},\vec{q})\ =\ \frac{\omega_{\slashed{i}}-\omega_{i}}{\omega_{\slashed{i}}+\omega_{i}}\,\epsilon^{\rm med}_i(\bar{\omega},\mathbf{q})\;,
\end{equation}
cf.~eq.~\eqref{EpsilonMed}. Although both mixing and oscillation contributions persist in the middle-shell approximation and are 
clearly identifiable, in agreement with the results of refs.~\cite{Dev:2014laa,Dev:2015wpa,Dev:2014wsa}, 
we see that the structure of the mixing contribution has been modified. In addition to the suppression by an 
additional factor of $\Delta M_{12}^2$, a relative sign has emerged between the $i=1$ and $i=2$ contributions. 
As a result, the modified mixing contribution is strongly suppressed when the deviations of the number densities
of the two flavours from equilibrium are similar. However, this is not always the case, for instance in scenarios 
of resonant $\ell$-genesis~\cite{Pilaftsis:2004xx,Pilaftsis:2005rv} (see also~refs.~\cite{Dev:2014laa,Dev:2015wpa,Dev:2014wsa}), 
where the lepton asymmetry is dominantly produced in a single flavour through the decays of heavy neutrinos 
of a particular family type. In such cases, both mixing and oscillation contributions will be present. Most 
significantly, we observe from eqs.~\eqref{Gexpanded} and~\eqref{afterinterference} that the pre-factors of these two distinct contributions 
to the lepton asymmetry carry exactly the same parametric dependence on the Yukawa couplings and mass 
splittings: $\sim \Delta M_{12}^2 R_{12}$, which is of order unity in the weakly-resonant or overlapping regime 
$\Gamma_i\ll \Delta M\ll \bar{M}$.  Finally, we remark that it remains to be seen how the mixing source is modified 
by interference in the case of more than two flavours. 

\paragraph{Explicit solution.} 
We now return to eq.~\eqref{eq:Gpoles} with the aim of finding the explicit solution for the 
deviations from equilibrium $\delta n^{ij}(t)$. The relevant transport equations for determining 
the functional form of the \emph{spectrally-dressed} number densities 
$\delta n^{ij}_{\rm dr}(t)$ can be written in the general form~\cite{Dev:2014wsa}
\begin{align}
\label{eq:transport}
\frac{\mathrm{d}\,\delta n^{ij}_{\rm dr}(t)}{\mathrm{d}t}\ &=\ \int\frac{\mathrm{d} p_0}{2\pi}\,\int\frac{\mathrm{d} p_0'}{2\pi}\;e^{-i(p_0-p_0')\tilde{t}}\theta(p_0)\theta(p_0')\nonumber\\
& \times\:\Bigl(-i\big[\bm{M}^2+\mathrm{Re}\,\bm{\Pi}_R,\bm{G}_\delta]_{\star}^{ij}\:
-\:\textstyle{\frac{1}{2}}\big\{\bm{\Pi}_<,\bm{G}_{\delta}\big\}_{\star}^{ij}\:+\:\textstyle{\frac{1}{2}}\big\{\bm{\Pi}_>,\bm{G}_\delta\big\}^{ij}_{\star}\Bigr)\;.
\end{align}
In the Markovian approximation, the commutators and anti-commutators appearing in eq.~\eqref{eq:transport} 
are defined as follows~\cite{Dev:2014wsa}:
\begin{subequations}
\begin{align}
[\bm{A},\bm{B}]_{\star}\ &\equiv\ \int_k \Big(\bm{A}(p,k)\cdot\bm{B}(k,p')\:-\:\bm{B}(p,k)\cdot\bm{A}(k,p')\Big)\;,\\
\{\bm{A},\bm{B}\}_{\star}\ &\equiv\ \int_k \Big(\bm{A}(p,k)\cdot\bm{B}(k,p')\:+\:\bm{B}(p,k)\cdot\bm{A}(k,p')\Big)\;,
\end{align}
\end{subequations}
in which we emphasize the order of the four-momenta.

In order to find the asymmetry at first order in $\Pi_{R(A)}$, we require the solution for the 
diagonals only at zeroth order. Thus, for the diagonals, we may work in terms of the tree-level 
$\bm{G}_\delta^0$, as obtained from eq.~\eqref{eq:freeprop}, yielding the following equation for the 
spectrally-free number densities
\begin{equation}
\frac{\mathrm{d}\delta n^{ii}}{\mathrm{d}t}\ =\ -\:\Gamma_i\,\delta n^{ii}\;,
\end{equation}
with solution
\begin{equation}
\delta n^{ii}(t)\ =\ e^{-\Gamma_i t}\,\delta n^{ii}(0)\;.
\end{equation}
For the off-diagonals, we may work in terms of the tree-level $\bm{G}_\delta^0$ in the time-derivative 
on the lhs and in all terms already at leading order in $\Pi_{R(A)}$ on the rhs, i.e.
\begin{align}
\label{eq:transport1}
&\frac{\mathrm{d}\,\delta n^{i\slashed{i}}(t)}{\mathrm{d}t}\ \supset\ \int\frac{\mathrm{d} p_0}{2\pi}\,\int\frac{\mathrm{d} p_0'}{2\pi}\;e^{-i(p_0-p_0')\tilde{t}}\theta(p_0)\theta(p_0')\nonumber\\&\qquad  \times\:\Bigl(-i\big[\mathrm{Re}\,\bm{\Pi}_R,\bm{G}_\delta^0]_{\star}^{i\slashed{i}}\:-\:
\tfrac{1}{2}\big\{\bm{\Pi}_<,\bm{G}_\delta^0\big\}_{\star}^{i\slashed{i}}\:+\:
\tfrac{1}{2}\big\{\bm{\Pi}_>,\bm{G}_\delta^0\big\}^{i\slashed{i}}_{\star}\Bigr)\;.
\end{align}
On the other hand, in the term
\begin{equation}
\label{eq:transport2}
\frac{\mathrm{d}\,\delta n^{i\slashed{i}}(t)}{\mathrm{d}t}\ \supset\ -\,i\,\int\frac{\mathrm{d} p_0}{2\pi}\,\int\frac{\mathrm{d} p_0'}{2\pi}\;e^{-i(p_0-p_0')\tilde{t}}\theta(p_0)\theta(p_0')\,\bigl[\bm{M}^2,\bm{G}_\delta\bigr]_{\star}^{i\slashed{i}}\;,
\end{equation}
we must use the resummed $\bm{G}_\delta$, since this will also contribute a term at first order 
in $\Pi_{R(A)}$. Proceeding in this manner, we find the equation for the off-diagonals
\begin{equation}
\label{eq:transoffdiag}
\frac{\mathrm{d}\delta n^{i\slashed{i}}}{\mathrm{d}t}\ =\ -\,i(\omega_i-\omega_\slashed{i})\,\delta 
n^{i\slashed{i}}\:-\:\bar{\Gamma}\,\delta n^{i\slashed{i}}\:-\:\frac{i}{2\bar{\omega}}\big(\Pi_R^{i\slashed{i}}(\omega_i)\,\delta n^{\slashed{i}\slashed{i}}\:-\:\Pi_A^{i\slashed{i}}(\omega_\slashed{i})\,\delta n^{ii}\big)\;,
\end{equation}
in which we have used the approximation $\omega_i\omega_\slashed{i}\approx \bar{\omega}^2$.

It is interesting to remark upon the origin of the order $\Pi_{R(A)}$ terms in eq.~\eqref{eq:transoffdiag}. 
The terms originating from the diagonal elements of the \emph{tree-level} $\bm{G}_\delta^0$ in 
eq.~\eqref{eq:transport1} are
\begin{equation}
\frac{\mathrm{d}\delta n^{i\slashed{i}}}{\mathrm{d} t}\ \supset\ -\,\frac{i}{2\bar{\omega}}\,\delta n^{\slashed{i}\slashed{i}}\,\Pi_R^{i\slashed{i}}(\omega_\slashed{i})\:+\:\frac{i}{2\bar{\omega}}\,\delta n^{ii}\,\Pi^{i\slashed{i}}_A(\omega_i)\;.
\end{equation}
Instead, the off-diagonal element of the \emph{resummed} $\bm{G}_\delta$ used in the commutator in 
eq.~\eqref{eq:transport2} yields the following terms:
\begin{equation}
\frac{\mathrm{d}\delta n^{i\slashed{i}}}{\mathrm{d} t}\ \supset\ -\,\frac{i}{2\bar{\omega}}\,\delta n^{ii}\bigl(\Pi_A^{i\slashed{i}}(\omega_i)\:-\:\Pi_A^{i\slashed{i}}(\omega_\slashed{i})\bigr)\:+\:\frac{i}{2\bar{\omega}}\,\delta n^{\slashed{i}\slashed{i}}\,\bigl(\Pi^{i\slashed{i}}_R(\omega_\slashed{i})\:-\:\Pi^{i\slashed{i}}_R(\omega_i)\bigr)\;.
\end{equation}
We see that the contribution from the resummed $\bm{G}_\delta$ in eq.~\eqref{eq:transport2}, which 
amounts to the interference of mixing and oscillation effects, swaps the arguments of the terms at 
first order in $\Pi_{R(A)}$.

The leading-order solutions to the off-diagonal equations in eq.~\eqref{eq:transoffdiag} have the form
\begin{equation}
\delta n^{i\slashed{i}}(t)\ =\ e^{-i(\omega_i-\omega_\slashed{i})t}e^{-\bar{\Gamma} t}\delta n_0^{i\slashed{i}}(0)\:+\:\frac{\Pi_A^{i\slashed{i}}(\omega_\slashed{i})}{\Delta M_{i\slashed{i}}^2}\,e^{-\Gamma_{i}t}\,\delta n^{ii}(0)\:-\:\frac{\Pi_R^{i\slashed{i}}(\omega_i)}{\Delta M_{i\slashed{i}}^2}\,e^{-\Gamma_{\slashed{i}}t}\,\delta n^{\slashed{i}\slashed{i}}(0)\;,
\end{equation}
where $\delta n^{i\slashed{i}}_0$ is the initial condition at zeroth order in $\Pi_{R(A)}$.
Next, we re-express this result in terms of the full initial condition for the off-diagonals 
$\delta n^{i\slashed{i}}(0)$:
\begin{equation}
\delta n^{i\slashed{i}}(0)\ =\ \delta n_0^{i\slashed{i}}(0)\:+\:\frac{\Pi_A^{i\slashed{i}}(\omega_\slashed{i})}{\Delta M_{i\slashed{i}}^2}\,\delta n^{ii}(0)\:-\:\frac{\Pi_R^{i\slashed{i}}(\omega_i)}{\Delta M_{i\slashed{i}}^2}\,\delta n^{\slashed{i}\slashed{i}}(0)\;.
\end{equation}
We then obtain the final form of the solution:
\begin{align}
\label{eq:expsols}
\delta n^{i\slashed{i}}(t)\ &=\ e^{-i(\omega_i-\omega_\slashed{i})t}e^{-\bar{\Gamma} t}\delta n^{i\slashed{i}}(0)\nonumber\\
&+\:\frac{\Pi_A^{i\slashed{i}}(\omega_\slashed{i})}{\Delta M_{i\slashed{i}}^2}\,\big(e^{-\Gamma_{i}t}-e^{-i(\omega_i-\omega_\slashed{i})t}e^{-\bar{\Gamma} t}\big)\,\delta n^{ii}(0)\nonumber\\
& -\:\frac{\Pi_R^{i\slashed{i}}(\omega_i)}{\Delta M_{i\slashed{i}}^2}\,\big(e^{-\Gamma_\slashed{i}t}
-e^{-i(\omega_i-\omega_\slashed{i})t}e^{-\bar{\Gamma} t}\big)\,\delta n^{\slashed{i}\slashed{i}}(0)\;.
\end{align}
Substituting eq.~\eqref{eq:expsols} into eq.~\eqref{eq:DeltaGint} for the off-diagonal components 
of $\bm{G}_\delta$, we obtain \emph{precisely} eq.~\eqref{GF12Expansion}, as derived in the Heisenberg 
picture. This remarkable agreement provides a significant illustration of both the self-consistency 
and complementarity of the Heisenberg- and interaction-picture approaches described in this article. 

Finally, we see from eq.~\eqref{eq:expsols} the reason for defining the pure oscillation and interference terms with respect to the \emph{time-dependent} number densities in eqs.~\eqref{eq:DeltaGint} and \eqref{dqdttime} and not with respect to their initial conditions, as appear in eqs.~\eqref{GF12Expansion} and \eqref{dqdtapprox}. In the weak-washout regime treated here, the explicit solution for the off-diagonals $\delta n^{i\slashed{i}}(t)$ in eq.~\eqref{eq:expsols} contains terms proportional to the initial deviation of the diagonal components $\delta n^{ii}(0)$ and $\delta n^{\slashed{i}\slashed{i}}(0)$, which appear comparable to the terms in the final line of eqs.~\eqref{eq:DeltaGint}  and \eqref{dqdttime}. On the other hand, in the strong-washout regime, whilst the solution for the off-diagonals $\delta n^{i\slashed{i}}(t)$ would no longer depend on the initial conditions, the final line in eq.~\eqref{dqdttime} would still be present.

\section{\label{sec:densitymatrix}Comparison with the density matrix approximation}

In the two previous sections, we have paid particular attention to the shell structure of the would-be heavy-neutrino propagators and emphasized the need to keep track of this complete structure in the equation for the asymmetry. However, in phenomenological studies and in order to obtain the final lepton asymmetry, we need first to solve the evolution equations for the number densities of the heavy neutrinos. In contrast to the equation for the asymmetry, for an almost degenerate mass spectrum, there is no need to keep track of the different
shells in the evolution of the heavy-neutrino number densities. 
Such a \emph{single shell} approximation applied to the heavy-neutrino transport equations will be referred to as the \emph{density matrix 
approximation} in what follows. In the Kadanoff-Baym formalism, this apparent disparity between the treatment of the equation for the asymmetry and the evolution equations of the heavy-neutrino number densities is actually a result of making a self-consistent loop-wise perturbative truncation of the leptonic and heavy-neutrino KB equations (see refs.~\cite{Dev:2014wsa} and~\cite{Millington:2012pf}).
 
Performing a Wigner-transform of the Kadanoff-Baym equations [eq.~\eqref{KBeqs_real}] and neglecting the gradient
terms, one arrives at the following equation for the Wigner transform of the Wightman propagators~\cite{Garbrecht:2011aw}:
\begin{align}
\label{KBWigner}
2q_0 \partial_t \bm{G}_<\:+\:i\bigl[\bm{M}^2,\bm{G}_<\bigr]\ =\ 
\tfrac12\bigl\{\bm{\Pi}_>,\bm{G}_<\bigr\}\:-\:\tfrac12\bigl\{\bm{\Pi}_<,\bm{G}_>\bigr\}\,,
\end{align}
where the term $\Re\, \bm{\Pi}_R$ [see eq.~\eqref{eq:transport}] is understood to have been absorbed into the mass 
matrix $\bm{M}^2$. In the density matrix approximation~\cite{Cirigliano:2009yt,Garbrecht:2011aw},
\begin{align}
\label{FlavorKBAnsatz}
\bm{G}_<(t,q)\ =\ \bm{n}(t,\vec{q})\,2\pi \,\delta(q^2-\bar{M}^2)\,,
\end{align}
where $\bar{M}^2=(M_1^2+M_2^2)/2$ and the density matrix (or, to be more precise, the matrix of densities) $\bm{n}(t,\vec{q})$ can be 
viewed as a dressed distribution function. Substituting eq.~\eqref{FlavorKBAnsatz} into eq.~\eqref{KBWigner},
we arrive at 
\begin{align}
2q_0 \partial_t \bm{n}\:+\:i\bigl[\bm{M}^2,\bm{n}\bigr]\ =\ 
\tfrac12\bigl\{\bm{\Pi}_>,\bm{n}\bigr\}\:-\:\tfrac12\bigl\{\bm{\Pi}_<,\bm{1}+\bm{n}\bigr\}\,.
\end{align}
Finally, integrating over $q_0$ and using the single-shell approximation in eq.~\eqref{FlavorKBAnsatz}, 
we obtain the kinetic equation for the matrix of densities 
\begin{align}
\label{DensMatrixEqs}
\partial_t \bm{n}\:+\:\frac{i}{2\bar\omega}\bigl[\bm{M}^2,\bm{n}\bigr]\ =\ 
\frac{1}{4\bar\omega}\bigl\{\bm{\Pi}_>,\bm{n}\bigr\}\:-\:
\frac{1}{4\bar\omega}\bigl\{\bm{\Pi}_<,\bm{1}+\bm{n}\bigr\}\,.
\end{align}

In equilibrium, $\bm{n}$ is time independent and, as follows from eq.~\eqref{DensMatrixEqs}, 
satisfies
\begin{align}
\label{DensMatrixEqsEq}
2i\bigl[\bm{M}^2,\bm{n}_{\rm eq}\bigr]\ =\ 
\bigl\{\bm{\Pi}_>,\bm{n}_{\rm eq}\bigr\}\:-\:\bigl\{\bm{\Pi}_<,\bm{1}+\bm{n}_{\rm eq}\bigr\}\,.
\end{align}
The Kubo-Martin-Schwinger relation (see e.g.~ref.~\cite{Lebellac:2000bj}) implies that $\bm{\Pi}_<=\widetilde{\bm{\Pi}}_\rho n_{\rm BE}$ in equilibrium,
where $n_{\rm BE}$ is the Bose-Einstein distribution function. Further, taking into account that 
$\widetilde{\bm{\Pi}}_\rho=\bm{\Pi}_>-\bm{\Pi}_<$, we conclude that 
$\bm{n}_{\rm eq}=\bm{1}\cdot n_{\rm BE}\,$ in the density matrix approximation, i.e.~that the equilibrium distribution function is diagonal
and, as expected, given by the Bose-Einstein distribution.
 
Writing $\bm{n}=\bm{n}_{\rm eq}+\delta \bm{n}$ and using eqs.~\eqref{DensMatrixEqs} and 
\eqref{DensMatrixEqsEq}, we obtain the kinetic equation for the deviation of the matrix of densities from its 
equilibrium value 
\begin{align}
\partial_t\delta\bm{n}\:+\:\frac{i}{2\bar{\omega}}\bigl[\bm{M}^2,\delta\bm{n}\bigr]\ =\ 
\frac{1}{4\bar{\omega}}\bigl\{\widetilde{\bm{\Pi}}_\rho,\delta\bm{n}\bigr\}\,.
\end{align}
Finally, introducing the matrix of effective decay widths 
\begin{align}
\label{GammaDef}
\bm{\Gamma}\ \equiv\ -\,\frac{\widetilde{\bm{\Pi}}_\rho}{2\bar \omega}\ =\ \frac{i\bm{\Pi}_{\rho}}{2\bar{\omega}}\,,
\end{align}
and using the standard trick $\bm{M^2}\rightarrow 2\bar{\omega}\,\bm{\omega}$, we arrive at 
the equation
\begin{align}
\label{eq:densmateq}
\partial_t\delta\bm{n}\:+\:i\bigl[\bm{\omega},\delta\bm{n}\bigr]\ =\ 
-\,\tfrac12\bigl\{\bm{\Gamma},\delta\bm{n}\bigr\}\,,
\end{align}
as obtained in ref.~\cite{Garbrecht:2011aw}.

It can readily be checked by substitution that the solution to eq.~\eqref{eq:densmateq} is
\begin{align}
\delta\bm{n}(t)\ =\ e^{-i\left(\bm{\omega}-\tfrac{i}{2}\bm{\Gamma}\right)t}\,
\delta \bm{n}(0)\,
e^{i\left(\bm{\omega}+\tfrac{i}{2}\bm{\Gamma}\right)t}\,.
\end{align}
Choosing, as in  ref.~\cite{Garbrecht:2011aw}, initial conditions of the form 
\begin{align}
\label{BjornInitCond}
\delta  \bm{n}(0)\ =\ 
\begin{pmatrix}
\delta n^{11}(0) & 0\\
0 & 0
\end{pmatrix}\,,
\end{align}
we obtain, to leading order in $\bm{\Gamma}$,
\begin{align}
\label{BjornSol}
\delta n^{i\slashed{i}}(t)\ \approx\ \frac{i}{2} \frac{\Gamma^{i\slashed{i}} }{(\omega_i-\omega_\slashed{i})
\:+\:\tfrac{i}{2}(\Gamma_i-\Gamma_\slashed{i})}\,\delta n^{11}(0) \Bigl(e^{-\Gamma_i t}\:-\: e^{-i(\omega_i-
\omega_\slashed{i})t}e^{-\bar{\Gamma}t}\Bigr)\,,
\end{align}
which is in exact agreement with the result of ref.~\cite{Garbrecht:2011aw}.
For general initial conditions, eq.~\eqref{BjornSol} takes the form
\begin{align}
\delta n^{i\slashed{i}}(t)\ &\approx\ 
\delta n^{i\slashed{i}}(0)\,e^{-i(\omega_i-\omega_\slashed{i})t}\,e^{-\bar\Gamma t}\nonumber\\
&+\ \frac{i}{2} \frac{\Gamma^{i\slashed{i}} }{(\omega_i-\omega_\slashed{i})
\:+\:\tfrac{i}{2}(\Gamma_i-\Gamma_\slashed{i})}\,\delta n^{ii}(0) \Bigl(e^{-\Gamma_i t}\:-\: e^{-i(\omega_i-
\omega_\slashed{i})t}e^{-\bar{\Gamma}t}\Bigr) \nonumber\\
&+\ \frac{i}{2} \frac{\Gamma^{i\slashed{i}} }{(\omega_i-\omega_\slashed{i})
\:+\:\tfrac{i}{2}(\Gamma_\slashed{i}-\Gamma_i)}\,\delta n^{\slashed{i}\slashed{i}}(0)
\Bigl(e^{-\Gamma_\slashed{i} t}\:-\: e^{-i(\omega_i-
\omega_\slashed{i})t}e^{-\bar{\Gamma}t}\Bigr)\,.
\end{align}
Multiplying this expression by $2\bar\omega$ and using again eq.~\eqref{GammaDef}, we find 
\begin{align}
\delta n^{i\slashed{i}}(t)\ &\approx\
\delta n^{i\slashed{i}}(0)\,e^{-i(\omega_i-\omega_\slashed{i})t}\,e^{-\bar\Gamma t}\nonumber\\
&-\ \frac{i}{2} \frac{\widetilde{\Pi}_{\rho}^{i\slashed{i}}}{\Delta M^2_{i\slashed{i}}}\,\delta n^{ii}(0) \Bigl(e^{-\Gamma_i t}\:-\: e^{-i(\omega_i-
\omega_\slashed{i})t}e^{-\bar{\Gamma}t}\Bigr) \nonumber\\
&-\ \frac{i}{2} \frac{\widetilde{\Pi}_{\rho}^{i\slashed{i}}}{\Delta M^2_{i\slashed{i}}}\,\delta n^{\slashed{i}\slashed{i}}(0)
\Bigl(e^{-\Gamma_\slashed{i} t}\:-\: e^{-i(\omega_i-
\omega_\slashed{i})t}e^{-\bar{\Gamma}t}\Bigr)\;,
\end{align}
which, in the density matrix approximation, is identical to eq.~\eqref{eq:expsols}.
In other words, for an almost degenerate mass spectrum, one can safely use the density matrix equations 
to compute the number density of the heavy neutrinos.

\section{\label{sec:vacuumcp}Comparison with the effective Yukawa approach}

As was emphasized in the preceding sections, it is necessary to account for the shell structure at the level of the equation for the asymmetry for \emph{all} mass spectra. This is necessary in order to capture the effect of mixing. In this section, we will compare the result from the present analysis with that of the semi-classical analysis of refs.~\cite{Dev:2014laa,Dev:2015wpa} and the KB analysis of ref.~\cite{Dev:2014wsa}. Therein, the effect of oscillations is captured by accounting for the dynamics of flavour coherences, encoded in the off-diagonal elements of the number density. The effect of mixing is accounted for by means of effective Yukawa couplings, following the work of refs.~\cite{Pilaftsis:1997jf,Pilaftsis:2003gt}. We briefly review the derivation of these effective Yukawa couplings for the present toy model below, as were treated in ref.~\cite{Garny:2009qn}. We reiterate that this approach was not followed in the preceding sections.

The heavy mixing scalars are unstable, and as such  they cannot appear as asymptotic  in- or out-states
of \emph{S}-matrix elements. Instead, their properties are defined by \emph{S}-matrix
elements for scattering of stable particles, $b$ and $\bar b$, mediated by the unstable ones~\cite{Veltman:1963th}.
Resumming the propagator of the intermediate heavy
states, we can represent two-body scattering processes as a sum of  resonant and non-resonant contributions.
The \CP-violating part of the resonant contribution can then be 
interpreted as a characteristic of the on-shell intermediate particle~\cite{Pilaftsis:1997jf,Plumacher:1997ru}.

The  amplitude of the $s$-channel two-body scattering process $bb\rightarrow
\bar b \bar b$ can be expressed as 
\begin{align}
\label{ScatteringVac}
\mathcal{M}_{bb\rightarrow \bar b\bar b}\ =\ \sum_{i,j}\Gamma_i^A
G^{ij}(s)\Gamma_j^B\,,
\end{align}
where $\Gamma_i^A$ and $\Gamma_j^B$ represent the vertices $\psi_i bb$ and 
$\psi_j \bar b\bar b$, including the wave functions of the initial and 
final states, and $G^{ij}$ are the full vacuum propagators obtained by resumming an 
infinite series of self-energy graphs~\cite{Pilaftsis:1997jf}. The resummation can be performed using the Schwinger-Dyson equation in vacuum:
\begin{align}
\label{SDVac}
[G^{-1}]^{ij}(p^2)\ =\ \left[p^2-M_i^2\right]\delta^{ij}-\Pi^{ij}(p^2)\,.
\end{align}
At one-loop level and in the on-shell scheme, the renormalized self-energy $\Pi^{ij}$ is given by~\cite{Garny:2009qn}
\begin{subequations}
\label{PiRenVacExpl}
\begin{align}
\Pi_{\rm ren}^{ii} \ &=\ \frac{H_{ii}}{16\pi^2}\left[
\ln \frac{|p^2|}{M_i^2}\:-\:\frac{p^2-M_i^2}{M_i^2}\:-\:i\pi\theta(p^2)
\right]\,,\\
\Pi_{\rm ren}^{i\slashed{i}}\ &=\ \frac{\Re H_{i\slashed{i}}}{16\pi^2}\biggl[
\frac{p^2-M_i^2}{M_\slashed{i}^2-M_i^2}\ln\frac{|p^2|}{M_\slashed{i}^2}
\:+\:\frac{p^2-M_\slashed{i}^2}{M_i^2-M_\slashed{i}^2}\ln \frac{|p^2|}{M_i^2}
\:-\:i\pi\theta(p^2)
\biggr]\,.
\end{align}
\end{subequations}
Inverting eq.~\eqref{SDVac}, we obtain the following result for the components of the renormalized 
resummed propagator:
\begin{subequations}
\label{GijRenVac}
\begin{align}
G^{ii}(p^2)\ &=\ +\,[G^{-1}]^{jj}(p^2)/\det [\bm{G}^{-1}(p^2)]\,,\\
G^{i\slashed{i}}(p^2)\ &=\ -\,[G^{-1}]^{i\slashed{i}}(p^2)/\det [\bm{G}^{-1}(p^2)]\,.
\end{align}
\end{subequations}
Because of the presence of   absorptive terms in eq.~\eqref{PiRenVacExpl}, the 
determinant of the inverse propagator in eq.~\eqref{GijRenVac} has 
two poles in the complex plane at 
\begin{align}
s_i\ \simeq\ M_i^2\:-\:iM_i\Gamma_i\,,
\end{align}
where $\Gamma_i=H_{ii}/16\pi M_i$ is the tree-level decay 
width of $\psi_i$. Expanding eq.~\eqref{GijRenVac} around the poles 
and substituting the leading expansion terms into eq.~\eqref{ScatteringVac},
we find \cite{Pilaftsis:2003gt}
\begin{align}
\label{Mrepr}
\mathcal{M}_{bb\rightarrow \bar b \bar b}\ \simeq\ 
\sum_i V_i^A(s)\frac{1}{s-s_i}V_i^B(s)\,,
\end{align}
where 
\begin{align}
\label{ViAB}
V_i^{A(B)}(s)\ &\equiv\ \Gamma_i^{A(B)}-\frac{[G^{-1}]^{i\slashed{i}}(s)}{[G^{-1}]^{\slashed{i}\slashed{i}}(s)}
\Gamma_{\slashed{i}}^{A(B)}
\ =\ \Gamma_i^{A(B)}+\frac{\Pi^{i\slashed{i}}(s)}{s-M_\slashed{i}^2-\Pi^{\slashed{i}\slashed{i}}(s)}
\Gamma_\slashed{i}^{A(B)}\,.
\end{align}
Equation \eqref{ViAB} can be used to define effective one-loop Yukawa couplings $\mathbf{h}^{(c)}_i$. Taking 
into account that the couplings in the vertices $\psi_ibb$ and $\psi_i\bar b\bar b$ differ 
by complex conjugation, we obtain 
\begin{subequations}
\label{EffVacuumCoupls}
\begin{align}
\mathbf{h}_i\ &\equiv\ h_i\:+\:\frac{\Pi^{i\slashed{i}}(s)}{s-M_\slashed{i}^2-\Pi^{\slashed{i}\slashed{i}}(s)}h_\slashed{i}\,,\\
\mathbf{h}^{c*}_i\ &\equiv\ h^*_i\:+\:\frac{\Pi^{i\slashed{i}}(s)}{s-M_\slashed{i}^2-\Pi^{\slashed{i}\slashed{i}}(s)}h^*_\slashed{i}\,,
\end{align}
\end{subequations}
which, as follows from eq.~\eqref{Mrepr}, are to be evaluated on the mass shell of the  $i$-th
quasi-particle~\cite{Pilaftsis:2003gt}. Using eqs.~\eqref{PiRenVacExpl} and the tree-level 
relation $H_{ii}=16\pi M_i\Gamma_i$, we obtain
\begin{subequations}
\begin{align}
\Pi_{i\slashed{i}}(M_i^2)\ &=\ -\,i\frac{\Re H_{i\slashed{i}}}{16\pi}\,,\\
\label{Pijj}
\Pi_{\slashed{i}\slashed{i}}(M_i^2)\ &=\ M_\slashed{i}\Gamma_\slashed{i}\left[\frac{1}{\pi}\left(
\ln\biggl(\frac{M_i^2}{M_\slashed{i}^2}\biggr)-\frac{M_i^2-M_\slashed{i}^2}{M_\slashed{i}^2}\right)-i\right]\,.
\end{align}
\end{subequations}
Note that, for $M_i\approx M_\slashed{i}$, each term in the round brackets in eq.~\eqref{Pijj} 
vanishes and therefore this contribution can be neglected. On the other hand, for 
$M_i\ll M_\slashed{i}$, the difference in the round brackets of eq.~\eqref{Pijj} only increases slowly 
with growing $M_\slashed{i}/M_i$ and is negligibly small compared to the $M_i^2-M_\slashed{i}^2$ term in 
the denominator of eq.~\eqref{EffVacuumCoupls}. Therefore, for practical purposes, it is sufficient to keep only the imaginary 
part of eq.~\eqref{Pijj}, i.e.~use $\Pi_{\slashed{i}\slashed{i}}(M_i^2)\approx -i M_\slashed{i}\Gamma_\slashed{i}$. In this way, we arrive at the effective Yukawa couplings
\begin{equation}
\mathbf{h}^{(c)}_i\ =\ h_i\bigg[1\:-(+)\: i\,\frac{H_{\slashed{i}\slashed{i}}}{32\pi}\bigg(1\:+\:\frac{H^*_{i\slashed{i}}}{H_{i\slashed{i}}}\bigg)\frac{1}{\Delta M_{i\slashed{i}}^2\:+(-)\:iM_\slashed{i}\Gamma_{\slashed{i}}}\bigg]\;.
\end{equation}
The self-energy contribution to the \CP-violating parameter in vacuum takes the form
\begin{align}
\label{EspilonVac}
\epsilon_i^{\rm vac}\ \equiv\ \frac{\Gamma_{\psi_i \rightarrow{b}{b}}
\:-\:\Gamma_{\psi_i \rightarrow\bar{b}\bar{b}}}{
\Gamma_{\psi_i \rightarrow{b}{b}}
\:+\:\Gamma_{\psi_i \rightarrow\bar{b}\bar{b}}}\,,
\end{align} 
and we find
\begin{align}
   \label{epsilonClassVac}
   \epsilon_i^{\rm vac}\ =\ 
   {\rm Im}\biggl(\frac{H_{i\slashed{i}}}{H^*_{i\slashed{i}}}\biggr)
   \frac{(M_i^2-M_\slashed{i}^2)M_\slashed{i}\Gamma_\slashed{i}}{(M_i^2 -M_\slashed{i}^2)^2+(M_\slashed{i} \Gamma_\slashed{i})^2}\,,
\end{align}
cf.~eq.~\eqref{EpsilonMed}.

Following ref.~\cite{Dev:2014wsa} (see Appendix~\ref{sec:noneq} for a comparison of conventions), the time-derivative of the asymmetry can be written in terms of the effective Yukawa couplings as
\begin{equation}
\frac{\mathrm{d}\eta}{\mathrm{d}t} \  \approx\ \int_q\theta(q_0)\Big[\mathbf{h}_i\mathbf{h}_{j}^*G_{\delta}^{0,\,ij}(t,q)\:-\:\mathbf{h}_i^{c*}\mathbf{h}_j^c G_{\delta}^{0,\,ij}(t,q)\Big]\widetilde{\Pi}_{\rho}(q)\;,
\end{equation}
where $G_{\delta}^{0,\,ij}(t,q)$ is the non-equilibrium part of the \emph{tree-level} propagator. As discussed in section~\ref{sec:densitymatrix}, the tree-level heavy-neutrino propagator may be written in the on-shell approximation (see ref.~\cite{Dev:2014wsa})
\begin{equation}
G_{\delta}^{0,\,ij}(t,q_0>0)\ =\ 2\pi\delta(q_0-\bar{\omega})\frac{1}{2\bar{\omega}}\delta n^{ij}(t,\mathbf{q})\;.
\end{equation}
The time-derivative of the asymmetry is then found to be
\begin{align}
\frac{\D\eta}{\D t}\ & \approx\  2 \sum_i \int_{\vec q}\frac{M_i}{\bar{\omega}} \,\delta n^{ii}(t,\vec{q}) 
\, \Gamma_i(\bar{\omega},\vec{q})\, \epsilon^{\rm vac}_i(\bar{\omega},\vec{q})\nonumber\\
&+\ 2\,\Im\,H_{12}\,\int_{\vec q}\frac{\widetilde{\Pi}_{\rho}(\bar{\omega},\mathbf{q})}{\bar{\omega}}\,\Im\,\delta n^{12}(t,\vec{q})\,.
\end{align}
With the exception of the additional factor of $\Delta M^2_{12}R_{12}$, which is of order unity 
in the weakly-resonant regime, the oscillation contribution resembles that appearing in 
eq.~\eqref{dqdttime}. On the other hand, the mixing contribution does not see the modifications 
that resulted in eq.~\eqref{dqdttime} from the interference terms in the final line of 
eq.~\eqref{eq:DeltaGint}. This conclusion is suggestive that the approach of 
refs.~\cite{Dev:2014laa,Dev:2015wpa,Dev:2014wsa}, although accounting successfully for both mixing 
and oscillation, may not fully capture the interference of these two effects. However, as identified earlier in section~\ref{sec:interaction}, the approximation used in (appendix A.1 of) ref.~\cite{Dev:2014wsa} does not hold in the weak-washout regime studied here, and it would be of interest to study the impact of these interference effects quantitatively in the strong-washout regime.

\section{\label{sec:pheno}Phenomenological implications}

In order to get a feeling for the relative size of the mixing, oscillation and interference  contributions, 
we now compute the asymmetry for various values of the degeneracy parameter 
\begin{align}
\label{eq:degenparam}
R\ \equiv\ \frac{M_2^2-M_1^2}{M_1\Gamma_1+M_2\Gamma_2}\,.
\end{align}
In the weak-washout regime, to which we limit ourselves in this work, the impact of the 
initial conditions on the final asymmetry is non-negligible. Therefore, in order to ensure 
that the produced asymmetry is of dynamical origin, we need to choose \C-symmetric initial 
conditions in our numerical analysis.  As is shown in refs.~\cite{Hohenegger:2013zia,
Hohenegger:2014cpa}, for a non-degenerate mass spectrum, the Lagrangian in eq.~\eqref{Lagrangian} 
is \C-symmetric if either $\Im\, H_{12}=0$ or $\Re\,H_{12}=0$. It is also automatically 
\C-symmetric for a degenerate mass spectrum. This can be summarized conveniently by forming  
the familiar basis-independent measure of both $C$- and $CP$-violation, the Jarlskog invariant 
\begin{equation}
\label{JarslkogInv}
J\ =\ \mathrm{Im}\,\Tr\,\bm{H}\bm{M}^3\bm{H}^{\mathsf{T}}\bm{M}\;,
\end{equation}
which vanishes when any of the $C$-conserving conditions are satisfied (see 
appendix~\ref{sec:syms} and ref.~\cite{Hohenegger:2014cpa} for more details).

Whereas the rhs of eq.~\eqref{dqdtapprox} 
vanishes in the  limit $\Im\, H_{12}=0$, it does not automatically vanish when $\Re\, H_{12}=0$
except in the absence of initial flavour coherences, i.e.~for $\delta  n^{12}(0)=0$. This implies that there are two ways to specify the initial 
conditions such that the asymmetry automatically vanishes if the Lagrangian is \C-conserving. 
The first possibility, studied in ref.~\cite{Hohenegger:2014cpa}, is to 
choose $\mathcal{K}^{12}=0$, i.e.~to set the leading oscillation term to zero. The second 
possibility is to require that the initial conditions be \C-symmetric, which, in the Heisenberg
picture, corresponds to choosing $G^{ij}_{\!\delta}(0,0)$ diagonal in the mass eigenbasis  
(see appendix~\ref{sec:syms} for more details). It can be shown by virtue of the constraints provided on the two-point functions by causality, unitarity, $CTP$ invariance and Hermiticity (see~ref.~\cite{Millington:2012pf}) that the source $K^{mn}$
has the same properties as the statistical 
one-loop self-energy in the thermal bath. Thus, it follows that \smash{$\mathcal{K}^{mn}=\mathcal{K}^{nm}$}.
Using this symmetry and requiring $G_{\!\delta}^{12}(0,0)$ to vanish, we obtain 
\begin{align}
\label{CSymmInitCond}
\mathcal{K}^{12}\ =\ \mathcal{K}^{21}\ =\ -\,\frac{\mathcal{K}^{11}
G_R^{11}(0)\,G_A^{12}(0)\:+\:\mathcal{K}^{22}\,G_R^{12}(0)
G_A^{22}(0)}{G_R^{11}(0)\,G_A^{22}(0)\:+\:G_R^{12}(0)\,
G_A^{12}(0)}\,,
\end{align}
where $0=0^+$ for the retarded and $0=0^-$ for the advanced propagator. The properties of the 
propagators also imply $G_A^{ij}(0)=G_R^{ji}(0)$, and therefore it is sufficient to 
consider for example only the retarded one. Expressed in terms of its Wigner transform, 
the retarded propagator takes the form 
\begin{align}
\label{GRAWigner}
G_{R}^{ij}(0)\ =\ 2\int^\infty_0\frac{\D q_0}{2\pi}\,\Re\,G_{R}^{ij}(q_0)\
\approx\ -\:2\,\Im\frac1{\Delta \Omega^2} \sum_{k\,=\,1}^{2}
\frac{(-1)^k}{2\omega_k} [\adj \bm{D}_R(\omega_k)]^{ij}\,,
\end{align}
where we have used eq.~\eqref{DRapprox} and Cauchy's theorem to  evaluate the 
integral approximately.
It follows from this expression that $G_R^{12}\propto \Pi_R^{12}\propto \Re H_{12}$ and similarly  
$G_A^{12}\propto \Pi_A^{12}\propto \Re H_{12}$. Thus, we find $\mathcal{K}^{12}=\mathcal{K}^{21} \propto 
\Re H_{12}$, such that, for a \C-symmetric choice of the initial conditions, the produced asymmetry 
automatically vanishes when either $\Im H_{12}=0$ or $\Re H_{12}=0$. In other words, although this is 
not immediately obvious because the $\psi_i$ are not necessarily on shell, for a \C-symmetric choice 
of the initial conditions, the mixing and oscillation sources of the asymmetry are proportional to the 
Jarlskog invariant in eq.~\eqref{JarslkogInv} (see ref.~\cite{Hohenegger:2014cpa} for a detailed discussion). As has been shown in 
refs.~\cite{Dev:2014laa,Dev:2015wpa,Dev:2014wsa}, in the strong-washout regime, in which the final asymmetry is known to be 
independent of the initial conditions, the solution of the kinetic equations automatically possesses
this property. For the toy model under consideration, we demonstrate this in appendix~\ref{sec:rate}.

\begin{figure}[h!]
\includegraphics[width=0.5\textwidth]{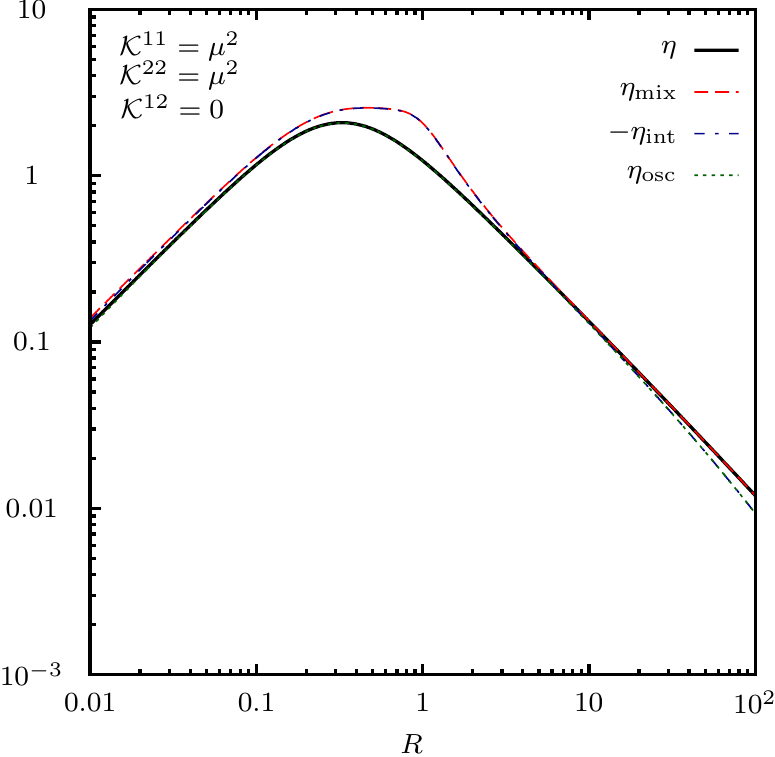}
\includegraphics[width=0.5\textwidth]{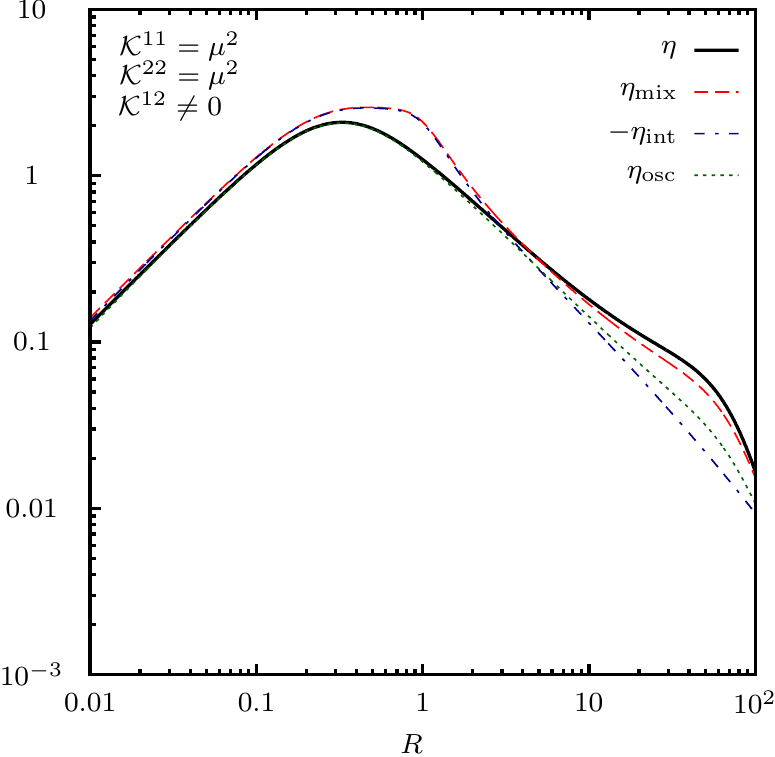}\\
\includegraphics[width=0.5\textwidth]{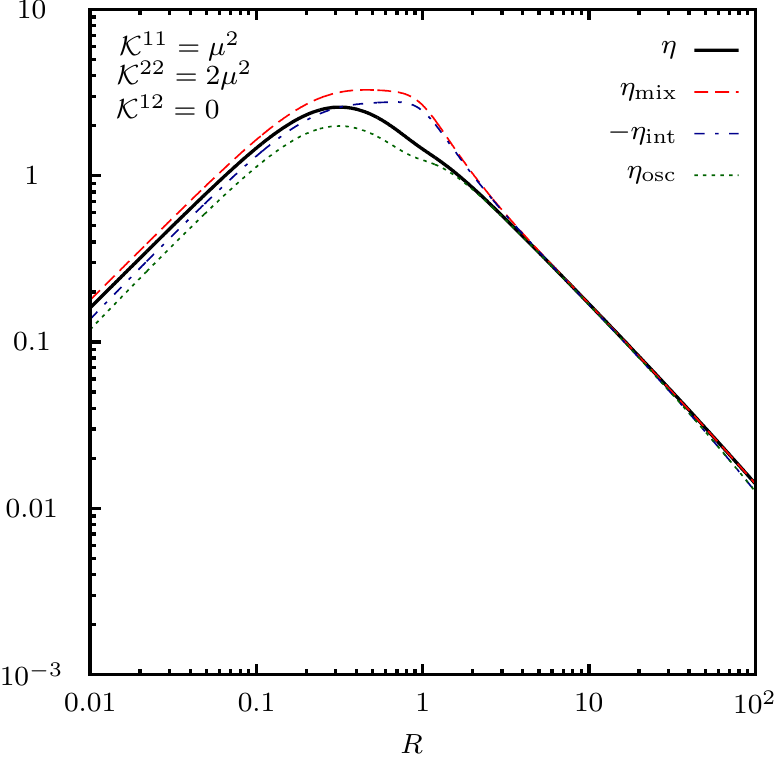}
\includegraphics[width=0.5\textwidth]{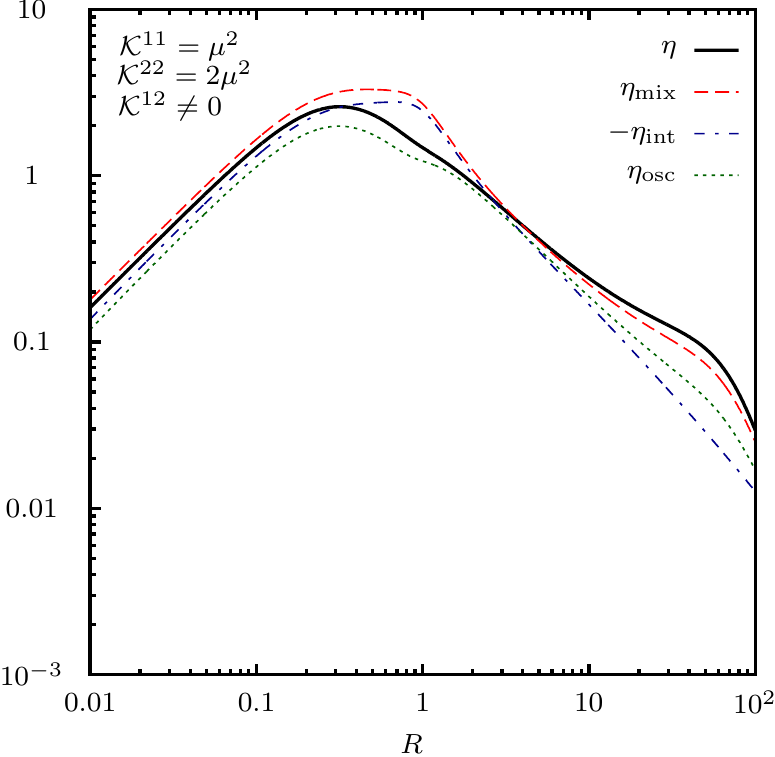}
\caption{\label{NumExample}Contributions of the mixing ($\eta_{\rm mix}$), oscillation
($\eta_{\rm osc}$) and interference ($\eta_{\rm int}$) sources  to the asymptotic value of 
the asymmetry as functions of the degeneracy parameter $R$ for various (\C-conserving) 
choices of the initial conditions and with ``Yukawa'' couplings $h_1=0.5\,\mu\exp(-i)$ and $h_2=-\,0.8\,\mu\exp(-2i/3)$.}
\end{figure} 

\begin{figure}[h!]
\includegraphics[width=0.5\textwidth]{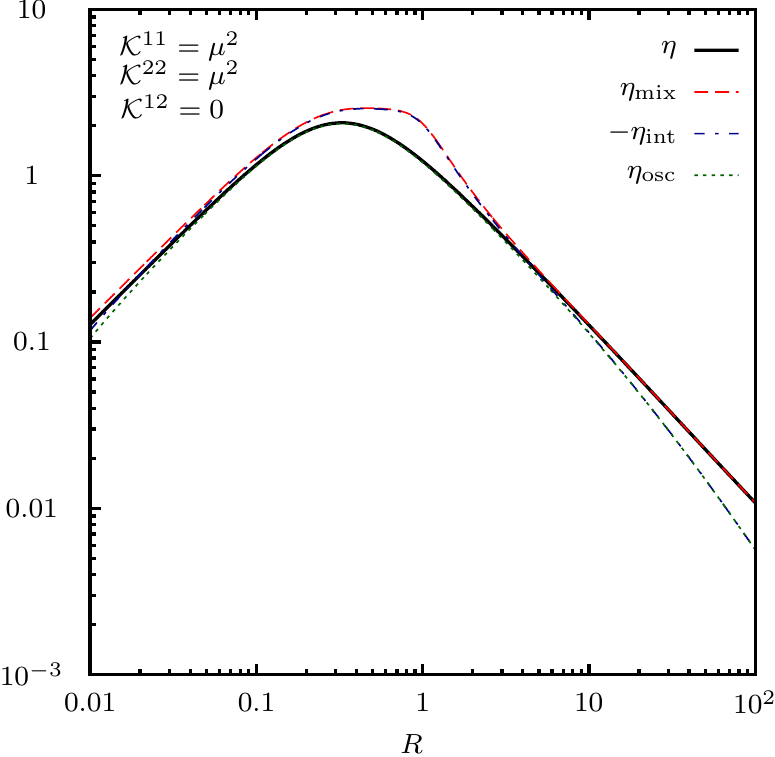}
\includegraphics[width=0.5\textwidth]{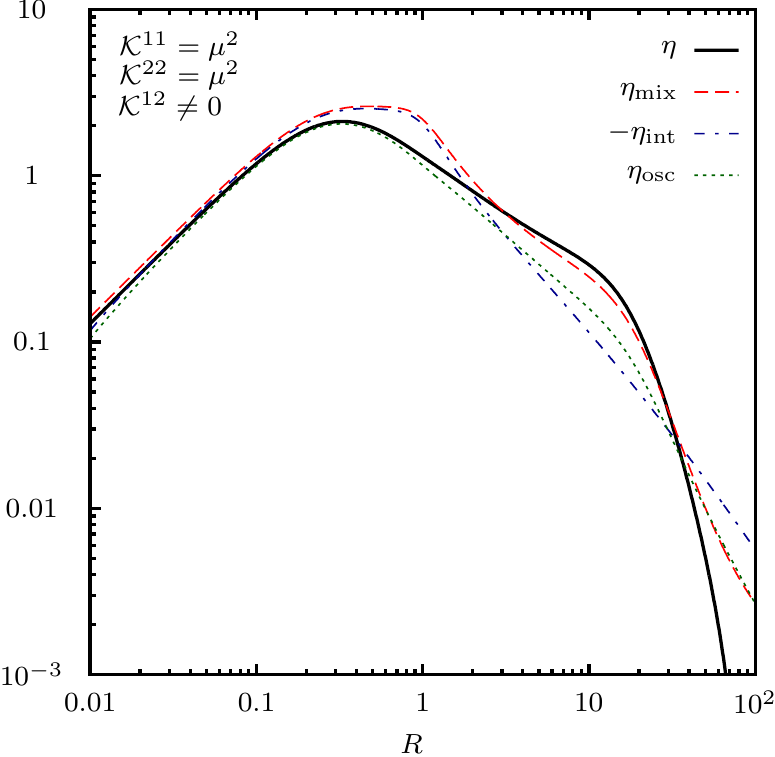}\\
\includegraphics[width=0.5\textwidth]{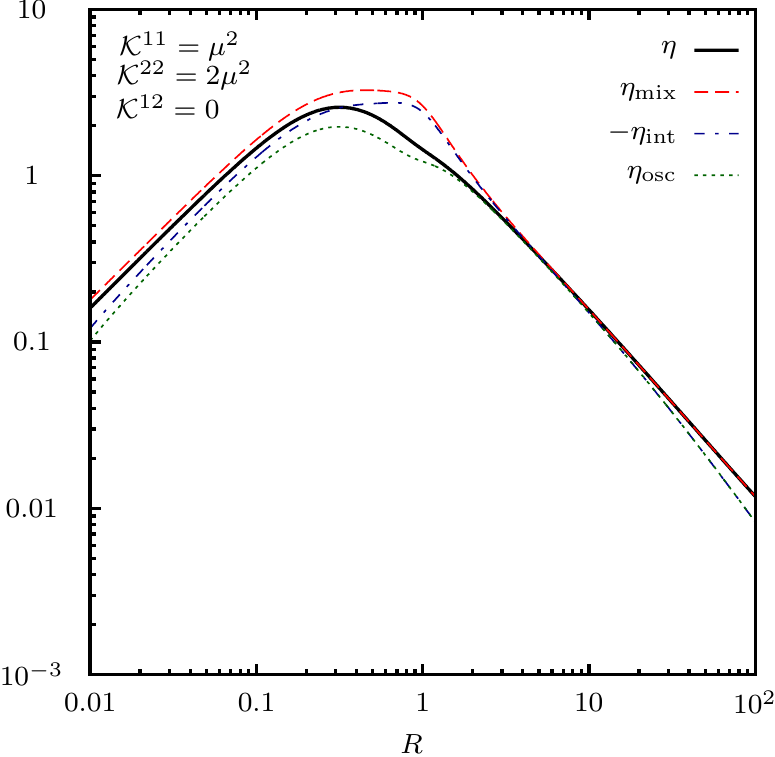}
\includegraphics[width=0.5\textwidth]{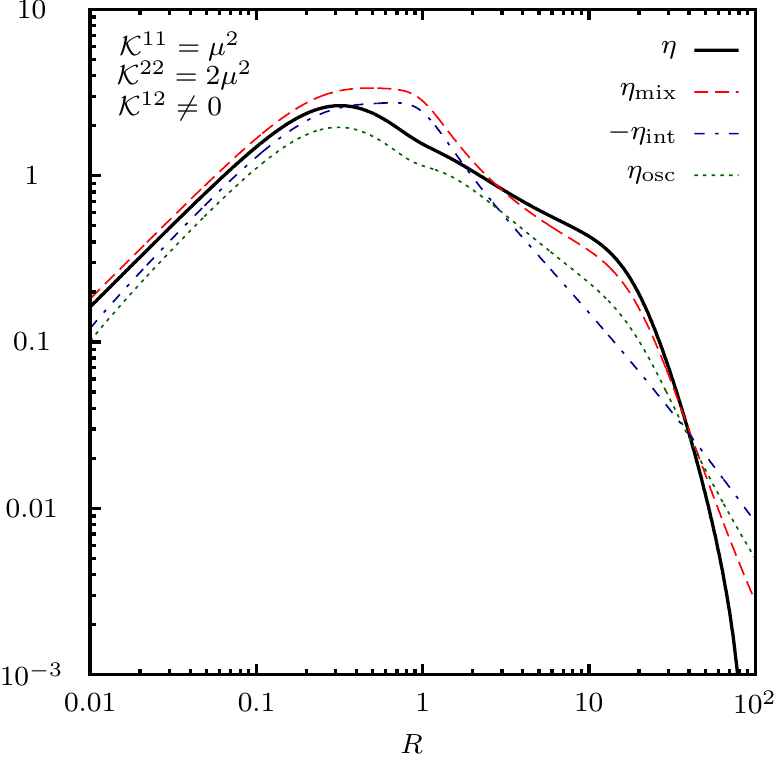}
\caption{\label{NumExampleBigYuk1}Contributions of the mixing ($\eta_{\rm mix}$), oscillation
($\eta_{\rm osc}$) and interference ($\eta_{\rm int}$) sources  to the asymptotic value of 
the asymmetry as functions of the degeneracy parameter $R$ for various (\C-conserving) 
choices of the initial conditions and with ``Yukawa'' couplings $h_1=\mu\exp(-i)$ and $h_2=-\,1.6\,\mu\exp(-2i/3)$.
Note that the deviation between the black and coloured curves at 
$R\sim 100$ is not related to the deviations in fig.~\ref{NumExample4}.}
\end{figure}

For the numerical analysis, it is more convenient to evaluate eq.~\eqref{SourceTerm} directly. 
Substituting the  Wigner transform of eq.~\eqref{DevFromEq} into eq.~\eqref{SourceTerm} and 
neglecting the sub-leading off-shell contributions, the time-derivative of the asymmetry takes the form
\begin{align}
\label{dqdtexact}
\frac{\D \eta}{\D t}\ &=\ 4\,\Im\, H_{12}\int_{\vec q} \frac{\epsilon_{ab}}{|\Delta\Omega^2|^2}\,
 \widetilde{\Pi}_{\rho}\bigl((\omega_a+\omega_b)/2,\vec{q}\bigr)\nonumber\\
&\times\ \Im\biggl[\frac{[\adj \bm{D}_R(\omega_a,\vec{q})]^{1m}}{2\Omega_a}\,\mathcal{K}^{mn}(\vec{q})\,
\frac{[\adj \bm{D}_A(\omega_b,\vec{q})]^{n2}}{2\Omega^*_b} 
e^{-i(\Omega_a-\Omega^*_b)t}\biggr]\,,
\end{align}
where we have restored the common momentum $\vec q$ and used the same notational
conventions as in eq.~\eqref{GdeltaInit}.  We emphasize that the numerical analysis of the present section is performed for the \emph{full} solution, without an expansion to a given order in $\Pi/\Delta M^2$. Were we to expand eq.~\eqref{dqdtexact} in powers 
of  $\Pi_{R(A)}$, we would recover eq.~\eqref{dqdtapprox}. We would also like to emphasize
that the leading order expansion [eq.~\eqref{dqdtapprox}] provides a very accurate approximation 
to eq.~\eqref{dqdtexact}, as has already been pointed out in ref.~\cite{Hohenegger:2014cpa}.
At the initial time surface $t=0$, the Heisenberg- [eq.~\eqref{GF12Expansion}] and interaction-picture
[eq.~\eqref{eq:DeltaGint}] propagators coincide. This can be used to extract from eq.~\eqref{dqdtexact} the mixing, oscillation, and 
interference sources, as identified in eq.~\eqref{eq:DeltaGint}. 
 
\begin{figure}[h!]
\includegraphics[width=0.5\textwidth]{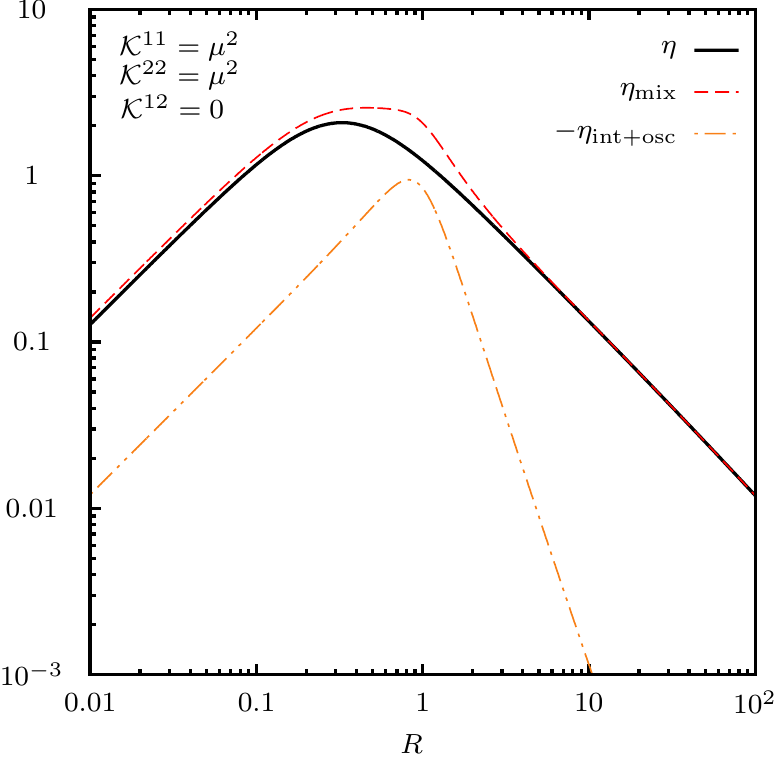}
\includegraphics[width=0.5\textwidth]{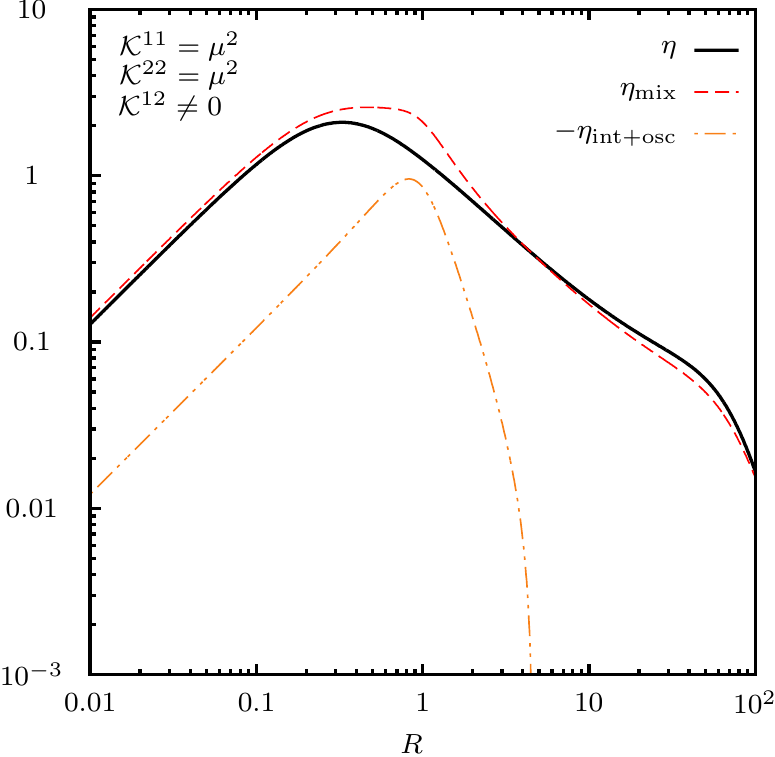}\\
\includegraphics[width=0.5\textwidth]{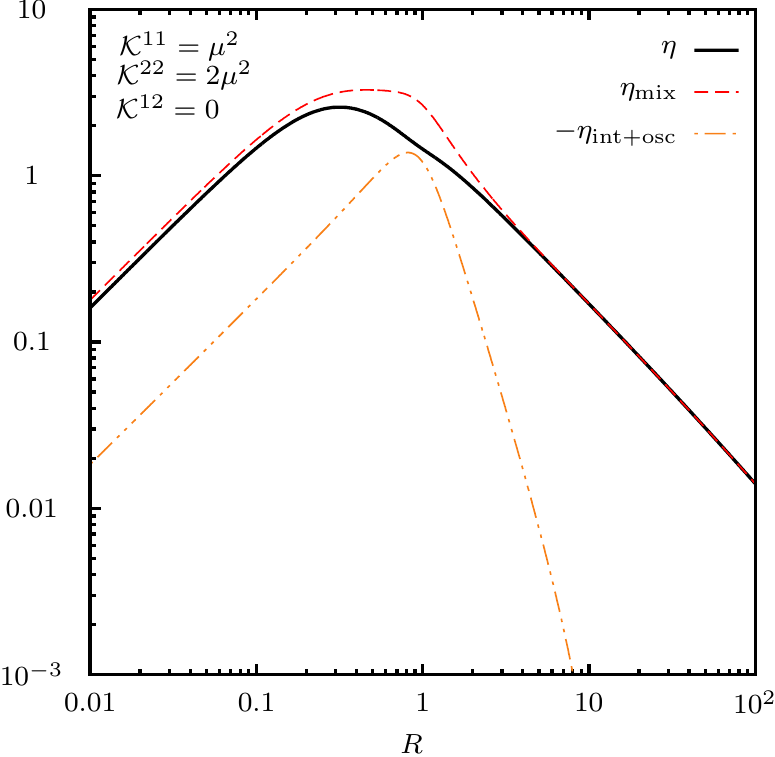}
\includegraphics[width=0.5\textwidth]{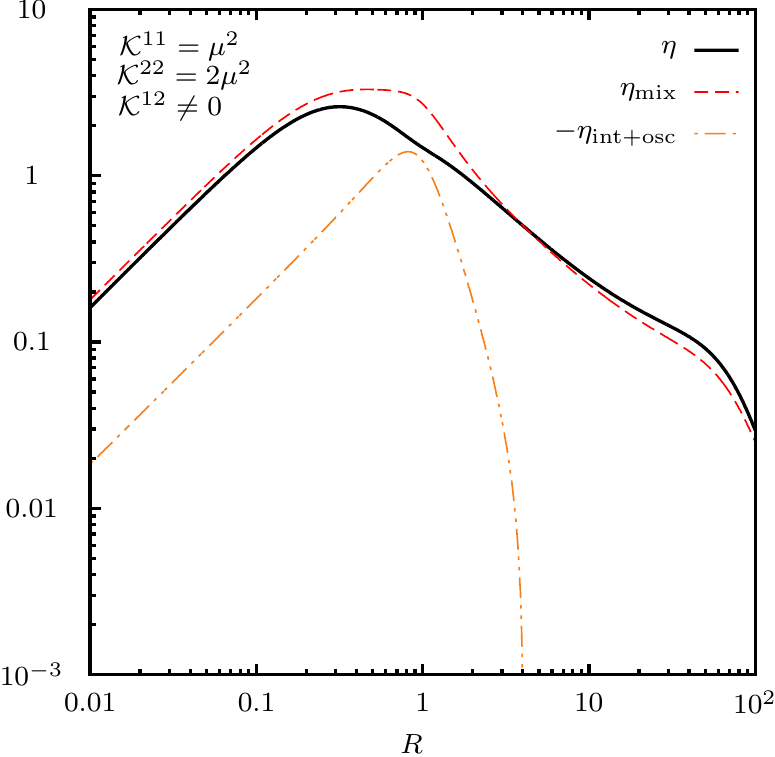}
\caption{\label{NumExample1}Numerical results interpreted in terms of the Boltzmann ``benchmark''
($\eta_{\rm mix}$) plus corrections ($\eta_{\rm osc+int}$) as functions of the degeneracy parameter $R$ 
for various (\C-conserving) choices of the initial conditions and with ``Yukawa'' couplings $h_1=0.5\,\mu\exp(-i)$ and $h_2=-\,0.8\,\mu\exp(-2i/3)$.}
\end{figure} 

For the numerical examples, following ref.~\cite{Hohenegger:2014cpa}, we choose  $T=\mu$ and $M_1=\mu$, where $\mu$ is the $\overline{\rm MS}$ 
renormalization scale. The second mass parameter $M_2$ can be expressed in terms 
of the degeneracy parameter $R$ in eq.~\eqref{eq:degenparam}. In figure~\ref{NumExample},  we plot the three contributions to the final asymmetry, as well as the total asymmetry itself, for four choices of initial conditions and with would-be Yukawa couplings $h_1=0.5\,\mu\exp(-i)$ and $h_2=-\,0.8\,\mu\exp(-2i/3)$. The same is plotted in figure~\ref{NumExampleBigYuk1} for larger ``Yukawa'' couplings  $h_1=\mu\exp(-i)$ and $h_2=-\,1.6\,\mu\exp(-2i/3)$. The red, dashed line indicates the contribution from the mixing source; the green, dotted line the contribution from the oscillation source; the blue, dash-dotted line the contribution from the interference terms; and the solid, black line the total asymmetry. By comparing figures~\ref{NumExample} and~\ref{NumExampleBigYuk1}, we see that the relative size of the various contributions is unaffected by the change in the ``Yukawa'' couplings. This can be understood in terms of their common parametric dependence on the 
couplings and mass splittings, as identified in section~\ref{sec:interaction}. A more significant 
effect is seen only for $R>1$ in the case of $\mathcal{K}^{12}\neq 0$ due to the additional dependence on the ``Yukawa'' couplings introduced via eq.~\eqref{CSymmInitCond}, as is necessary in order to specify $C$-symmetric initial conditions in the weak-washout regime.

\begin{figure}[t!]
\includegraphics[width=0.5\textwidth]{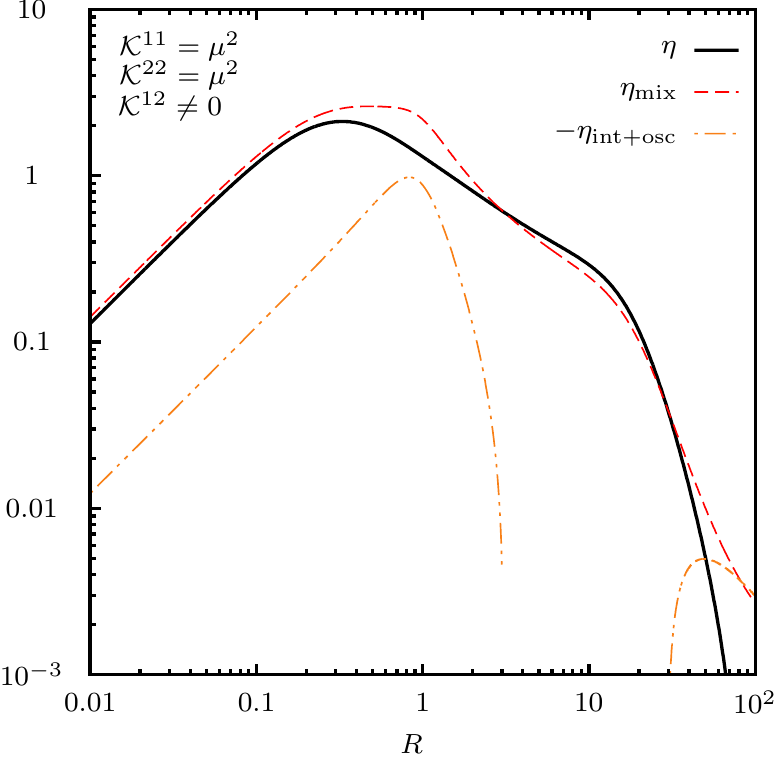}
\includegraphics[width=0.5\textwidth]{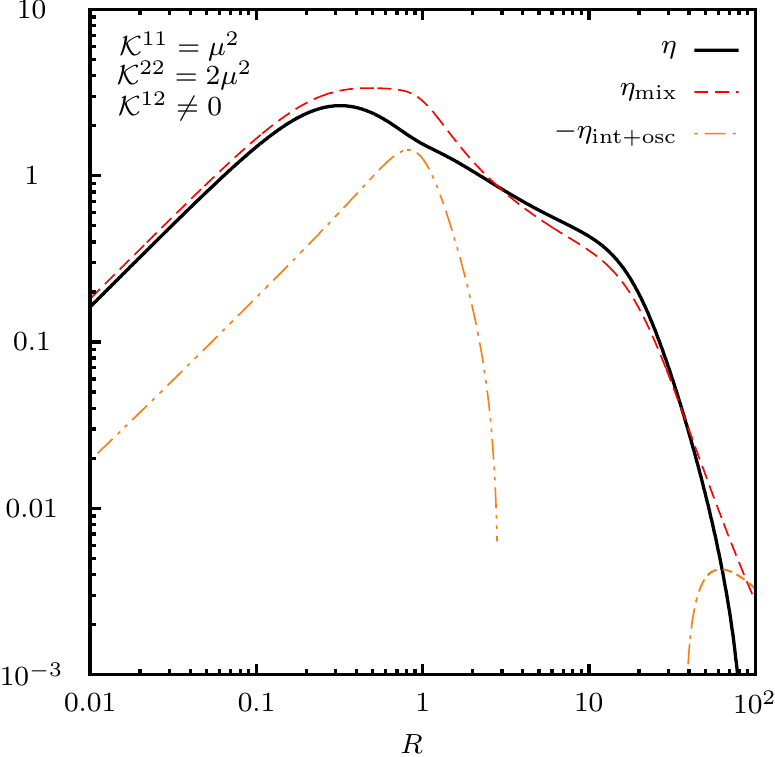}\\
\includegraphics[width=0.5\textwidth]{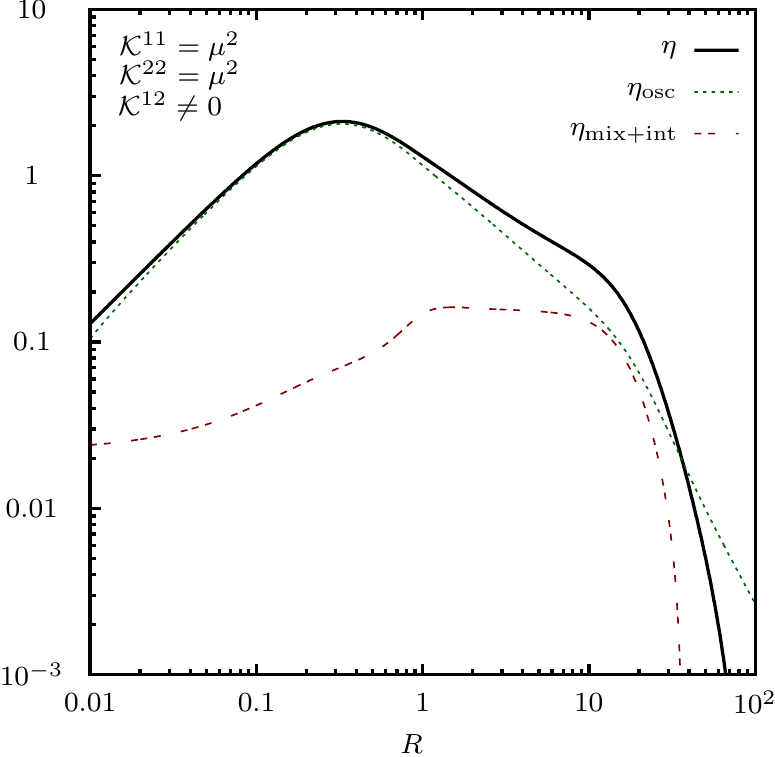}
\includegraphics[width=0.5\textwidth]{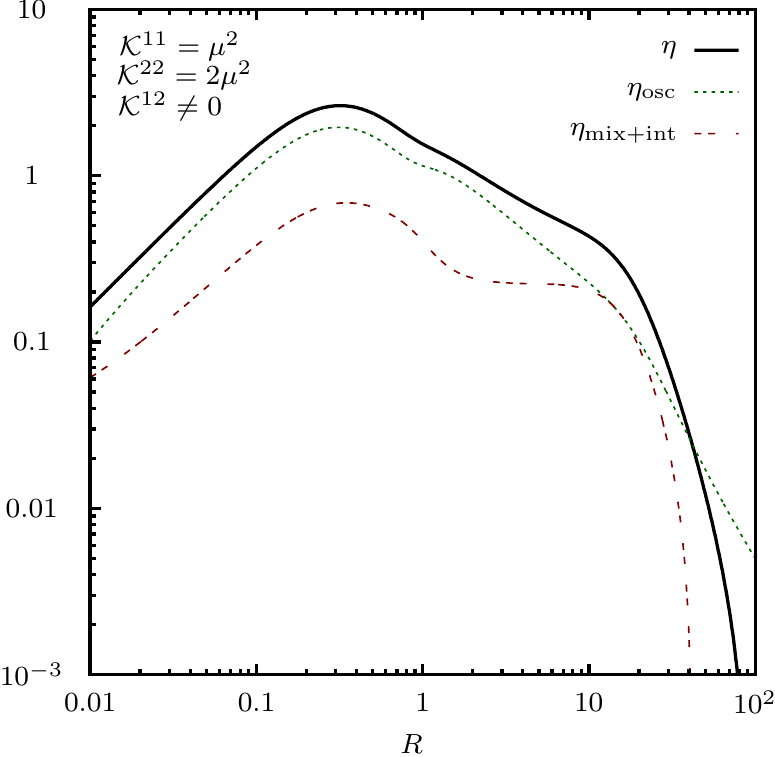}
\caption{\label{NumExampleBigYuk2}Numerical results interpreted in terms of the Boltzmann ($\eta_{\rm mix}$) and density matrix ($\eta_{\rm osc}$) ``benchmarks'' as functions of the degeneracy parameter $R$ 
for various (\C-conserving) choices of the initial conditions and with ``Yukawa'' couplings $h_1=\mu\exp(-i)$ and $h_2=-\,1.6\,\mu\exp(-2i/3)$. The oscillation contribution (dotted green line) is in excess of the total asymmetry (solid black line) for large $R$ due to a change of sign in the mixing plus interference terms, which is not visible on the plot.}
\end{figure}

\noindent The numerical results can be interpreted in two ways:
\begin{enumerate}
\item[(i)] On the one hand, taking the Boltzmann approximation (effective ``Yukawa'' couplings but diagonal number densities) as the ``benchmark'', one would think of the sum of the \emph{oscillation} (green, dotted lines)
and interference contributions (blue, dash-dotted lines) as the correction to the original approximation. With this interpretation in mind, we plot in figure~\ref{NumExample1} the total asymmetry (solid, black line) versus the mixing contribution (red, dashed line) and the sum of the oscillation and interference terms (orange, dash-dotted line). In agreement with expectations, the correction from the oscillation and interference terms is large for $R\sim 1$ and is very small for hierarchical  mass spectra. 
Interestingly, it is also small for quasi-degenerate mass spectra. All in all, we find that 
the Boltzmann approximation, which is equivalent to the mixing contribution (red, dashed line), agrees well with the 
total asymmetry, with the exception of the region $R\sim 1$.

\item[(ii)] Instead, taking the density matrix approximation (tree-level ``Yukawa'' couplings but off-diag\-onal 
number densities) as the ``benchmark'', one would think of the sum of the \emph{mixing} (red, dashed lines) 
and the interference terms (blue, dash-dotted lines) as the correction to the original approximation. With 
this interpretation in mind, we plot in figure~\ref{NumExample2} 
\begin{figure}[h!]
\includegraphics[width=0.5\textwidth]{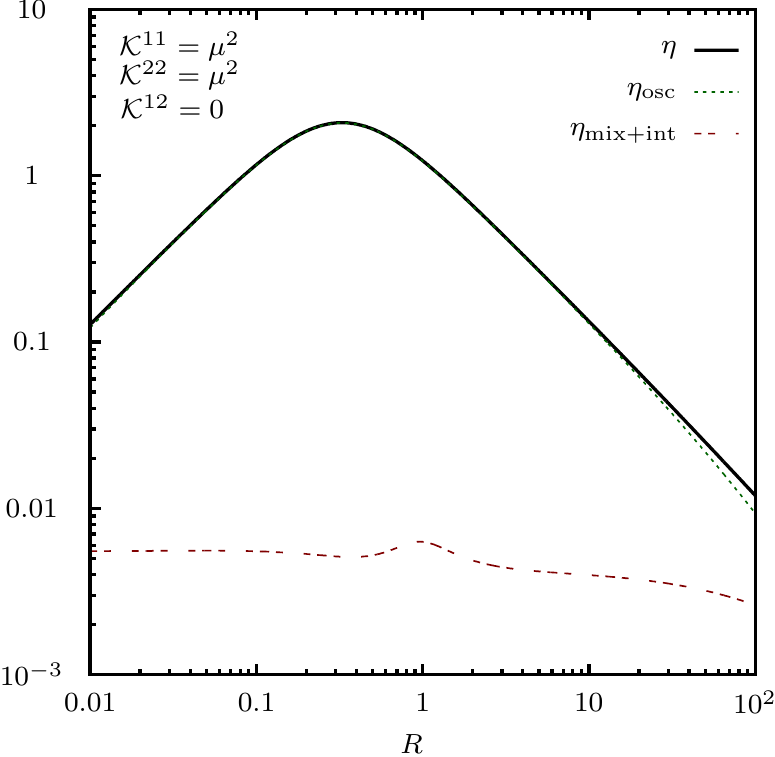}
\includegraphics[width=0.5\textwidth]{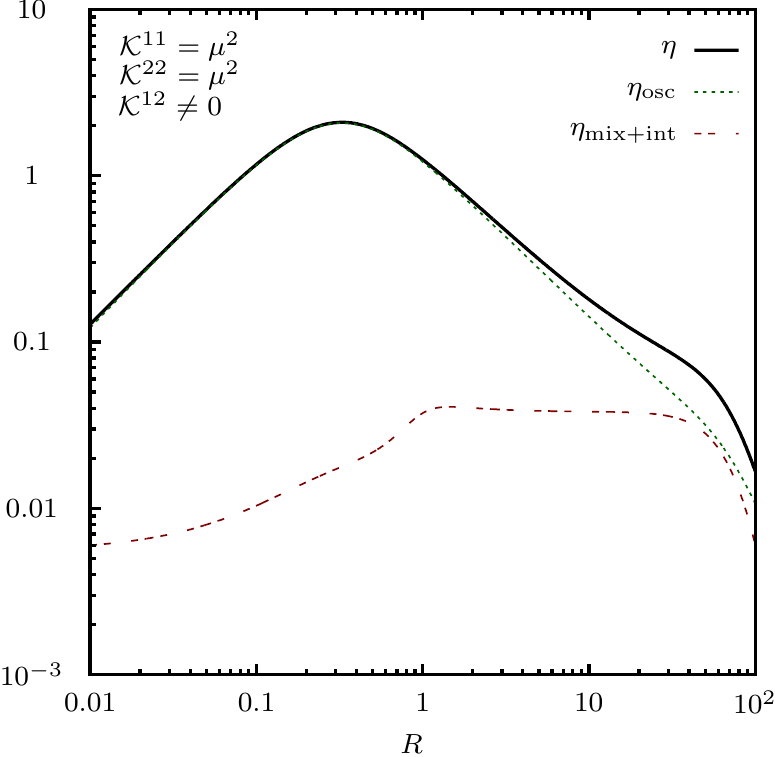}\\
\includegraphics[width=0.5\textwidth]{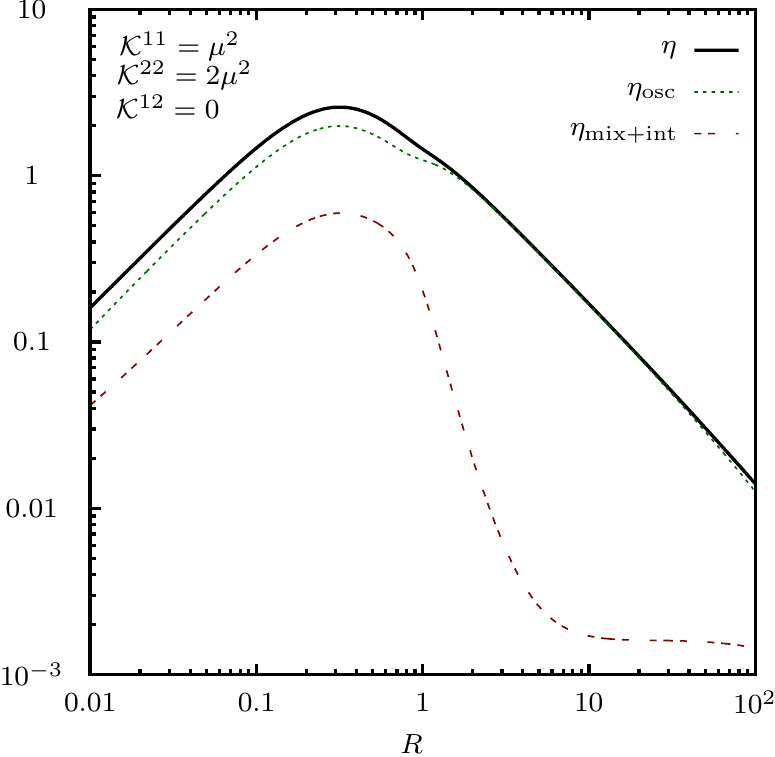}
\includegraphics[width=0.5\textwidth]{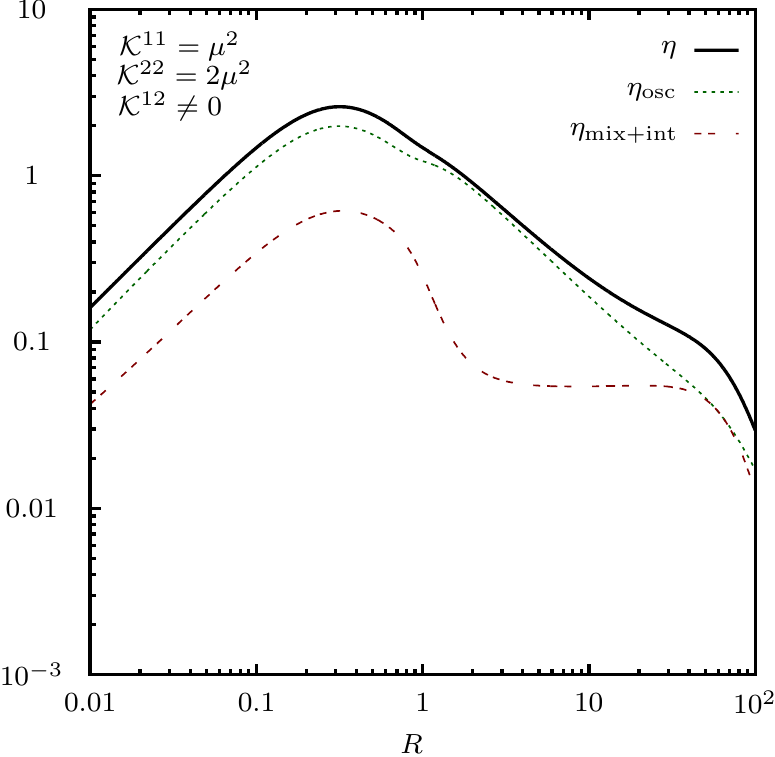}
\caption{\label{NumExample2}Numerical results interpreted in terms of the density matrix ``benchmark''
($\eta_{\rm osc}$) plus corrections ($\eta_{\rm mix+int}$) as functions of the degeneracy parameter $R$ 
for various (\C-conserving) choices of the initial conditions and with ``Yukawa'' couplings $h_1=0.5\,\mu\exp(-i)$ and $h_2=-\,0.8\,\mu\exp(-2i/3)$.}
\end{figure} 
the total asymmetry (solid, black line) 
versus the oscillation contribution (green, dotted line) and the sum of the mixing and interference terms 
(brown, dash-dotted line). In this case, we see that the density matrix approximation, which is equivalent 
to the oscillation contribution (green, dotted line), agrees well with the total asymmetry when the number 
densities of the two flavours are of similar magnitudes (upper two panels of figure~\ref{NumExample2}). 
On the other hand, when the number densities of the two flavours are not similar  (lower 
two panels of figure~\ref{NumExample2}), as is enforced in the weak-washout regime by choosing 
differing initial conditions for the two flavours, we see that the density matrix approximation underestimates 
the total asymmetry for smaller $R$. This observation can be understood from the analytic results given 
in eqs.~\eqref{afterinterference} and \eqref{epsilontilde} for the effective \CP-violating parameter. 
Specifically, with the density matrix approximation as the benchmark, the interference terms can be 
seen as a modification to the mixing source. This modification introduces a relative sign between the 
contribution to the asymmetry from the two flavours. Hence, when the deviations from equilibrium of the 
two flavours are similar, the mixing contribution is strongly suppressed. On the other hand, when this is 
not the case, the cancellation is no longer exact and both the oscillation and mixing sources contribute additively to the 
asymmetry, leading to an underestimate if one were to neglect one or other of these sources. 
In spite of the possibility that the interference terms may not be captured fully in the analysis of 
refs.~\cite{Dev:2014laa, Dev:2015wpa, Dev:2014wsa}, we nevertheless see as marked an enhancement 
in the present analysis up to a maximal factor of two for certain values of the parameters.
It remains to be seen the extent to which 
the interference terms modify the final asymmetry for more-realistic phenomenological models in the 
strong-washout regime and for an expanding background. However, since the mixing, 
oscillation and interference terms all share common parametric dependence upon the ``Yukawa'' couplings 
and mass splittings, one can reasonably anticipate that all three effects may be of relevance for such models 
in the weakly-resonant (or overlapping) regime $\Gamma_i \ll \Delta M \ll \bar{M}$. 
We would like to emphasize that, for large $R$, the oscillation source in figure~\ref{NumExample2}
cannot be directly associated with the results of section~\ref{sec:densitymatrix}, because the latter are 
only applicable for $\omega_2\simeq \omega_1$, which is only fulfilled for quasi-degenerate mass spectra.
\begin{figure}[t!]
\includegraphics[width=0.5\textwidth]{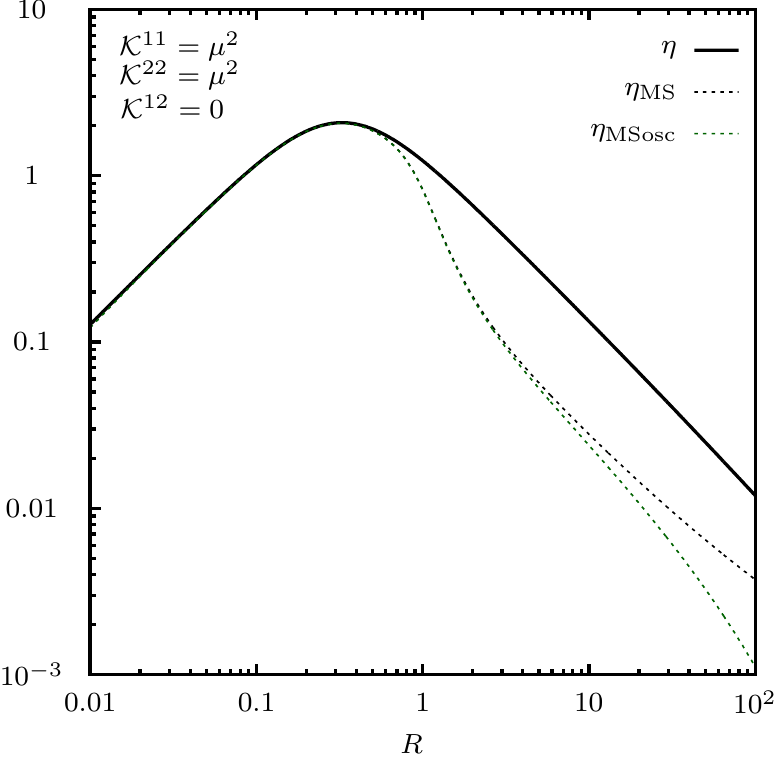}
\includegraphics[width=0.5\textwidth]{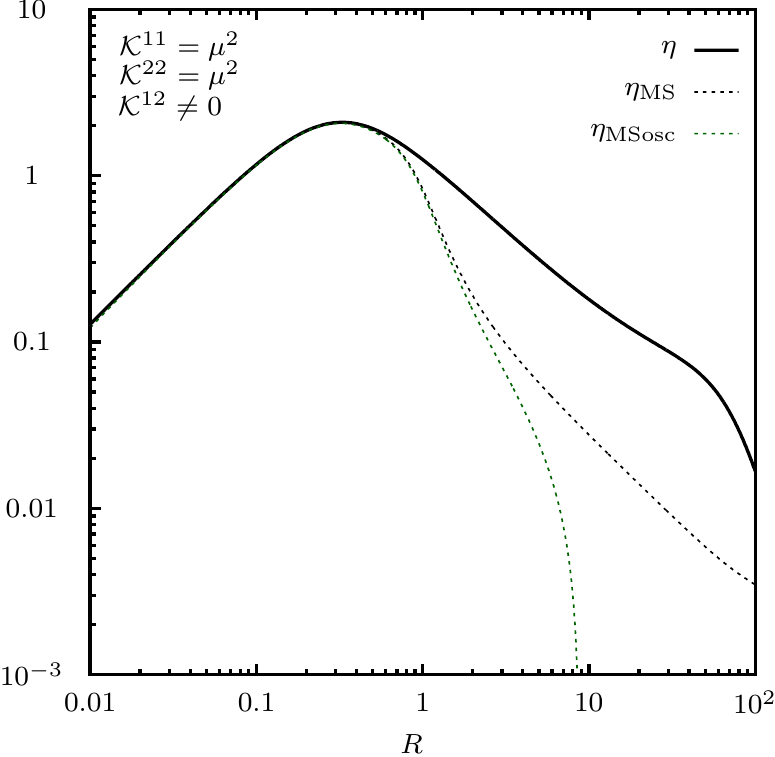}\\
\includegraphics[width=0.5\textwidth]{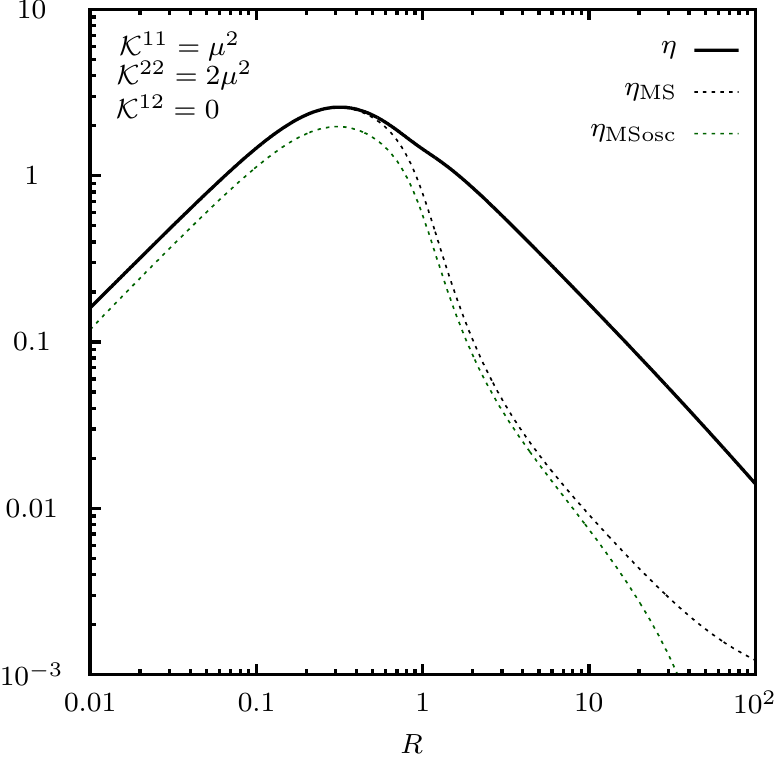}
\includegraphics[width=0.5\textwidth]{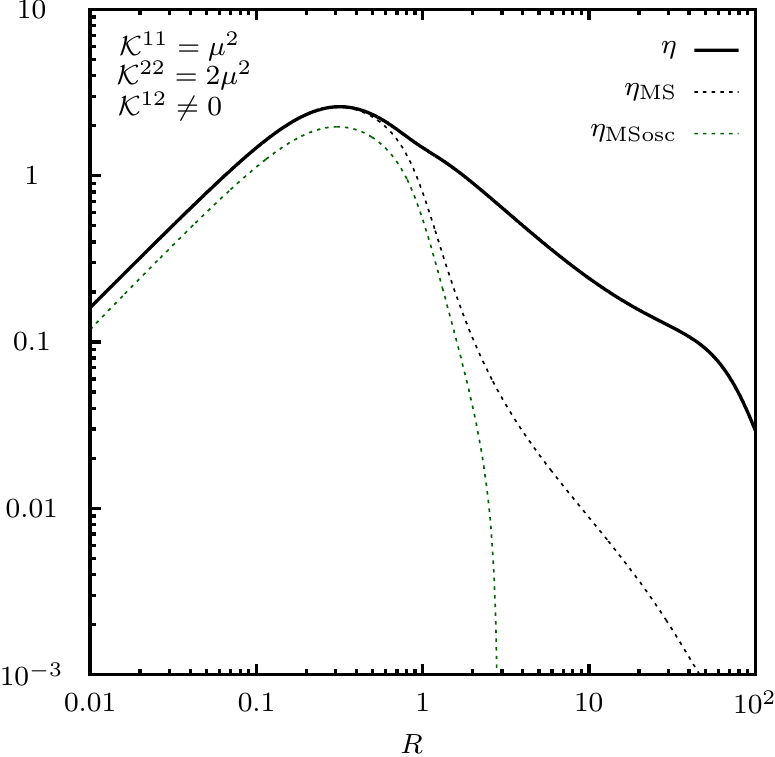}
\caption{\label{NumExample4}Numerical results for the total asymmetry computed from eq.~\eqref{dqdtexact} 
using the complete shell structure  (solid black line), and the total  asymmetry 
and the oscillation contribution (dotted black green lines respectively) for the middle-shell approximation $\omega_1=\omega_2=\bar{\omega}$
as functions of the degeneracy parameter $R$.}
\end{figure} 
This point is illustrated in figure~\ref{NumExample4}, where we compare the total asymmetry computed 
taking into account the full shell structure to the total asymmetry and the oscillation
contribution computed using the middle-shell approximation $\omega_1=\omega_2=\omega$ in 
eq.~\eqref{dqdtexact}. As expected, for hierarchical mass spectra the results differ by orders of magnitude.
\end{enumerate}
For completeness, we also plot in figure~\ref{NumExampleBigYuk2}
the same comparisons for the larger choice of ``Yukawa'' couplings and for the initial condition $\mathcal{K}^{12}\neq 0$. 
We draw attention to the bottom right panel of this figure, where the underestimate of the lepton asymmetry provided by 
the oscillation source alone is very apparent. The maximum factor of two enhancement in the asymmetry can also be seen
clearly when both rather than only one (oscillation or mixing) source are accounted for.

\section{\label{sec:conc}Conclusions and outlook}

For a hierarchical mass spectrum of heavy Majorana neutrinos, baryogenesis 
via leptogenesis can be  studied in detail  using conventional flavour-diagonal
Boltzmann equations. On the other hand, for a mildly  quasi-degenerate mass 
spectrum, the Boltzmann approximation is insufficient, and there is  ongoing 
work on the first-principles derivation of kinetic- and systematically-improved
Boltzmann equations capable of fully accounting 
for all relevant effects, in particular the resonant enhancement of \CP-violating 
parameters and the oscillation between different flavours. In practice, it is 
necessary to find consistent approximation schemes in order to render the solution 
of these equations tractable for the purpose of performing numerical scans of the 
available parameter space. 

The mixing of particle flavours and the oscillations between them are two physically distinct and identifiable phenomena, as is known from the neutral $K$, $D$, $B$ and $B_s$ systems~\cite{Agashe:2014kda}.
In this work, using Kadanoff-Baym equations, we 
confirm (in the weak-washout regime) that mixing and oscillations indeed provide two distinct 
sources of lepton asymmetry, which can be readily 
identified by means of the shell structure of the resummed heavy-neutrino
propagators. The mixing contributions correspond  to the usual \CP-violation in 
decay and live  on the mass shell of the corresponding quasi-particles with energy $\omega_i$. 
Instead, flavour oscillations between the heavy neutrinos and interference between 
mixing and oscillation can be identified with an ``oscillation shell'' of energy 
$\bar{\omega}=(\omega_1+\omega_2)/2$.

Historically, leptogenesis was first studied in the Boltzmann approximation, i.e.~using diagonal 
number densities with the transition amplitudes computed in vacuum. From the perspective of
this approximation, one would 
think of the sum of the \emph{oscillation} and interference contributions as the correction. In agreement with expectations,
 this correction is large when the difference of the 
masses is comparable to the decay widths and is very small for hierarchical mass spectra. 
Interestingly, it is also small for quasi-degenerate mass spectra. Within the past decade, 
a  lot of work has been devoted to the re-analysis of resonant leptogenesis within the density matrix 
formalism, i.e.~taking into account off-diagonals of the matrix of number densities. From this perspective, one would instead think
of the sum of the \emph{mixing} and interference terms as the 
correction. When the number densities of the two flavours are of similar 
size, this correction is small. On the other hand, when the number densities of the two 
flavours are not similar, the correction becomes sizable even for quasi-degenerate 
mass spectra, i.e.~in the parameter range where it originally was thought to be small. We find that the mixing and oscillation sources are of the same sign, contributing additively to the final asymmetry  up to a factor of two, in agreement with the conclusions of refs.~\cite{Dev:2014laa, Dev:2015wpa, Dev:2014wsa}. However, we also find that the interference terms may lead to a suppression of the contribution from mixing. Hence, it would be of interest to perform an equivalent analysis for a
realistic phenomenological model in the strong-washout regime and for an expanding background.

By comparing the Heisenberg- and interaction-picture Kadanoff-Baym equations,
we have found \emph{identical} results, illustrating the self-consistency and 
complementarity of these two significantly different approaches. We note that 
this exact agreement relied upon the interference between the mixing and 
oscillation contributions in the transport equations for the number densities.
Whereas the Heisenberg-picture Kadanoff-Baym equations are very 
useful for studying qualitative features of the regimes that cannot be addressed 
by either Boltzmann (not applicable for quasi-degenerate mass spectra) or density 
matrix (not applicable for hierarchical mass spectra) approaches, their use in phenomenological 
studies is severely limited by the difficulty of solving them. 
The fact that the 
interaction-picture Kadanoff-Baym equations, which are much easier to solve, 
are identical to the Heisenberg-picture ones means that we now have a convenient 
tool that allows us to treat hierarchical and quasi-degenerate mass spectra on 
an equal footing, thereby performing parameter scans over a large range.

\acknowledgments

P.M.~would like to express a debt of gratitude to Bhupal Dev, Marco Drewes, Bj\"{o}rn Garbrecht, 
Apostolos Pilaftsis and Daniele Teresi for many fruitful, enjoyable and illuminating 
discussions in this area. The authors would like to thank Bj\"{o}rn Garbrecht and Daniele Teresi for careful reading of 
the manuscript and helpful comments and suggestions. 
The work of P.M.~was supported in part by a University Foundation 
Fellowship (TUFF) from the Technische Universit\"{a}t M\"{u}nchen and the Science and Technologies Facilities Council (STFC) under Grant No.~ST/L000393/1. A.K.~and  H.V.~acknowledge financial support by the European Union under Grant 
No.\ PITN-GA-2011-289442 (FP7 Initial Training Network ``Invisibles''). All 
three authors acknowledge support from the Deutsche Forschungsgemeinschaft 
(DFG) under Grant No.\ EXC-153 (Excellence Cluster ``Universe'').

\begin{appendix}

\section{\label{sec:noneq}Non-equilibrium field theory}

The purpose of this appendix is two fold: firstly, it is intended to provide a brief 
outline of the background to the Heisenberg- and interaction-picture realizations of the 
Kadanoff-Baym approach to transport phenomena; secondly, it serves as a collection of the 
definitions and notational conventions of the  various two-point functions and self-energies 
that appear in the body of this manuscript. In addition and in order to aid the comparison 
of this work with the existing literature, we identify the correspondence of the conventions 
employed herein with those appearing elsewhere.

In the study of transport phenomena, we are interested in the statistical or ensemble 
expectation values (EEVs) of operators evaluated at a \emph{given} time. Specifically, 
in the Schr\"{o}dinger picture, these EEVs take the generic form
\begin{equation}
\braket{\bullet}\!(t)\ \equiv\ \mathcal{Z}^{-1}\,\mathrm{tr}\,\rho(t)\,\bullet\;,
\end{equation}
where $\rho(t)$ is the quantum-statistical density operator and $\mathcal{Z}=\mathrm{tr}\,\rho$ 
is the partition function. This is in stark contrast to scattering-matrix theory, where we 
are instead interested in the overlap of states evaluated at \emph{different} times: 
specifically, \emph{in} and \emph{out} asymptotic states. This \emph{in-out} formalism 
naturally lends itself to a path-integral description, leading to a time integral that 
runs from the infinitely-distant past to the infinitely-distant future. On the other hand, 
EEVs contain the overlap of states evaluated at the \emph{same} time, i.e.~two \emph{in} 
(or two \emph{out}) states.  

\paragraph{Closed-time path.}

In order to define a path-integral representation of EEVs, 
we must deform a contour in the complex-time plane that takes us from the \emph{in} state 
to the \emph{out} state and back again. This construction gives rise to the so-called 
\emph{in-in} or closed-time path (CTP) formalism due to Schwinger and Keldysh~\cite{Schwinger:1960qe,Keldysh:1964ud}.

The CTP contour comprises two branches: one running forwards in time, which we refer to as 
the time-ordered branch, and one running backwards in time, which we refer to as the 
anti-time-ordered branch. On this contour, we may introduce a path-ordering operator 
$\mathrm{T}_{\mathcal{C}}$. Given two field operators with times $x^0$ and $y^0$ both 
lying on the time-ordered branch, the path ordering reduces to the usual time ordering. 
When both times lie instead on the anti-time-ordered branch, the path ordering corresponds
to anti-time ordering. Finally, times lying on the time-ordered branch are, for the 
purposes of path ordering, always `earlier' than those on the anti-time-ordered branch. 
As a consequence of our ability to place field operators on either of the two branches, 
the CTP formalism leads to a doubling of degrees of freedom. The need for the latter can 
be understood as follows: we need sufficient degrees of freedom 
to build both the statistical ensemble and excitations 
within it.

In the same way that expectation values of operators can be written in any one of the three 
equivalent pictures of quantum mechanics, viz.~the Schr\"{o}dinger, interaction (Dirac) 
and Heisenberg pictures, so too can the corresponding operator-level representation of 
the CTP formalism. These three pictures are coincident at a boundary time $\tilde{t}_i$, i.e.
\begin{equation}
\rho_H(;\tilde{t}_i)\ =\ \rho_I(\tilde{t}_i;\tilde{t}_i)\ =\ \rho_S(\tilde{t}_i;\tilde{t}_i)\;,
\end{equation}
where we have indicated Heisenberg-, interaction- (Dirac-) and Schr\"{o}dinger-picture 
operators by subscripts $H$, $I$ and $S$, respectively. In the Heisenberg-picture, field 
operators are time-\emph{dependent}, evolving with the full Hamiltonian $H_H(\tilde{t};\tilde{t}_i)$, 
whereas the density operator is time-\emph{independent}, encoding the initial conditions 
at the time $\tilde{t}_i$. Here, following ref.~\cite{Millington:2012pf}, we have indicated 
the implicit dependence on the boundary time $\tilde{t}_i$ by means of a semi-colon. On the 
other hand, in the interaction picture, both the field and density operators are time-\emph{dependent}, 
evolving respectively under the influence of the free and interaction parts of the Hamiltonian 
$H_I^0(;\tilde{t}_i)$ and $H_I^{\rm int}(\tilde{t};\tilde{t}_i)$.

\paragraph{Pinch singularities.} It is well-known that perturbation theory breaks down in the 
Heisenberg-picture realization of non-equilibrium field theory as a result of so-called pinch 
singularities or secular terms. At the level of the perturbation series, this pathological 
behaviour arises from ill-defined products of Dirac delta functions with identical arguments. 
Their origin can be understood in terms of the Fermi golden rule: for systems in which 
time-translational invariance is broken, the relevant expansion parameter is the product of the 
coupling, $g$ say, and the time $t$ over which the interactions have been permitted to take place. 
Thus, for $t>1/g$, the perturbation series will not converge. As an example, we may consider the 
exponential approach to equilibrium governed by a decay rate $\Gamma\propto g$: an expansion of 
$e^{-\Gamma t}$ in powers of the coupling exists only for $t<1/g$.  In the CTP formalism, the 
time over which the interactions have been permitted to take place corresponds to the length of 
the CTP contour. In the Heisenberg-picture realization, this contour is extended to infinity, 
thereby leading to the emergence of pinch singularities out of equilibrium. On the other hand, 
it has been shown in ref.~\cite{Millington:2012pf} that the CTP contour is necessarily of finite length 
in the interaction picture and, as a result, a well-defined perturbation theory does indeed exist. 
The contour is bounded from the left by the initial (boundary) time $\tilde{t}_i$ and from the 
right by the final time $\tilde{t}$, at which the EEV is calculated. One is then led to 
introduce the concept of a \emph{macroscopic} time $t=\tilde{t}-\tilde{t}_i$, where the tilde 
notation, which we have hitherto not qualified, is reserved for the \emph{microscopic} times 
of the operators.

\paragraph{Ensemble expectation value.} In the Heisenberg picture, the ensemble expectation 
value (EEV) is written in bra-ket notation as:
\begin{equation}
\braket{\bullet}_0\ =\ \mathrm{tr}\,\rho_H(;\tilde{t}_i)\,\bullet\;,
\end{equation}
where the Heisenberg-picture density operator $\rho_H(;\tilde{t}_i)$ is evaluated at the 
\emph{macroscopic time} $t=\tilde{t}_i-\tilde{t}_i = 0$.
In the interaction picture, the EEV is written as
\begin{equation}
\braket{\bullet}_t\ =\ \mathrm{tr}\,\rho_I(\tilde{t};\tilde{t}_i)\,\bullet\;,
\end{equation}
where the interaction-picture density operator $\rho_I(\tilde{t};\tilde{t}_i)$ is instead 
evaluated at the \emph{macroscopic time} $t=\tilde{t}-\tilde{t}_i\neq 0$. Hereafter, we 
suppress the dependence of both Heisenberg- and interaction-picture operators, as well as 
all two-point functions and self-energies, on the boundary time $\tilde{t}_i$.

\paragraph{Coordinate conventions.}
The coordinate-, Wigner- and double-momentum-space representations of the various two-point 
functions are, following ref.~\cite{Millington:2012pf}, distinguished only by the form of 
their arguments. Interaction-picture two-point functions are distinguished from their 
Heisenberg-picture counterparts by a superscript $0$, in the case of the tree-level two-point 
functions, and explicit dependence on the microscopic time $\tilde{t}$, in the case of 
self-energies and the resummed two-point functions. Wherever possible, coordinate-space 
variables are denoted by the lower-case Roman characters $x,y,\dots$; and their Fourier-conjugates 
by the four-momenta $p,p',\dots$. The central and relative coordinates are denoted by the 
upper-case Roman characters $X$ and $R$, respectively, where
\begin{equation}
R^{\mu}\ =\ x^{\mu}\:-\:y^{\mu}\;,\qquad X^{\mu}\ =\ (x^{\mu}\:+\:y^{\mu})/2\;.
\end{equation}
Finally, the characters $q$ and $Q$ are reserved for the central and relative momenta
\begin{equation}
q^{\mu}\ =\ (p^{\mu}\:+\:p^{\mu}{}')/2\;,\qquad Q^{\mu}\ =\ p^{\mu}\:-\:p^{\mu}{}'\;.
\end{equation}
These conventions are summarized in table~\ref{tab:arguments}.
\begin{table}
\begin{center}
\begin{tabular}{| l || c | c |}
\hline
 & Heisenberg Picture & Interaction Picture \\
 \hline\hline
Coordinate Space & $(x,y)$ & $(x,y,\tilde{t})$ \\
Wigner Space & $(q,X)$ & $(q,X,\tilde{t})$\\
Double-Momentum Space & $(p,p')$ & $(p,p',\tilde{t})$\\
\hline
\end{tabular}
\end{center}
\caption{\label{tab:arguments} The form of the arguments of the various resummed two-point 
functions and self-energies, indicating whether they belong to the Heisenberg or interaction 
picture and if they are expressed in the coordinate-, Wigner- or double-momentum-space 
representation. The interaction picture is consistently identified by the appearance of an 
explicit dependence on the microscopic time $\tilde{t}$.}
\end{table}

\paragraph{Double Fourier and Wigner transforms.}
The double Fourier transform $f(p,p')$ of a function $f(x,y)$ is defined as follows:
\begin{align}
f(p,p')\ &\equiv \mathcal{F}_{y}[\mathcal{F}_{x}[f(x,y)](p)](-p')\ \equiv \mathcal{F}_{x}[\mathcal{F}_{y}[f(x,y)](-p')](p)\nonumber\\& \equiv\ \int_{-\infty}^{+\infty}\!\mathrm{d}^4x\int_{-\infty}^{+\infty}\!\mathrm{d}^4y\;e^{-ip\cdot x}\,e^{ip'\cdot y}\,f(x,y)\;.
\end{align}
We emphasize the relative sign in the exponent of the right-most $y$-dependent kernel. This is 
chosen such that translational invariance $f(x,y)=f(x-y)$ corresponds to the conservation of 
four-momentum $p=p'$.

The Wigner transform $f(q,X)$ of a function $f(x,y)$ is defined as follows:
\begin{equation}
f(q,X)\ \equiv\ \mathcal{F}_R[f(x,y)](q)\ \equiv\ \int_{-\infty}^{+\infty}\!\mathrm{d}^4R\;e^{iq\cdot{R}}\,f(x,y)\;.
\end{equation}
It may also be written in terms of an inverse transform of the double-momentum representation $f(p,p')$:
\begin{equation}
\label{WignerFromDouble}
f(q,X)\ \equiv\ \mathcal{F}^{-1}_{Q}[f(p,p')](X)\ \equiv\ \int_{-\infty}^{+\infty}\!
\frac{\mathrm{d}^4 Q}{(2\pi)^4}\;e^{-iQ\cdot X}\,f(p,p')\;.
\end{equation}

\paragraph{CTP propagators.} In the present discussion of the relevant two-point functions, 
we denote by the upper-case Roman character $G$ the conventions of
ref.~\cite{Hohenegger:2014cpa} and those used throughout the body of this article. Those 
denoted by the upper-case Greek character $\Delta$ follow the conventions of 
refs.~\cite{Dev:2014wsa,Millington:2012pf}. Parenthesized names correspond to the nomenclature 
of refs.~\cite{Dev:2014wsa,Millington:2012pf} and are placed in the text immediately following the 
corresponding nomenclature of ref.~\cite{Hohenegger:2014cpa}. Table~\ref{tabcomp} provides a summary of the relation between these conventions.

The CTP propagator of the would-be heavy neutrinos of the model in eq.~\eqref{Lagrangian} 
is defined as
\begin{equation}
G^{[0,]ij}_{\mathcal{C}}(x,y[,\tilde{t}])\ \equiv\ \braket{\mathrm{T}_{\mathcal{C}}[\psi_{H[I]}^i(x)\psi_{H[I]}^j(y)]}_{0[t]}\;.
\end{equation}
Objects appearing in brackets ($[]$) correspond to the interaction-picture definitions. For 
times $x^0$ and $y^0$ on the time-ordered branch, $G_{\mathcal{C}}^{[0,]ij}(x,y[,\tilde{t}])$ 
is equal to the time-ordered (Feynman) propagator $G_{\mathrm{T}}^{[0]ij}(x,y[,\tilde{t}])$. 
For times $x^0$ and $y^0$ on the anti-time-ordered branch, $G_{\mathcal{C}}^{[0,]ij}(x,y[,\tilde{t}])$ 
is equal to the anti-time-ordered (Dyson) propagator $G_{\overline{\mathrm{T}}}^{[0,]ij}(x,y[,\tilde{t}])$.
When $x^0$ is on the time-ordered branch and $y^0$ is on the anti-time-ordered branch, 
$G_{\mathcal{C}}^{[0,]ij}(x,y[,\tilde{t}])$ is equal to the negative-frequency Wightman propagator 
$G_{<}^{[0,]ij}(x,y[,\tilde{t}])$. On the other hand, when $x^0$ is on the anti-time-ordered branch 
and $y^0$ is on the time-ordered branch, $G_{\mathcal{C}}^{[0,]ij}(x,y[,\tilde{t}])$ is equal to the 
positive-frequency Wightman propagator $G_{>}^{[0,]ij}(x,y[,\tilde{t}])$. Of the four aforementioned 
propagators, only two are independent.

Rather than working in terms of path ordering, we may also represent the doubling of degrees 
of freedom by means of a covariant $\mathbb{SO}(1,1)$ notation~\cite{Jordan:1986ug,Calzetta:1986cq,Calzetta:1986ey}, 
with the CTP propagator transforming as a rank-2 tensor. However, in order to avoid proliferation of 
sub- and superscripts, we do not employ this notation in this article.

\paragraph{(Anti)commutator functions.} 
The spectral (Pauli-Jordan) function and the statistical (Hadamard) propagator are defined as follows:
\begin{subequations}
\begin{align}
G^{[0,]ij}_{\rho}(x,y[,\tilde{t}])\ &=\ i\braket{[\psi_{H[I]}^i(x),\,\psi_{H[I]}^j(y)]}_{0[t]}\ 
\equiv\ -\,\Delta^{[0,]ij}(x,y)\;,\\
G^{[0,]ij}_F(x,y[,\tilde{t}])\ &=\ \tfrac{1}{2}\braket{\{\psi_{H[I]}^i(x),\,\psi_{H[I]}^j(y)\}}_{0[t]}\ 
\equiv\ \tfrac{1}{2}\,i\Delta_1^{[0,]ij}(x,y[,\tilde{t}])\;.
\end{align}
\end{subequations}
The subscript $F$, indicating the statistical (Hadamard) propagator, should not be confused 
with the same subscript used in \cite{Dev:2014wsa, Millington:2012pf} to indicate the 
time-ordered (Feynman) propagator.

We draw attention 
to the fact that the Wigner transform of the spectral (Pauli-Jordan) function $G_{\rho}(p,X)$ 
differs from the object $\widetilde{G}(p,X)$ appearing in ref.~\cite{Hohenegger:2008zk} by an 
overall factor of $i$. It is for this reason that we have chosen to identify the Wigner 
representation only by the form of the arguments. Specifically, we have
\begin{equation}
\label{eq:Gtilderho}
G^{ij}_{\rho}(q,X[,\tilde{t}])\ =\ \mathcal{F}_R[G^{ij}_{\rho}(x,y[,\tilde{t}])](q)\ 
\equiv\ i\,\widetilde{G}^{ij}_{\rho}(q,X[,\tilde{t}])\;.
\end{equation}

\paragraph{Causal functions.} The retarded and advanced propagators are defined in terms 
of the spectral (Pauli-Jordan) function as follows:
\begin{subequations}
\begin{gather}
G_R^{ij}(x,y[,\tilde{t}])\ = \ \theta(x^0-y^0)G^{ij}_{\rho}(x,y[,\tilde{t}])\ 
\equiv\ -\,\Delta_R^{ij}(x,y[,\tilde{t}])\;,\\
G_A^{ij}(x,y[,\tilde{t}])\ = \ -\,\theta(y^0-x^0)G^{ij}_{\rho}(x,y[,\tilde{t}])\ 
\equiv\ -\,\Delta_A^{ij}(x,y[,\tilde{t}])\;,
\end{gather}
\end{subequations}
from which we may obtain the identity
\begin{equation}
G_{\rho}^{ij}(x,y[,\tilde{t}])\ =\ G_R^{ij}(x,y[,\tilde{t}])\:-\:G_A^{ij}(x,y[,\tilde{t}])\;.
\end{equation}
In addition, the Hermitian (principal-part) propagator is defined via
\begin{align}
G^{ij}_h(x,y[,\tilde{t}])\ &=\ \textstyle{\frac{1}{2}}\,\big(G_R^{ij}(x,y[,\tilde{t}])+G_A^{ij}(x,y[,\tilde{t}])\big)
\nonumber\\& =\ \tfrac{1}{2}\,\mathrm{sign}(x^0-y^0)\,G_{\rho}^{ij}(x,y[,\tilde{t}])\ \equiv
\ -\,\Delta_{\mathcal{P}}^{ij}(x,y[,\tilde{t}])\;.
\end{align}
Note that the superscript $0$ does not appear in the interaction-picture cases, since the 
above identities hold at any order in perturbation theory.

\paragraph{Wightman propagators.} The absolutely-ordered Wightman propagators are defined as follows:
\begin{subequations}
\begin{gather}
G_>^{[0,]ij}(x,y[,\tilde{t}])\ =\ \braket{\psi^i_{H[I]}(x)\psi^j_{H[I]}(y)}\ \equiv\ i\Delta^{[0,]ij}_>(x,y[,\tilde{t}])\;,\\
G_<^{[0,]ij}(x,y[,\tilde{t}])\ =\ \braket{\psi^j_{H[I]}(y)\psi^i_{H[I]}(x)}\ \equiv\ i\Delta^{[0,]ij}_<(x,y[,\tilde{t}])\;.
\end{gather}
\end{subequations}
These may also be written in terms of the spectral (Pauli-Jordan) function and statistical (Hadamard) propagator:
\begin{equation}
G_{\gtrless}^{ij}(x,y[,\tilde{t}])\ =\ G_F^{ij}(x,y[,\tilde{t}])\:\mp\:\frac{i}{2}\,G^{ij}_{\rho}(x,y[,\tilde{t}])\;,
\end{equation}
yielding the identities
\begin{subequations}
\begin{gather}
G_{\rho}^{ij}(x,y[,\tilde{t}])\ =\ iG_>^{ij}(x,y[,\tilde{t}])\:-\:iG_<^{ij}(x,y[,\tilde{t}])\;,\\
G_{F}^{ij}(x,y[,\tilde{t}])\ =\ \tfrac{1}{2}\,\big(G_>^{ij}(x,y[,\tilde{t}])\:+\:G_<^{ij}(x,y[,\tilde{t}])\big)\;.
\end{gather}
\end{subequations}

\paragraph{Time-ordered propagators.} The time-ordered (Feynman) and anti-time-ordered (Dyson) 
propagators do not feature in the body of this article. However for completeness, they are defined as
\begin{equation}
G_{\mathrm{T}(\overline{\mathrm{T}})}^{ij}(x,y[,\tilde{t}])\ =\ \theta(x^0-y^0)G_{>(<)}^{ij}(x,y[,\tilde{t}])\:+\:\theta(y^0-x^0)G_{<(>)}^{ij}(x,y[,\tilde{t}])\;.
\end{equation}

\paragraph{Self-energies.} We follow the sign convention of ref.~\cite{Hohenegger:2014cpa} for 
the definition of the self-energies, such that a positive dispersive self-energy correction 
corresponds to a positive shift in the mass-squared. For example, in the Markovian approximation, 
we denote the inverse of the momentum-space resummed retarded (advanced) propagator by
\begin{equation}
D_{R(A)}^{ij}(p)\ \equiv\ p^2\delta^{ij}\:-\:[M^2]^{ij}\:-\:\Pi^{ij}_{R(A)}(p)\;,
\end{equation}
where we have adopted the notation $D_{R(A)}^{ij}(p)$ from ref.~\cite{Dev:2014wsa}.
This inverse appears in ref.~\cite{Hohenegger:2014cpa} as $\Omega_{R(A)}^{ij}(p)$ and in 
refs.~\cite{Dev:2014wsa,Millington:2012pf} as $\Delta^{-1}_{R(A)}(p)$.

The various self-energies satisfy identities analogous to those identified above for the 
two-point functions. In the case of the analogue of the spectral function,
\begin{equation}
\Pi^{ij}_{\rho}(x,y[,\tilde{t}])\ =\ i\Pi_>^{ij}(x,y[,\tilde{t}])\:-\:i\Pi_<^{ij}(x,y[,\tilde{t}])\;,
\end{equation}
we have introduced the real-valued distribution
\begin{equation}
\label{eq:pitilde}
\widetilde{\Pi}_{\rho}(p,p'[,\tilde{t}])\ \equiv\ -\,i\Pi_{\rho}(p,p'[,\tilde{t}])\ =\ -\,i\mathcal{F}_{y}[\mathcal{F}_{x}[\Pi_{\rho}(x,y[,\tilde{t}])](p)](-p')\;.
\end{equation}

\begin{table}
\begin{center}
\begin{tabular}{| l | c |}
\hline
Statistical/Hadamard & $G_F^{ij}\ \equiv \frac{1}{2}i\Delta_1^{ij}$\\
Spectral/Pauli-Jordan & $G_{\rho}^{ij}\ \equiv\ -\,\Delta^{ij}$\\
Retarded (Advanced) & $G_{R(A)}^{ij}\ \equiv\ -\,\Delta_{R(A)}^{ij}$\\
Wightman & $G_{\gtrless}^{ij}\ \equiv i\Delta_{\gtrless}^{ij}$\\
Hermitian (Principal-part) & $G_{h}^{ij}\ \equiv\ -\,\Delta_{\mathcal{P}}^{ij}$\\\hline
\multirow{3}{*}{Self-energies} & $\Pi^{ij}_{\gtrless}\ \longleftrightarrow\ i\Pi^{ij}_{\gtrless}$ \\ & $\Pi_F^{ij}\ \longleftrightarrow\ \frac{1}{2}i\Pi^{ij}_1$\\  &$\Pi^{ij}_{R(A)}\ \longleftrightarrow\ -\,\Pi^{ij}_{R(A)}$ \\\hline
\end{tabular}
\end{center}
\caption{\label{tabcomp}Comparison of the notations for the various two-point functions and self-energies used 
in ref.~\cite{Hohenegger:2014cpa} (lhs) versus refs.~\cite{Dev:2014wsa,Millington:2012pf} 
(rhs).}
\end{table}

\paragraph{CTP Schwinger-Dyson equation.}

The Schwinger-Dyson equation of the CTP formalism may be derived systematically from the 
2PI CJT effective action~\cite{Cornwall:1974vz} (see also refs.~\cite{Norton:1974bm,Jordan:1986ug,
Calzetta:1986cq,Calzetta:1986ey}), which is defined via the Legendre transform ($\hbar=1$)
\begin{equation}
\label{eq:2PI}
\Gamma[\phi,G]\ =\ -\,i\ln\mathcal{Z}[J,K]\:-\:\int_{\mathcal{C}_{[t]},x}J(x)\phi(x)\:-\:\frac{1}{2}\int_{\mathcal{C}_{[t]},x,y}K(x,y[,\tilde{t}])\,\big(\phi(x)\phi(y)\:+\:G_{\mathcal{C}}(x,y[,\tilde{t}])\big)\;,
\end{equation}
where, for simplicity, we consider here the case of single real scalar field $\phi$. The contour 
integral has the explicit form
\begin{equation}
\int_{\mathcal{C}_{[t]},x}\ \equiv\ \int\!\mathrm{d}^3\mathbf{x}\;
\bigg[\int_{-\infty[\tilde{t}_i]+i\epsilon}^{+\infty[\tilde{t}]+i\epsilon}\mathrm{d}x^0\:-\:\int_{-\infty[\tilde{t}_i]-i\epsilon}^{+\infty[\tilde{t}]-i\epsilon}\mathrm{d}x^0\bigg]\;,
\end{equation}
where $\epsilon=0^+$ and the two terms correspond to the two branches of the CTP contour.

In order to build the generating functional $\mathcal{Z}[J,K]$, we start from the partition 
function $\mathcal{Z}\ =\ \mathrm{tr}\,\rho$, which is picture- and, in the absence of external 
sources, time-independent. A path-integral representation of the partition function can be derived 
by perturbing the system with the introduction of an external test source $J$. Note that the 
presence of this external source means that the density operator depends explicitly on time in 
all pictures. Proceeding in the Heisenberg picture, we insert into the partition function complete 
sets of eigenstates of the Heisenberg picture field operator $\ket{\phi(\mathbf{x}),x^0}$ and 
construct the path integral in the standard text-book fashion (see e.g.~ref.~\cite{Millington:2012pf}). 
The density operator gives rise to a term
\begin{equation}
\braket{\phi(\mathbf{x}),x^0-i\epsilon|\rho_H(\tilde{t})|\phi(\mathbf{x}),x^0+i\epsilon}_J\;,
\end{equation}
which we expand in terms of the field eigenvalues and a series of poly-local 
sources~\cite{Calzetta:1986cq,Calzetta:1986ey}:
\begin{equation}
\braket{\phi(\mathbf{x}),x^0-i\epsilon|\rho_H(\tilde{t})|\phi(\mathbf{x}),x^0+i\epsilon}_J\: =\: \exp\big(i\mathcal{K}[\phi[,\tilde{t}]]\big)\: =\: \exp\bigg[i\!\int_{\mathcal{C}_{[\tilde{t}]},x,y}\phi(x)K(x,y,\tilde{t})\phi(y)\bigg]\;,
\end{equation}
where, assuming a Gaussian density operator, we have kept only the bi-local source $K(x,y,[,\tilde{t}])$, 
as appeared in the 2PI effective action [eq.~\eqref{eq:2PI}]. We note that the path integral is a 
$c$-number and, as such, we are free to interpret it in \emph{any} picture.

By varying the 2PI effective action with respect to the resummed CTP propagator $G_{\mathcal{C}}$, 
we obtain the Schwinger-Dyson equation
\begin{equation}
\label{eq:SDint}
G_{\mathcal{C}}^{-1}(x,y[,\tilde{t}])\ = \ D_{\mathcal{C}}^0(x,y)\:+\:K(x,y[,\tilde{t}])\:-\:\Pi_{\mathcal{C}}(x,y[,\tilde{t}])\;,
\end{equation}
where
\begin{equation}
D_{\mathcal{C}}^0(x,y)\ =\ \delta^{4}_{\mathcal{C}}(x,y)\big(-\,\Box_x\:-\:M^2\big)
\end{equation}
is the Klein-Gordon operator and $\Pi_{\mathcal{C}}(x,y[,\tilde{t}])$ 
is the CTP self-energy, whose structure is analogous to that of the CTP propagator. For instance,
when $x^0$ ($y^0$) is on the time-ordered branch and $y^0$ ($x^0$) is on the anti-time-ordered branch, 
$\Pi_{\mathcal{C}}(x,y[,\tilde{t}])$ is equal to $i\Pi_{<(>)}(x,y[,\tilde{t}])$ (with the $i$ conventions detailed above). The contour 
delta function $\delta^{4}_{\mathcal{C}}(x-y)$ coincides with the usual Dirac delta function if 
$x^0$ and $y^0$ lie on the same branch of the CTP contour and is zero otherwise.

The unique inverse of the Klein-Gordon operator is constrained by the Hermiticity properties and 
CPT invariance of the action, as well as unitarity and causality (see ref.~\cite{Millington:2012pf}). 
In addition, one must provide a boundary condition, which corresponds to the EEV of the normal-ordered 
product of fields. In the path-integral representation, the latter is encoded in the bi-local source. 
In this way, the tree-level propagators of the Heisenberg picture encode the initial conditions, 
whereas those of the interaction picture encode the current state of the system. In each case, the 
bi-local source $K(x,y,\tilde{t}_i[\tilde{t}])$ must be proportional to Dirac delta functions that 
lie respectively on the initial and final time surfaces, i.e.
\begin{equation}
K(x,y,\tilde{t}_i[\tilde{t}])\ \propto\ \delta(x_0-\tilde{t}_i[\tilde{t}])\,\delta(y_0-\tilde{t}_i[\tilde{t}])\;.
\end{equation}
It is the differing physical content of the tree-level propagators, which marks the main distinction 
between the Heisenberg- and interaction-picture realizations of non-equilibrium field theory.

\section{\label{sec:syms}Discrete symmetry transformations}

In this appendix, we revisit the properties of the theory in eq.~\eqref{Lagrangian} under 
the discrete symmetry transformations of parity $P$, time-reversal $T$ and charge-conjugation 
$C$. In particular, as identified in ref.~\cite{Hohenegger:2014cpa}, we emphasize the 
relevance of these properties to the specification of $C$-symmetric initial conditions both 
in the Heisenberg- and interaction-picture realizations. The provision of $C$-symmetric 
initial conditions for the two-point functions (in the case of the Heisenberg picture) 
or the number densities (in the case of the interaction picture) ensures that any 
non-zero asymmetry generated in the weak-washout regime arises dynamically and 
vanishes in the $C$-conserving limit of the theory, as it should. The latter 
considerations are, of course, irrelevant in the strong-washout regime, since the 
final asymmetry is independent of the initial conditions.

\paragraph{CPT transformations.}
In the presence of flavour mixing, it is necessary to consider generalized discrete symmetry 
transformations, which, in an arbitrary flavour basis, contain additional transformations in 
flavour space (see e.g.~ref.~\cite{Dev:2014laa}). For the model in eq.~\eqref{Lagrangian}, we have the 
following transformation properties under the generalized parity $P$, time-reversal $T$ and 
charge-conjugation $C$ transformations:
\begin{enumerate}
 \item[a)] \textit{Parity.} Under the linear transformation $P$, the scalar fields transform as
 \cite{Hohenegger:2013zia}
\begin{subequations}
\begin{align}
b(x^0,\mathbf{x})^P\ &=\ \beta_P\,b(x^0,-\mathbf{x})\;,\\
\bar{b}(x^0,\mathbf{x})^P\ &=\ \beta_P^*\,\bar{b}(x^0,-\mathbf{x})\;,\\
\psi_i(x^0,\mathbf{x})^P\ &=\ \pm\,\psi_i(x^0,-\mathbf{x})\;,
\end{align}
\end{subequations}
where the complex phase $\beta_P$ satisfies $|\beta_P|^2=1$.

\item[b)] \textit{Time-reversal.} Under the anti-linear transformation $T$, the scalar fields 
transform as \cite{Hohenegger:2013zia}
\begin{subequations}
\begin{align}
b(x^0,\mathbf{x})^T\ &=\ \beta_T\,b(-x^0,\mathbf{x})\;,\\
\bar{b}(x^0,\mathbf{x})^T\ &=\ \beta_T^*\,\bar{b}(-x^0,\mathbf{x})\;,\\
\label{eq:genT}
\psi_i(x^0,\mathbf{x})^T\ &=\ U_{ji}\,\psi_i(-x^0,\mathbf{x})\;,
\end{align}
\end{subequations}
where the complex phase $\beta_T$ satisfies $|\beta_T|^2=1$ and $\bm{U}$ is an 
orthogonal transformation in flavour space, i.e.~$U_{ik}U_{jk}=U_{ki}U_{kj}=\delta_{ij}$.

\item[c)] \textit{Charge-conjugation.} Under the linear transformation $C$, the scalar 
fields transform as \cite{Hohenegger:2013zia}
\begin{subequations}
\begin{align}
b(x)^C\ &=\ \beta_C\,\bar{b}(x)\;,\\
\bar{b}(x)^C\ &=\ \beta_C^*\,b(x)\;,\\
\label{eq:genC}
\psi_i(x)^C\ &=\ U_{ij}\,\psi_j(x)\;,
\end{align}
\end{subequations}
where the complex phase $\beta_C$ satisfies $|\beta_C|^2=1$. In order for the Lagrangian 
to be invariant under $C\!PT$, the same orthogonal transformation $\bm{U}$ must appear in 
both the generalized $T$ transformation in eq.~\eqref{eq:genT} and the generalized $C$ 
transformation in eq.~\eqref{eq:genC}. This orthogonal transformation $\bm{U}$ may be either a 
rotation or a reflection in flavour space, having the general form
\begin{equation}
U\ =\ \begin{pmatrix} \cos(\alpha) & -\sin(\alpha) \\ \sin(\alpha) & \cos(\alpha) 
\end{pmatrix}\qquad \mathrm{or}\qquad  U\ =\ \begin{pmatrix} \cos(\alpha) & \sin(\alpha) 
\\ \sin(\alpha) & -\cos(\alpha) \end{pmatrix}\;,
\end{equation}
or an arbitrary product of the rotations and reflections (a product of a rotation and a reflection is still a reflection).
\end{enumerate}
The Lagrangian in eq.~\eqref{Lagrangian} is invariant under $C$ so long as we can find a 
phase $\beta_C$ and transformation $\bm{U}$ such that the mass matrix $\bm{M}^2$ and 
Yukawa couplings $\bm{h}$ satisfy
\begin{subequations}
\label{eq:Cinv}
\begin{align}
\label{eq:Cinv1}
U_{mi}M^2_{mn}U_{nj}\ &= \ M^2_{ij}\;,\\
\label{eq:Cinv2}
\beta_C^2\,U_{ki}\,h_k\ &=\ h_i^*\;.
\end{align}
\end{subequations}
In order to analyze the constraint on the Yukawa couplings provided by eq.~\eqref{eq:Cinv2}, 
it is convenient to introduce the dyadic product $H_{ij}\ \equiv \ h_ih_j^*$.
The second condition eq.~\eqref{eq:Cinv2} may then be recast in the more convenient form
\begin{equation}
U_{mi}\,H_{mn}\,U_{nj}\ =\ H_{ij}^*\;,
\end{equation}
in which the phase $\beta_C$ of the complex scalar field has been eliminated.

In the mass eigenbasis, where $\bm{M}^2$ is diagonal, the first condition in
eq.~\eqref{eq:Cinv1} can be satisfied for $M_1^2\neq M_2^2$ only for rotations and 
reflections through angles of $\alpha=0$ or $\pi$. If $\bm{U}$ is a rotation, 
$C$-invariance follows if $H_{12}=H_{12}^*$, i.e.~$\mathrm{Im}\,H_{12}=0$. On 
the other hand, if $\bm{U}$ is a reflection, $C$-invariance follows if 
$H_{12}=-\,H_{12}^*$, i.e.~$\mathrm{Re}\,H_{12}=0$.

Under a general flavour rotation $\bm{O}$ through an angle $\theta$, $\mathrm{Im}\,H_{12}$ is unchanged, since $\bm{O}$
is orthogonal. Instead, $\mathrm{Re}\,H_{12}$ transforms as
\begin{equation}
\mathrm{Re}\,H_{12}\ \longrightarrow\ \mathrm{Re}\,H_{12}'\ =
\ \cos(2\theta)\,\mathrm{Re}\,H_{12}\:+\:\frac{1}{2}\,\sin(2\theta)\,(H_{11}-H_{22})\;.
\end{equation}
Therefore, in the degenerate limit $M_1^2=M_2^2$, the resulting $O(2)$ invariance 
of the free theory (see e.g.~ref.~\cite{Dev:2014laa}) means that we may always rotate 
through an angle
\begin{equation}
\theta\ =\ \frac{1}{2}\,\mathrm{arctan}\,\frac{H_{12}+H_{21}}{H_{22}-H_{11}}
\end{equation}
to a basis in which $\mathrm{Re}\,H_{12}=0$. Thus, the Lagrangian is also \C-conserving
in this case.

The above observations may be conveniently summarized by forming the Jarlskog invariant
\begin{equation}
J\ =\ 2\,\mathrm{Im}\,H_{12}\,\mathrm{Re}\,H_{12}\,M_1\,M_2\,\Delta M^2\;,
\end{equation}
which vanishes when any of the $C$-conserving conditions are satisfied, 
viz.~$\mathrm{Re}\,H_{12}=0$, $\mathrm{Im}\,H_{12}=0$ or $M_1=M_2$. Rotating 
to a general flavour basis, the Jarlskog invariant may be written in the form
\begin{equation}
J\ =\ \mathrm{Im}\,\Tr\,\bm{H}\bm{M}^3\bm{H}^{\mathsf{T}}\bm{M}\;,
\end{equation}
providing the familiar basis-independent measure of both $C$- and $CP$-violation.

As identified above, in the case that $\mathrm{Re}\,H_{12}=0$, the Lagrangian is 
invariant under $C$ transformations that include a \emph{reflection} in flavour 
space. In the mass eigenbasis, the orthogonal transformation $\bm{U}$ is necessarily 
diagonal. The permitted reflections are therefore about 
angles of $0$ and $\pi$, and we see that the two flavours must transform with 
\emph{opposite} phases under $C$, i.e.
\begin{equation}
\label{Reflections}
\psi_1(x)^C\ = \ \pm\,\psi_1(x)\;,\qquad \psi_2(x)^C\ =\ \mp\,\psi_2(x)\;,
\end{equation}
and, correspondingly, \emph{opposite} phases under $T$.
On the other hand, in the case that $\mathrm{Im}\,H_{12}=0$, the Lagrangian is 
invariant under $C$ transformations that include a \emph{rotation} in flavour 
space. In the mass eigenbasis, the orthogonal transformation $\bm{U}$ is now 
necessarily isotropic.  The permitted rotations are about angles of $0$ and $\pi$, 
and we see that the two flavours must transform with \emph{equal} phases under $C$, i.e.
\begin{equation}
\label{Rotations}
\psi_1(x)^C\ = \ \pm\,\psi_1(x)\;,\qquad \psi_2(x)^C\ =\ \pm\,\psi_2(x)\;,
\end{equation}
and, correspondingly, \emph{equal} phases under $T$. As identified in 
ref.~\cite{Hohenegger:2014cpa}, the import of this observation is significant for 
the specification of $C$-symmetric initial conditions.

\paragraph{Heisenberg picture.} In order to derive properties of  the Wightman 
propagators under \C-conjugation, we use their definition in the form 
\begin{subequations}
\begin{align}
G^{ij}_>(x,y)&=\Tr[\rho\,\psi^i(x)\psi^j(y)]\,,\\
G^{ij}_<(x,y)&=\Tr[\rho\,\psi^j(y)\psi^i(x)]\,,
\end{align}
\end{subequations}
where $\rho$ is the density matrix. We \emph{define} charge-conjugated propagators
as propagators with only the fields, but not the density matrix, conjugated. This definition
corresponds to the intuitive definition  that replaces particles with 
antiparticles. The Wightman propagators then transform under generalized 
$C$-conjugation as 
\begin{equation}
G^{ij}_{\gtrless}(x,y)\ \xrightarrow{C}\ U_{im}\,G^{mn}_{\gtrless}(x,y)\,U_{jn}\;.
\end{equation}
For the case of flavour rotations with $\alpha=0$ or $\pi$, we may readily verify that the 
propagators are automatically $C$-symmetric. This is consistent with the fact that if 
$\Im H_{12}=0$ then no asymmetry can be produced irrespective of the value of the propagators 
at the initial time surface. On the other hand, for the case of flavour reflections  with $\alpha=0$ 
or $\pi$, the propagators
are \C-symmetric only if their off-diagonals vanish at the initial time surface. As we demonstrate 
in section~\ref{sec:pheno}, the requirement of vanishing off-diagonals makes the produced asymmetry 
proportional to $\Re H_{12}$. In other words, once we impose \C-symmetric initial 
conditions on the propagators, the produced asymmetry automatically vanishes if the Lagrangian
is \C-symmetric, as one would expect.

\paragraph{Interaction picture.} In the interaction picture, we may begin by fixing the 
transformation properties of the free field operators under generalized discrete symmetry 
transformations in Fock space directly~\cite{Dev:2014laa}. The matrix of number densities
is defined by
\begin{align}
\label{ndef}
n^{ij}=\langle a^\dagger_j a_i \rangle= \Tr(\rho\,a^\dagger_j a_i)\,.
\end{align}
Similarly to the Heisenberg picture, we \emph{define} \C-transformation such
that it transforms the creation and annihilation operators but not the density 
matrix. 

As follows from eq.~\eqref{Rotations}, under rotations $a_1\xrightarrow{C} \pm a_1$ and 
$a_2\xrightarrow{C} \pm a_2$. The additional phase cancels in eq.~\eqref{ndef} and therefore
the matrix of number densities is automatically \C-symmetric. This is in agreement with 
the observation that once $\Im H_{12}=0$ no asymmetry can be generated 
irrespective of the choice of the initial conditions. On the other hand, for reflections
$a_1\xrightarrow{C} \pm a_1$ and $a_2\xrightarrow{C} \mp a_2$ such that off-diagonals of $n_{ij}$
acquire a relative sign. The condition of \C-invariance therefore implies that the 
matrix of number densities must be diagonal at the initial time surface. 

\section{\label{sec:rate}Rate equations in the radiation-dominated universe} 

Full treatment of leptogenesis would require solving the kinetic equations for all momentum 
modes. However, such an analysis is technically demanding and has been performed only in a 
handful of works (see e.g.~ref.~\cite{HahnWoernle:2009qn}). Instead, one usually 
assumes kinetic equilibration of the mixing fields and approximates kinetic equations for 
the distribution functions by so-called rate equations for the corresponding number densities.
In this appendix, we derive the rate equations in the radiation-dominated universe and 
re-derive the strong-washout approximation formulas presented in ref.~\cite{Garbrecht:2014aga}.

\paragraph{Source and washout terms.} Decays $\psi_i\rightarrow bb$  as well as inverse decays 
$\bar b\bar b \rightarrow \psi_i$ increase the asymmetry by two units, wheres decays 
$\psi_i\rightarrow \bar b \bar b$  and inverse decays $b b \rightarrow \psi_i$ decrease the 
asymmetry by two units. Using the results of appendix~E in ref.~\cite{Garny:2009qn}, we obtain, 
in agreement with this physical picture, the time-derivative of the asymmetry
\begin{align}
\label{detadt}
\frac{\D \eta}{\D t}\ &=\ \int \D\Pi^3_\vec{q}\, \D\Pi^3_\vec{p}\, \D\Pi^3_\vec{k}\;(2\pi)^4\delta^{4}(k-p-q)\nonumber\\
&\times
\ \Bigl\{H_{ij}\Bigl([1+n^{ij}(k)]n_{\bar b}(p)n_{\bar b}(q)-n^{ij}(k)[1+n_{\bar b}(p)][1+n_{\bar b}(q)]\Bigr)\nonumber\\
&-\ H_{ij}^*\Bigl([1+n^{ij}(k)]n_b(p)n_b(q)-n^{ij}(k)[1+n_b(p)][1+n_b(q)]\Bigr)\Bigr\}\,,
\end{align}
where $\D\Pi^3_{\vec q}\equiv \frac{\D^3\mathbf{q}}{(2\pi)^3}\frac{1}{2E}$ is the Lorentz-invariant phase-space element. One can easily recognize the usual structure of the gain and loss terms in 
eq.~\eqref{detadt}. 

It is common to approximate the rhs of eq.~\eqref{detadt} by the difference of the source and washout
terms. The source term is defined as the rhs with distribution functions of the complex field set 
to equilibrium, $n_b=n_{\bar b}=n_{\rm eq}$. By detailed balance, the contribution of the equilibrium 
part of $n^{ij}$ vanishes, and we are left with the source term
\begin{align}
S\ =\ 4\,\Im H_{12}\int \D\Pi^3_{\mathbf{k}} \, \widetilde{\Pi}_\rho(k)\,\Im\, \delta n^{12}(k)\,,
\end{align}
where the function $\widetilde{\Pi}_\rho$ is defined in eq.~\eqref{PirhoDef}. The assumption of kinetic equilibrium 
amounts to 
\begin{align}
\label{KineticEq}
\delta n^{12}(k)\ \approx\ \frac{\delta N^{12}}{N_{\rm eq}}\,n_{\rm eq}(k)\,,
\end{align}
where $N$ denotes the total particle number density. It is furthermore common to approximate $n_{\rm eq}$ by the Boltzmann distribution. Substituting 
eq.~\eqref{KineticEq} into eq.~\eqref{SourceTerm} and approximating $\widetilde{\Pi}_\rho$ by its 
low-temperature limit, we then find 
\begin{align}
S\ =\ \frac{\Im H_{12}}{4\pi \bar M}\,\frac{K_1(\bar M/T)}{K_2(\bar M/T)}\,\Im\, \delta N^{12}\,,
\end{align}
where $K_1$ and $K_2$ are modified Bessel functions of the second kind.

The washout term is defined by setting the density matrix of the mixing fields to its equilibrium 
form, $\bm{n}=\bm{1}\cdot n_{\rm eq}\,$. After some straightforward algebra we find 
\begin{align}
W\ &=\ \sum_i H_{ii} \int \D\Pi^3_\vec{q}\, \D\Pi^3_\vec{p}\, \D\Pi^3_\vec{k}\;(2\pi)^4\delta^4(k-p-q)\nonumber\\
&\times\ [n_b(p)-n_{\bar b}(p)][n_b(p)+n_{\bar b}(p)-n_{\rm eq}(k)]\,.
\end{align}
Here, we again use Boltzmann statistics for the distribution functions, $n_b\approx\exp(-(E-\mu)/T)$ and 
$n_{\bar b}\approx\exp(-(E+\mu)/T)$, where $\mu$ is the chemical potential. Neglecting further the 
(quantum-statistical) term $n_{\rm eq}(k)$ and expanding to the first order in $\mu$, we obtain
\begin{align} 
W\ \approx\ -\,2 \eta\, \bar{\Gamma}\, \frac{\bar M^2}{T^2}K_1(\bar M/T)\,.
\end{align}

\paragraph{Generalization to radiation-dominated universe.}

In the expanding universe, one can recast the kinetic equation for the asymmetry in the form similar to 
eq.~\eqref{detadt} by using the co-moving number densities \cite{Kartavtsev:2008fp}:
\begin{align} 
\frac1{a}\,\frac{\D\eta}{\D t}\ =\ \frac{\Im H_{12}}{4\pi \bar M}\frac{K_1(\bar M/T)}{K_2(\bar M/T)}\,\Im\, \delta N^{12}\:
-\:2 \eta\, \bar{\Gamma}\, \frac{\bar M^2}{T^2}K_1(\bar M/T)\,,
\end{align}
where $t$ is now the conformal time, $a$ is the scale factor, and $\eta$ and $\delta N^{12}$ are the 
comoving number densities. In the radiation-dominated universe $a(t)=a_R t\,$, where $a_R$ is constant. 
Following ref.~\cite{Garbrecht:2011aw}, we choose 
$a_R= M_{Pl}\,(45/4\pi^3 g_*)^\frac12$
with $M_{Pl}$ being the Planck mass, and $g_*$ the effective number of massless degrees of freedom. 
For this choice, $z\equiv \bar M/T=\bar M t$. The resulting rate equation for the asymmetry reads 
\begin{align}
\label{etaRateEq}
\frac{\D\eta}{\D z}\ =\ \frac{a_R z}{\bar M^2}\left[
\frac{\Im H_{12}}{4\pi \bar M}\frac{K_1(z)}{K_2(z)}\,\Im\, \delta N^{12}\:
-\: 2 \eta\, \bar{\Gamma}\, z^2 K_1(z)
\right]\,.
\end{align}

\paragraph{Strong-washout approximation.}
It follows from eq.~\eqref{etaRateEq} that the larger the washout parameter
\begin{align}
\kappa\ \equiv\ \frac{a_R \bar \Gamma}{\bar M^2}\,,
\end{align}
the more asymmetry is washed out by the inverse decay processes.
If $\kappa\gg 1$, then one speaks of the strong-washout regime. 
In the strong-washout regime, the final 
asymmetry does not depend on the initial conditions and most of the asymmetry is produced 
after the temperature drops below the mass of the heavy decaying particle (see e.g.~ref.~\cite{Buchmuller:2004nz}). For $T< \bar M$, the 
density matrix $\bm n$ is suppressed at momenta $|\vec q|> \bar M$, such that, in eq.~\eqref{DensMatrixEqs}, 
we can approximate $\bar\omega$  by $\bar M$. Integrating eq.~\eqref{DensMatrixEqs} over the 
phase space and switching to the co-moving number densities, we obtain in this approximation 
\begin{align}
\label{dndz}
\frac{\D \delta \bm{N}}{\D z}\:+\:\bm{1}\,\frac{\D N_{\rm eq}}{\D z}\:+\:i\frac{a_Rz}{\bar{M}^3}[\bm{M}^2,\delta \bm{N}]
\ =\ \frac{a_Rz}{\bar{M}^3}\{\widetilde{\bm{\Pi}}_\rho,\delta \bm{N}\}
\end{align}
(see refs.~\cite{Garbrecht:2011aw,Garbrecht:2014aga} for the details of the derivation). In 
ref.~\cite{Iso:2014afa}, it was proposed that, in a strong-washout regime, one can obtain an 
approximate solution for $\delta \bm{N}$ at late times by neglecting the derivative of 
$\delta \bm{N}$. This solution reads 
\begin{align}
\delta N^{12}\  \approx\ \frac{\Re H_{12}\Tr\bm{H} (2\bar M \bar \Gamma-
i\Delta M^2_{12})\,{\bar M}^2}{H_{11}H_{22}(\Delta M^2_{12})^2+16{\bar M}^2{\bar \Gamma}^2
\det \Re \bm{H}} \frac{\bar M}{a_R z} \frac{\D N_{\rm eq}}{\D z} 
\ \approx \
\frac{-i}{8\pi}\frac{\Re H_{12}}{\Delta M^2_{12}}\frac{{\bar\Gamma}^2}{\Gamma_1\Gamma_2}
\frac{1}{\kappa z}\frac{\D N_{\rm eq}}{\D z}\,,
\end{align}
where the second approximate equality is valid for $|\Delta M^2_{12}|\gg \bar M\bar\Gamma$. It 
is interesting to note that the solution is proportional to $\Re H_{12}$. This implies that 
the resulting source term automatically vanishes if either of the \C-symmetry conditions,
$\Im H_{12}=0$ or $\Re H_{12}=0$, is fulfilled.

\end{appendix}


\providecommand{\href}[2]{#2}\begingroup\raggedright

\end{document}